\documentclass[prx,%
 reprint,
superscriptaddress,
 amsmath,amssymb,
 aps,
floatfix,
]{revtex4-2}

\usepackage{hyperref}
\usepackage{graphicx}% Include figure files
\usepackage{xcolor}
\usepackage{dcolumn}% Align table columns on decimal point
\usepackage{bm}% bold math
\usepackage{physics}
\usepackage{tkz-euclide}
\usetikzlibrary{shapes.misc}
\usetikzlibrary{decorations.pathmorphing,decorations.pathreplacing}
\usepackage[caption=false]{subfig}
\usepackage{slashed}
\begin{document}

\preprint{APS/123-QED}

\title{Theory of Three-Photon Transport Through a Weakly Coupled Atomic Ensemble}% Force line breaks with \\
%\thanks{A footnote to the article title}%

\author{YangMing Wang}
 \email{ywan8652@uni.sydney.edu.au}
\affiliation{
 ARC Centre of Excellence for Engineered Quantum Systems, \\School of Physics, The University of Sydney, Sydney, NSW 2006, Australia
}
\affiliation{Sydney Quantum Academy, Sydney, NSW, Australia}

\author{No\'e Demazure}
\affiliation{
 ARC Centre of Excellence for Engineered Quantum Systems, \\School of Physics, The University of Sydney, Sydney, NSW 2006, Australia
}
\affiliation{
Department of Physics, ENS Paris-Saclay, University Paris-Saclay, 91190 Gif-sur-Yvette, France
}

\author{Sahand Mahmoodian}
 \email{sahand.mahmoodian@sydney.edu.au}
\affiliation{
 ARC Centre of Excellence for Engineered Quantum Systems, \\School of Physics, The University of Sydney, Sydney, NSW 2006, Australia
}
\affiliation{Institute for Photonics and Optical Sciences (IPOS), School of Physics, The University of Sydney, NSW 2006, Australia}

%\collaboration{CLEO Collaboration}%\noaffiliation

\date{\today}% It is always \today, today,
             %  but any date may be explicitly specified

\begin{abstract}
Understanding multi‐photon interactions in non-equilibrium quantum systems is an outstanding challenge in quantum optics. In this work, we develop an analytical and diagrammatic framework to explore three‐photon interactions in atomic ensembles weakly coupled to a one-dimensional waveguide, and complement the analysis initiated in our accompanying work~[Y. Wang and S. Mahmoodian,\cite{WM2025}]. Taking advantage of the weak coupling, we use our diagrammatic framework to perform perturbation theory and calculate the leading-order contributions to the three-photon wavefunction, which would otherwise be intractable. We then compute the outgoing photon wavefunction of a resonantly driven atomic ensemble, with photon-photon interactions truncated up to three photons. Our formulation not only captures the individual transmission of photons but also isolates the connected $S$-matrix elements that embody genuine photon-photon correlations. Through detailed analysis, we obtain the analytic expressions of the connected third-order correlation function and the third-order electric-field-quadrature cumulant, which reveal non-Gaussian signatures emerging from the interplay of two- and three-photon processes. We also calculate the optical depth where non-Gaussian photon states can be observed. Numerical simulations based on a cascaded master equation validate our analytical predictions on a small-scale system. These results provide a formalism to further explore non-equilibrium quantum optics in atomic ensembles and extend this to the regime of non-Gaussian photon transport.
%\begin{description}
%\item[Structure]
%You may use the \texttt{description} environment to structure your abstract;
%use the optional argument of the \verb+\item+ command to give the category of each item. 
%\end{description}
\end{abstract}

%\keywords{Suggested keywords}%Use showkeys class option if keyword
                              %display desired
\maketitle

%\tableofcontents

\section{Introduction}

Developing theoretical descriptions of non-equilibrium quantum systems is one of the main challenges in physics~\cite{Ferioli2021,Strack2013,Eisert2014}. In many-body quantum transport, the interaction between the quantum particles and the effect of the transport medium on the particles creates entanglement among the particle states. In quantum optics, one prominent transport medium is an ensemble of two-level atoms coupled to a one-dimensional continuum of optical modes~\cite{Johnson2019,Pennetta2022}, where the atoms function as localized scatterers that induce strong nonlinearities between  propagating photons. The collective coupling of an atomic ensemble to the single electromagnetic mode introduces intricate entanglement in the photon states. 

Significant theoretical effort has been put into studying the interplay between photons and atomic ensembles, revealing rich physics emerging from light–matter interactions in one dimension ~\cite{Forero2025,Sheremet2023,Das2018}. Early work introduced exact $S$-matrix approaches to capture nontrivial photon–photon correlations mediated by single or few atoms in systems with ideal one-dimensional coupling \cite{Rupasov1984,ShenFan2007,Xu2015,Yudson1988,Yudson1983}. Subsequent studies expanded these ideas by applying master-equation and input–output formalisms to multi-atom systems~\cite{Caneva2015}, thereby elucidating collective phenomena such as superradiance, subradiance, and selective radiance~\cite{Garcia2017}—even in the presence of imperfect coupling and thus significant photon loss. More recent investigations have incorporated techniques like  mean-field theory~\cite{Agarwal2024}, Green’s function methods~\cite{Lang2020,Pletyukhov2012,Poshakinskiy2016}, and numerical simulation~\cite{Manzoni2017} to address disorder and many-body localization~\cite{Fayard2021}, and demonstrated that photon loss can play a constructive role in generating correlated quantum states~\cite{Chang2014}. Very recent studies investigated the superradiant phase transition in driven-dissipative atomic ensembles in free space~\cite{Agarwal2024,Ruostekoski2025}, as well as a phase separation between saturated and unsaturated regions in atomic ensembles, both in free space and in cavity settings~\cite{Kusmierek2024,Chang2025}.

Photon transport in atomic ensembles typically features many atoms weakly coupled to an optical mode. Although initially investigated for their linear optical response~\cite{Hammerer2010}, ensembles of moderate optical depth have a nonlinear response that can modify the photon statistics of transmitted light~\cite{Mahmoodian2018,Prasad2020}. A very recent experiment claimed to observe stable non-Gaussian correlations in the steady-state light emitted by a driven-dissipative dense atomic ensemble~\cite{Ferioli2024}. Here, a state is defined as \textit{Gaussian} if its Wigner function is a multivariate Gaussian, which is equivalent to saying that its expectation value of arbitrary product of creation/annihilation operators should satisfy Isserlis's or Wick's theorem~\cite{Shi2018,Cardin2024}. Generally speaking, a theoretical formalism describing non-Gaussian photon correlations in dilute or dense atomic ensembles is lacking. While developing a full theoretical description of 3D dense atomic ensembles remains challenging, as a first step, the physics of dilute ensembles can be modeled using the Maxwell–Bloch equations~\cite{Chang2025,McCall1967,Hammerer2010,Gross1982}---a minimal model featuring an ensemble as an array of atoms unidirectionally coupled to a one-dimensional continuum of photon modes. Moreover, such an effective unidirectional coupling model can also be achieved in a 1D atomic chain with bidirectional coupling with photon modes and large position disorder~\cite{Kusmierek2024}. To study non-Gaussian correlations in such systems, one could equivalently employ the complete cascaded master equation. However, for large atom numbers, computing higher-order correlation functions using the cascaded master equation—even under improved mean-field theory based on cumulant expansions~\cite{Mahmoodian2023}— can become computationally intractable. 

In this work, we adopt a scattering-theory approach to analyze a simplified model consisting of a weakly driven, dilute ensemble of two-level atoms chirally coupled to a waveguide. This simplified model can be an effective description of ensembles as trapped atomic clouds~\cite{Ferioli2021,Araujo2016,Glicenstein2020}, atoms coupled to a nanofibre~\cite{Nieddu2016,Vetsch2010,Prasad2020, Cordier2023}, or Mossbauer x-ray systems with weak coupling~\cite{Lohse2025, Salditt2020}. This framework enables us to investigate the emergence of non-Gaussianity in the steady-state output of such systems beyond the limitations of master-equation-based methods. Starting from the exact $n$-photon single-atom $S$-matrix, we isolate its connected components, which capture genuine two- and three-photon interaction effects. In the weak-coupling regime, these atom-mediated interactions can be treated perturbatively on top of individual elastic photon propagation. We construct the full $S$-matrix for the atomic ensemble via a diagrammatic expansion, order by order in the coupling strength. This formalism yields analytic expressions for:
\begin{enumerate}
    \item The outgoing three-photon wavefunction. 
    \item The connected third-order photon correlation function $g^{(3)}_c$.
    \item The third-order electric-field quadrature cumulant $\langle{:} \Delta  \hat{X}_\theta(x_1) \Delta  \hat{X}_\theta(x_2) \Delta  \hat{X}_\theta(x_3) {:}\rangle$.
\end{enumerate}
When $g^{(3)}_c$ is non-vanishing, it indicates the presence of non-Gaussian intensity correlation in the scattered light. On the other hand, if a state of light is Gaussian at the level of its electric field, then its correlation functions should satisfy Isserlis/Wick's theorem~\cite{Shi2018,Hackl2021}. The violation of the theorem by any observable of the outgoing state indicates non-Gaussian field correlations. Our calculations show that at sufficiently large optical depth (OD), the outgoing photons exhibit regions of both non-Gaussian correlations and anti-correlations in time. To assess the experimental observability of these non-Gaussian features, we identify a parameter regime in which the signal strength is appreciable and the relative error of the perturbative theory remains small.

This paper is outlined as follows: In Sec.~\ref{sec:model}, we introduce the model Hamiltonian and explain how the Weyl transformation allows photon loss to be incorporated into the scattering framework. In Sec.~\ref{sec:analy meth}, we revisit the Yudson--Bethe-Ansatz representation techniques and derive the connected $S$-matrix elements for single-atom scattering. We also introduce a diagrammatic representation for each connected component of the $S$-matrix. In Sec.~\ref{subsec:two-photon}, we show that the photon wavefunction for two interacting photons can be obtained recursively. However, due to the increased complexity, this recursive approach is not applicable in the three-photon case. In Sec.~\ref{subsec:diag exp}, we instead demonstrate that the diagrams representing connected $S$-matrices  can be concatenated together to construct a perturbative description of three-photon transport through the entire ensemble. In Sec.~\ref{sec:results}, we apply this analytic method to compute $g_c^{(3)}(x_1,x_2,x_3)$ and $\langle{:}\Delta  \hat{X}_\theta(x_1)\Delta  \hat{X}_\theta(x_2)\Delta  \hat{X}_\theta(x_3){:}\rangle$. We also estimate the relative error of our method and evaluate the expected experimental signal strength. 

\section{The Model} \label{sec:model}

In this section, we introduce the model describing chiral photon transport through a dissipative atomic array coupled to a nanofibre. We show that the unidirectional nature of the coupling allows the many-atom scattering problem to be decomposed into a sequence of single-atom, two-channel scattering events, and we then show how the use of a Weyl transformation enables a mapping of each scattering event onto an effective single-channel model. This sets the stage for the analytic treatment of photon transport in terms of exact few-photon $S$-matrices derived from Yudson's representation, which we develop in the next section.

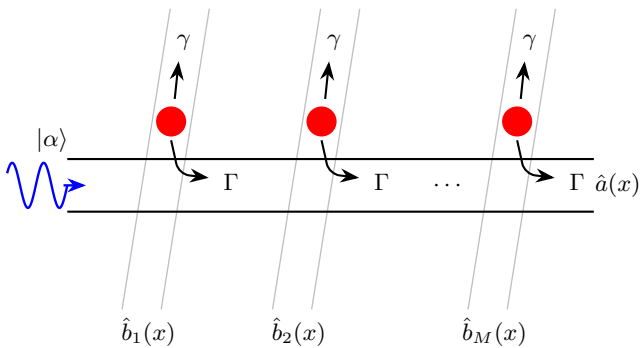
\begin{figure}[t!]
    \centering
    \begin{tikzpicture}[>=Stealth]
  % Define constants for spacing and dimensions
  \def\xdist{2cm} % Horizontal distance between atoms
  \def\dotsize{0.2cm} % Size of the quantum atoms (red dots)
  \def\linewidth{0.8pt} % Width of the waveguide lines
  \def\arrowsize{0.25cm} % Size of the arrow heads
  
  % Calculate total width (for 3 atoms plus ellipsis)
  \def\totalwidth{3*\xdist + 1cm}
  
  % Draw the diagonal gray lines (field modes)
  \foreach \i in {1,2} {
    \draw[lightgray, line width=0.5pt] (\i*\xdist -0.95\xdist - 0.3\xdist, -2cm) -- 
                                  (\i*\xdist -0.95\xdist + 0.3\xdist, 2cm);
    \draw[lightgray, line width=0.5pt] (\i*\xdist  - 0.43\xdist-0.3\xdist, -2cm) -- (\i*\xdist - 0.43\xdist+0.3\xdist, 2cm);                              
  }

\draw[lightgray, line width=0.5pt] (3.3*\xdist -0.95\xdist - 0.3\xdist, -2cm) -- (3.3*\xdist -0.95\xdist + 0.3\xdist, 2cm);
\draw[lightgray, line width=0.5pt] (3.3*\xdist  - 0.43\xdist-0.3\xdist, -2cm) -- (3.3*\xdist - 0.43\xdist+0.3\xdist, 2cm);     
  
  % Draw the horizontal waveguide lines
  \draw[line width=\linewidth] (0, 0) -- (\totalwidth, 0);
  \draw[line width=\linewidth] (0, -0.7cm) -- (\totalwidth, -0.7cm);
  
  % Draw the red dots (quantum atoms)
  \foreach \i in {1,2} {
    \fill[red] (\i*\xdist-0.6\xdist, 0.5cm) circle (\dotsize);
  }

  \fill[red] (3.3*\xdist-0.6\xdist, 0.5cm) circle (\dotsize);
  
  % Draw the gamma arrows (upward, free space coupling)
  \foreach \i in {1,2} {
    \draw[-{Stealth[length=\arrowsize]}, line width=0.8pt] 
        (\i*\xdist - 0.57\xdist, 0.8cm) -- (\i*\xdist - 0.57\xdist+0.05\xdist, 1.3cm);
    \node at (\i*\xdist - 0.45\xdist, 1.55cm) {$\gamma$};
  }

  \draw[-{Stealth[length=\arrowsize]}, line width=0.8pt] 
        (3.3*\xdist - 0.57\xdist, 0.8cm) -- (3.3*\xdist - 0.57\xdist+0.05\xdist, 1.3cm);
    \node at (3.3*\xdist - 0.45\xdist, 1.55cm) {$\gamma$};
  
  % Draw the Gamma arrows (downward into waveguide, waveguide coupling)
  \foreach \i in {1,2} {
    \draw[-{Stealth[length=\arrowsize]}, line width=0.8pt, rounded corners] 
        (\i*\xdist - 0.6\xdist, 0.25cm) -- (\i*\xdist - 0.6\xdist + 0.1cm, -0.2cm) -- 
        (\i*\xdist - 0.6\xdist + 0.5cm, -0.25cm);
    \node at (\i*\xdist - 0.3\xdist + 0.5cm, -0.3cm) {$\Gamma$};
  }

  \node at (3.3*\xdist - 0.3\xdist + 0.5cm, -0.3cm) {$\Gamma$};

  \draw[-{Stealth[length=\arrowsize]}, line width=0.8pt, rounded corners] 
        (3.3*\xdist - 0.6\xdist, 0.25cm) -- (3.3*\xdist - 0.6\xdist + 0.1cm, -0.2cm) -- 
        (3.3*\xdist - 0.6\xdist + 0.5cm, -0.25cm);
  
  % Add the field mode labels at the bottom
  \foreach \i in {1,2} {
    \node at (\i*\xdist - 0.9\xdist, -2.3cm) {$\hat{b}_{\i}(x)$};
  }
  \node at (3.3*\xdist - 0.9\xdist, -2.3cm) {$\hat{b}_{M}(x)$};
  
  % Add the waveguide mode label on the left
  \node at (\totalwidth+0.3\xdist, -0.35cm) {$\hat{a}(x)$};
  
  % Add the ellipsis in the middle
  \node at (\totalwidth -1.9\xdist, -0.35cm) {$\cdots$};

   % Add sine wave on the left side of the waveguide
  \draw[blue, line width=0.9pt] plot[domain=-0.8:0, samples=100] 
    (\x, {0.3cm*sin(15*\x r) - 0.35cm});
  
  % Add an arrow at the end of the sine wave connecting to waveguide
  \draw[-{Stealth[length=\arrowsize]}, blue, line width=0.9pt] 
    (-0.05, -0.35cm) -- (0.25, -0.35cm);

  \node at (-0.2, 0.25cm) {$\ket{\alpha}$};

    \end{tikzpicture}
    % \captionsetup{justification=justified, singlelinecheck=false}
    \caption{\label{fig: sketch of system} An array of $M$ chirally coupled two-level atoms (depicted as red circles) driven by an external coherent field $\ket{\alpha}$ producing a strongly correlated output photon state $|\text{out}\rangle$. Each atom couples to the waveguide (black line) with a decay rate $\Gamma = \beta\Gamma_{\text{tot}}$ and to its own external loss channel with a decay rate $(1-\beta)\Gamma_{\text{tot}}$ (gray line). Without loss of generality, for theoretical convenience we model each loss channel as an auxiliary waveguide.}
    
\end{figure}

Our model considers unidirectional photon transport through an ensemble of $M$ atoms coupled to a waveguide.  The special case of a nanofibre geometry is depicted in Fig.~\ref{fig: sketch of system}, where the atoms are trapped near the fibre such that they couple to its optical mode via the evanescent field. Each atom weakly couples to the propagating channel inside the waveguide with rate $\Gamma=\beta \Gamma_{\rm tot}$, and couples to its own loss channel with rate $\gamma=(1-\beta)\Gamma_{\rm tot}$, where $\Gamma_{\rm tot}$ is the overall decay rate of an atom. Here, $\beta$ is a dimensionless coupling constant quantifying the relative coupling strength between a single atom and the waveguide. Weak coupling implies that $\beta \ll 1$. In typical experimental setups with atomic ensembles, obtaining large couplings is challenging, and  atoms coupled to nanofibres~\cite{Corzo2019,Beguin2014,Vetsch2010,LeKien2004,Prasad2020} typically have $\beta \sim 10^{-2}$. Even though $\beta$ is small, the optical depth $\rm{OD}=4\beta M$ of the entire atomic array is significant ($\sim \mathcal{O}(1)$). The spacing between atoms is greater or equal to one wavelength of the driving laser, so the collective emission via unguided optical modes outside the waveguide is negligible~\cite{Garcia2017}. 

We are interested in the output state of light after the system is driven to steady state at low saturation. Due to the unidirectional property of photon transport in chiral waveguides, the calculation of the scattering problem in a lossy atomic array of $M$ atoms can be converted into $M$ iterations of a two-channel single-atom scattering problem. Specifically, the output state of the scattering event on the $m$th atom is the input state of the $(m+1)$th atom, as illustrated in Fig.~\ref{fig: sketch of system}. The dynamics of the previous atoms in the array are not influenced by the dynamics of successive atoms. In systems with unidirectional coupling, when computing the steady-state properties with a continuous drive, the time delay between atoms is arbitrary and does not play a role in photonic observables ~\cite{Gardiner1993}. In the context of scattering theory, this means that we can set the distance $x_m-x_{m+1}$ between each pair of atoms to be large such that the input and output state of the individual scattering events on the atom $n$ are considered at asymptotic time $t=-\infty$ and $t=\infty$, respectively. The propagation phase between atoms, in the Markovian limit, reduces to a global phase factor and thus has no influence on the dynamics~\cite{Gardiner1993, Pichler2015PRA, Lodahl2017}. Moreover, in each iteration of the single-atom scattering problem, the atom’s position can be freely chosen due to the unidirectional and Markovian nature of the transport. Without loss of generality, we set the position of the atom under study to $x=0$.

The scattering event at the $m$th atom can be locally described by the following Hamiltonian:
\begin{equation} \label{eq:model ham}
  \begin{aligned}
    \hat{H}_m &= \int_{-\infty}^{\infty} d x \Biggl\{ \,
    \hat{a}^\dagger(x) (-i\partial_x) \hat{a}(x)  \\
    &+  \hat{b}_m^\dagger(x)(-i\partial_x) \hat{b}_m(x) \\
    &+ \delta(x)\left[ \hat{\sigma}_m^+ (\sqrt{\Gamma}\hat{a}(x)+\sqrt{\gamma}\hat{b}_m(x)) + \rm{h.c.}\right] \Biggr\},
  \end{aligned}
\end{equation}
where $\hat{a}^\dagger(x)$ creates a photon in the waveguide at position $x$, $\hat{b}_m^\dagger(x)$ creates a photon in the loss channel of the $m$th atom, and $\hat{\sigma}_m^\pm$ are Pauli operators for the $m$th two-level atom. ``h.c." denotes the Hermitian conjugate. Here, we use natural units with $\hbar=v_g=1$ , where $v_g$ is the group velocity of the waveguide. The operators $\hat{a}(x)$, $\hat{b}_m(x)$ and $\hat{\sigma}_m^\pm$ obey the following commutation relations:
\begin{align*}
    [\hat{a}(x),\hat{a}^\dagger(x')]=\delta(x-x')&,\quad [\hat{\sigma}_m^+,\hat{\sigma}_n^-]=\delta_{mn}\hat{\sigma}_m^z  \\
    [\hat{b}_m(x),\hat{b}_n^\dagger(x')]&=\delta_{mn}\delta(x-x'), \\
\end{align*}
where $\delta(x)$ is Dirac delta function and $\delta_{mn}$ is Kronecker delta function. The Hamiltonian (\ref{eq:model ham}) commutes with the excitation number operator,
\begin{equation*}
  \hat{N}_m = \int_{-\infty}^{\infty} d x \left[ \hat{a}^\dagger(x) \hat{a}(x) + \hat{b}_m^\dagger(x) \hat{b}_m(x) + \frac{1}{2}(\hat{\sigma}_m^z+1) \right],
\end{equation*}
indicating that different excitation-number manifolds decouple. A practical method for analyzing the two-channel single-atom scattering problem at the $m$th atom involves transforming it into an effective one-channel problem using the Weyl basis~\cite{Yuds2008,Culver2021,ShenFan2007}. In the prior studies of non-chiral lossless systems, this transformation has been employed to express odd and even channels as linear combinations of forward and backward propagating waves \cite{ShenFan2007}. In our present work for the chiral system, we use this transformation to decompose the propagating and loss channel into their respective odd and even components (see Fig.~\ref{fig: sketch of system}). This approach introduces new field operators defined by
\begin{eqnarray} \label{eq:Weyl_Transformation}
\hat{a}_m^{(e)}(x) =&& \sqrt{\beta}\,\hat{a}(x) + \sqrt{1-\beta}\,\hat{b}_m(x),\nonumber\\
\hat{a}_m^{(o)}(x) =&& \sqrt{1-\beta}\,\hat{a}(x) - \sqrt{\beta}\,\hat{b}_m(x).
\end{eqnarray}
These operators effectively create linear combinations of the original propagating and loss channels. In this basis, the Hamiltonian [cf. Eq.~(\ref{eq:model ham})] decomposes into even and odd sectors:
\begin{equation} \label{eq:Weyl_Hamiltonian}
\begin{aligned}
\hat{H}_m &= \hat{H}_m^{(e)} + \hat{H}_m^{(o)},\\[1mm]
\hat{H}_m^{(e)} &= \int_{-\infty}^{\infty} dx\, \Biggl\{ \hat{a}_m^{\dagger(e)}(x) \bigl(-i\partial_x\bigr) \hat{a}_m^{(e)}(x) \\
&\quad\quad + \delta(x)\sqrt{\Gamma_{\rm tot}}\Bigl[\hat{\sigma}_m^+\,\hat{a}_m^{(e)}(x) + \mathrm{h.c.}\Bigr] \Biggr\},\\[1mm]
\hat{H}_m^{(o)} &= \int_{-\infty}^{\infty} dx\, \hat{a}_m^{\dagger(o)}(x) \bigl(-i\partial_x\bigr) \hat{a}_m^{(o)}(x).
\end{aligned}
\end{equation}
The odd sector, $\hat{H}_m^{(o)}$, describes free propagation without interaction, thus confining all scattering effects to the even sector. This formulation reduces the problem to an exactly solvable one-channel single-atom scattering system with coupling strength $\Gamma_{\rm tot}$~\cite{Yuds1985}. It is noteworthy that the parameter $\beta$ does not appear explicitly in the decomposed Hamiltonians in (\ref{eq:Weyl_Hamiltonian}), but only manifests in the basis transformation (\ref{eq:Weyl_Transformation}). This reflects the fact that these Hamiltonians encode the lossless scattering of photons, while the waveguide-atom coupling efficiency is entirely captured through the basis transformation between the physical channels and the even-odd sectors. 

To solve the two-channel scattering problem, we first decompose the incoming state at the $m$th atom into even and odd subspaces, solve the scattering in the interacting (even) sector, and subsequently recombine this solution with the odd component, which merely accumulates a phase factor. For the critical step of solving the even-sector scattering, we employ the $S$-matrix $S_{p_1 \dots p_n,k_1 \dots k_n}$ derived from Yudson's representation~\cite{Yuds1985}. In the following section, we elaborate on this powerful analytical approach which builds upon the Bethe Ansatz technique, providing the foundation for our multi-photon transport calculations.

\section{\label{sec:analy meth} Analytical Method}

In this section, we establish the analytical framework for multi-photon transport in chiral waveguide QED systems. We begin by revisiting the Bethe Ansatz method and Yudson's representation, which provide exact solutions for single-atom scattering problems with Fock-state inputs. We then introduce connected $S$-matrix elements that isolate genuine multi-photon interactions from individual scattering processes. This formalism serves as the foundation for our subsequent diagrammatic treatment of photon transport through atomic arrays.

\subsection{Revisiting Bethe's Ansatz and Yudson's Representation}

The even sector Hamiltonian $\hat{H}_m^{(e)}$ preserves the number of excitations in the ``atom+field'' system. The $n$-particle eigenstate can be constructed by Bethe Ansatz~\cite{Bethe1931,Thacker1981,Yuds2008}:
\begin{equation}
\begin{aligned}
  &\ket{\vec{\lambda}} = C(\vec{\lambda}) \int d^n \vec{x} \prod_{l < j} \left(1 + \frac{i \Gamma_{\rm tot} \, \text{sgn}(x_l - x_j)}{\lambda_l - \lambda_j+i\Gamma_{\rm tot}}\right) \\
  &\times \prod_{j=1}^n \exp(i \lambda_j x_j) f(\lambda_j, x_j) \left[\hat{a}_m^{\dagger(e)}(x_j)-\frac{\sqrt{\Gamma_{\rm tot}}}{\lambda_j}\delta(x_j)\hat{\sigma}_m^+ \right] \ket{0}, \\
  \\
  & \text{where } f(\lambda, x)=\frac{\lambda-(i\Gamma_{\rm tot}/2)\text{sgn}(x)}{\lambda+i\Gamma_{\rm tot}/2},
\end{aligned}
\end{equation}
and where $C(\vec{\lambda})$ is a normalization factor, $\text{sgn}(x)=\{-1 \text{ for } x<0;0 \text{ for } x=0;1 \text{ for } x>0\}$, $\Pi_{l<j}$ equals unity in the one-photon case, and $x_j$ and $\lambda_j$ are the position and rapidity of the $j$th photon. The state $\ket{\vec{\lambda}}=\ket{\lambda_1\cdots \lambda_n}$ represents a configuration of $n$ wavenumbers or rapidities. These rapidities $\{\lambda_j\}$ are the solutions of the Bethe equations and can be grouped into ``strings'' in the form~\cite{Thacker1981}:
\begin{equation*}
  \lambda_j^l = \Lambda_l-i\Gamma_{\rm tot}(n_l+1-2j),\quad j=1,2\cdots n_l.
\end{equation*} 
The principal rapidity $\Lambda_l$ is the real part of a given string $l$ and $n_l \leq n$ is the number of the rapidities forming the string $l$. Each string configuration corresponds to a composite excitation of the system. The rapidities can either be real or form complex conjugate pairs. Real rapidities are associated with spatially extended states, whereas complex conjugate pairs indicate the presence of a bound state \cite{Mahmoodian2020PRX}. In these bound states, the probability amplitude decays exponentially with the spatial separation between the constituent excitations.  In the thermodynamic limit where the system size $L\rightarrow\infty$, $\{\Lambda_l\}$ takes independent arbitrary values in $\mathbb{R}$~\cite{Yuds1984}. Since the Bethe states form a complete set of eigenstates for Hamiltonian $\hat{H}_m^{(e)}$, the time evolution of an arbitrary initial state can be written as 
\begin{equation}\label{eq:psi sum config}
    \ket{\Psi(t)}=\sum_{\text{\tiny strings}}\int d^n\vec{\Lambda} \exp[-iE(\vec{\lambda})t]\ket{\vec{\lambda}}\bra{\vec{\lambda}}\ket{\text{in}}. 
\end{equation}
Unfortunately, the above expression is far from being practical due to the sum over all possible string configurations. The number of summands, which is a set of the $n$-fold integrals over $\vec{\Lambda}$, grows exponentially with $n$. 
 
Yudson \cite{Yuds1985} made a remarkable observation that the expression (\ref{eq:psi sum config}) can be represented as a $n$-fold integral (\ref{eq:yuds rep}) over a product path $(\gamma_1,\gamma_2\cdots,\gamma_n)$ in $\mathbb{C}^n$ with only one string configurations. Here, $\gamma_j$ is a path parallel to the real axis in the complex plane for $\lambda_j$, and satisfies two conditions: (i) $\operatorname{Im}\gamma_{j+1}-\operatorname{Im}\gamma_j>\Gamma_{\rm tot}$. (ii) $\operatorname{Im}\gamma_j>-\Gamma_{\rm tot}/2$. The contributions from the other non-trivial string configurations are encoded in the poles of $\lambda_j$'s. We then have,
\begin{equation}\label{eq:yuds rep}
  \ket{\Psi(t)}=\int d^n\vec{\lambda} \exp[-iE(\vec{\lambda})t]\ket{\vec{\lambda}}(\vec{\lambda}\ket{\text{in}}. 
\end{equation}
Here, $|\vec{\lambda})$, denoted with a parenthesis instead of a ket, is an auxiliary state determined by the incoming state $\ket{\text{in}}$. The substitution $\bra{\vec{\lambda}}\ket{\text{in}}\to (\vec{\lambda}\ket{\text{in}}$ eliminates certain removable poles via contour integration, simplifying the expression. For the problem of $n$-photon scattering from a ground state two-level atom, any incoming state can be represented as a superposition of the basic states
\begin{equation} \label{eq:basic state}
\ket{\text{in,}\vec{x}} =\prod_{j=1}^n \hat{a}_m^{\dagger(e)}(x_j)\,\ket{0},\; 0>x_1>x_2>\dots>x_n.
\end{equation}
Substituting the initial state (\ref{eq:basic state}) into Eq. (\ref{eq:yuds rep}), performing the integrals over $\vec{\lambda}$, and taking the limit at asymptotic time $t\rightarrow \infty$, we obtain the $S$-matrix element in the sector $y_1 > y_2 > \cdots > y_n$ as given by~\cite{Yuds1985}:
\begin{equation}
\begin{aligned}\label{eq:real Sm}
  &\hat{S}_{y_1\dots y_n,x_1\dots x_n}=\theta(\xi_1\geq y_1 \geq \cdots \xi_n\geq y_n)\frac{1}{n!} \sum_{P}\\
  & \prod_{j=1}^n \Biggl\{\delta(y_{P_j}-\xi_j) 
   -\Gamma_{\rm tot}\theta(\xi_j>y_{P_j})\exp[\frac{\Gamma_{\rm tot}}{2}(y_{P_j}-\xi_j)]\Biggr\},
\end{aligned}
\end{equation}
where $\xi_j=x_j+t$ is the light cone coordinates. The ``long $\theta$'' function becomes 1 if the input inequality is true, elsewhere it is zero. The sum above is performed over all the permutations $P$, as the permutations of integers $\{1,2,\dots,n\}$, constrained by the condition $\{P_j\geq j-1,j=2,\cdots,n\}$. This condition requires that in a legitimate permutation $P$, the $j$th element must be greater or equal to $j-1$. For example, with $n=3$, the legitimate permutations in the summand are $\{1,2,3\},\{1,3,2\},\{2,1,3\},\{3,1,2\}$. The permutations with $P_3=1$ are eliminated. 

Unfortunately, the scattering matrix (\ref{eq:real Sm}) is not directly applicable to the problem of multi-atom scattering because the domain of the integral over the incoming coordinates $x_j$'s is constrained by $\theta(\xi_1\geq y_1 \geq \cdots \xi_n\geq y_n)$ and $\theta(\xi_j>y_{P_j})$. Two $\theta$ functions create a complicated, permutation-dependent integration region that is not feasible to iterative calculations for multiple atoms.
This obstacle can be avoided by Fourier transforming the $S$-matrix element $S_{y_1\dots y_n,x_1\dots x_n}$ to momentum space $S_{p_1\dots p_n,k_1\dots k_n}$. This Fourier transformation is not straightforward and relies on using the special property of the $P$ permutations. The details of the derivation is given in the Appendix~\ref{apx:n photon S matrix}.

\subsection{Connected $S$-matrix element} \label{subsec:connected S matrix}

One of the cornerstones of quantum field theory is the \textit{cluster decomposition principle}~\cite{WeinbergQFT}. This principle asserts that widely separated scattering processes should be independent of each other. In other words, if two sets of particles are sufficiently far apart, their scattering processes factorize into independent terms in the $S$-matrix. We refer these processes as \textit{disconnected} processes. This requirement is essential for ensuring that local interactions do not produce unphysical, long-range correlations.

In practice, the full $S$-matrix encompasses not only genuine multi-particle interactions but also non-interesting contributions arising from the independent scatterings of subgroups (clusters) of particles. To extract the truly intrinsic $n$-particle effects, we isolate the part of the $S$-matrix that does not factorize into lower-order interactions. This is what we refer to as the \textit{connected} part of the $S$-matrix. The connected $S$-matrix element is defined recursively: We first define $\hat{S}_{p_1p_2,k_1k_2}^C$ for two-photon scattering, then use $\hat{S}_{p_1p_2,k_1k_2}^C$ to define $\hat{S}_{p_1p_2p_3,k_1k_2k_3}^C$ for three photon scattering. The aim is to subtract from the full matrix element all contributions that can be described as products of simpler, independent processes.

In the photon scattering experiments, photons can either independently scatter off atoms or interact with other photons via the atoms. The $S$-matrix element encodes all the information of photon-atom, photon-photon interactions. Isolating the connected part of the $S$-matrix enables us to precisely extract the effects of $n$-body photon-photon interactions, which are characterized by a single overall momentum-conservation delta function.

We formalize this idea by introducing Weinberg’s definition~\cite{WeinbergQFT} of the connected $S$-matrix element (but with a different normalization convention compatible with (\ref{eq:real Sm})), which systematically removes the contributions from disconnected processes, thereby highlighting the genuine interactions among $n$ photons.

The two-photon connected element is,
\begin{equation}
  \hat{S}_{p_1p_2,k_1k_2}^C = \hat{S}_{p_1p_2,k_1,k_2} - \frac{1}{2!}\left(\hat{S}_{p_1,k_1}\hat{S}_{p_2,k_2} + \hat{S}_{p_2,k_1}\hat{S}_{p_1,k_2}\right).
  \end{equation}
The three-photon connected element is,
\begin{eqnarray}  \label{eq:def for S connected}
  &&\hat{S}_{p_1p_2p_3,k_1k_2k_3}^C = \hat{S}_{p_1 p_2 p_3,k_1 k_2 k_3} \nonumber\\
  && - \frac{1}{3!}\left(\hat{S}_{p_1,k_1}\hat{S}_{p_2 p_3,k_2 k_3}^C + \text{permutations}\right) \nonumber\\
  && - \frac{1}{3!}\left(\hat{S}_{p_1,k_1}\hat{S}_{p_2,k_2}\hat{S}_{p_3,k_3} + \text{permutations} \right).
\end{eqnarray}
Here, the added permutations are among all the incoming photon momenta $k_j$. We note that for a single photon, the connected part is simply the scattering matrix-element itself $\hat{S}^C_{p,k} = \hat{S}_{p,k}$

To bridge the gap between the abstract formalism and its physical interpretation, we now introduce a diagrammatic representation for single-atom scattering, that will serve as the foundation for our subsequent analysis for atomic arrays in Sec. \ref{subsec:diag exp}. The diagrams are drawn according to the rules:
\begin{enumerate}
    \item Draw \( N \) horizontal ``photon lines'' with incoming momenta labeled by $k$ on the left-hand side and outgoing momenta labeled by $p$ on the right hand side. 
    \item Draw filled black dots called ``vertices'' on the photon lines represents the atoms that mediate the photon-photon interaction. The black dots in the same column are associated with the same atom.
    \item Draw vertical wavy lines called an ``interaction line'' connecting $n$ vertices represents the $n$-photon connected part of $S$-matrix.
\end{enumerate}

We present connected diagrams for one-, two-, and three-photon scattering processes. First for a single photon,
  \begin{align*}
    &\hat{S}^C_{p,k}=\left(1-\frac{i\Gamma_{\rm tot}}{p+i\frac{\Gamma_{\rm tot}}{2}}\right)\delta(k-p).
  \end{align*}

\begin{center}
  \begin{tikzpicture}[scale=1]
    % Horizontal lines
    \draw[thick] (-1, 0) node[above right]{$k$} -- (1, 0) node[above left]{$p$}; % Middle line
    \filldraw[black] (0, 0) circle (2pt);
  \end{tikzpicture}
\end{center}
  \vskip 0.5ex
Now for two photons,  
  \begin{small}
  \begin{equation*}
    \hat{S}_{p_1p_2,k_1k_2}^C=\frac{i\Gamma_{\rm tot}^2}{2\pi
  }\frac{(p_1+p_2+i\Gamma_{\rm tot})\delta(p_1+p_2-k_1-k_2)}{(p_1+i\frac{\Gamma_{\rm tot}}{2})(p_2+i\frac{\Gamma_{\rm tot}}{2})(k_1+i\frac{\Gamma_{\rm tot}}{2})(k_2+i\frac{\Gamma_{\rm tot}}{2})}.
  \end{equation*}
  \end{small}

  \begin{center}
  \begin{tikzpicture}[scale=1]
    % Horizontal lines
    \draw[thick] (-1, 1) node[below right]{$k_1$} -- (1, 1) node[below left]{$p_1$}; 
    \draw[thick] (-1, 0) node[above right]{$k_2$} -- (1, 0) node[above left]{$p_2$}; % Middle line
    \draw[thick, decorate,
      decoration={
        snake,
        amplitude=1mm,
        segment length=4mm
      }] (0, 1) -- (0, 0); % Bottom line
    \filldraw[black] (0, 0) circle (2pt);
    \filldraw[black] (0, 1) circle (2pt);
  \end{tikzpicture}
\end{center}
And finally for three photons,  
  \begin{small}
  \begin{eqnarray*}
    &&\hat{S}_{p_1p_2p_3,k_1k_2k_3}^C =-\frac{i\Gamma_{\rm tot}^3}{3!\pi^2}\sum_{\hat{\sigma}(\{k_j\})\hat{\sigma}(\{p_j\})}\frac{1}{(p_1+p_2-k_1+i \frac{\Gamma_{\rm tot}}{2})}   \\
    && \frac{\delta(p_1+p_2+p_3-k_1-k_2-k_3)}{(p_1+i\frac{\Gamma_{\rm tot}}{2})(k_1+i\frac{\Gamma_{\rm tot}}{2})(k_2+i\frac{\Gamma_{\rm tot}}{2})(k_3+i\frac{\Gamma_{\rm tot}}{2})}.
  \end{eqnarray*}
  \end{small}
  \vskip 0.5ex

  \begin{center}
  \begin{tikzpicture}[scale = 1]
    % Horizontal lines
    \draw[thick] (-1, 1) node[below right]{$k_1$} -- (1, 1) node[below left]{$p_1$}; 
    \draw[thick] (-1, 0) node[below right]{$k_2$}-- (1, 0) node[below left]{$p_2$}; % Middle line
    \draw[thick] (-1, -1) node[above right]{$k_3$}-- (1, -1) node[above left]{$p_3$}; % Bottom line
    
    \draw[thick, decorate,
      decoration={
        snake,
        amplitude=1mm,
        segment length=4mm
      }]
      (0,1) -- (0,-1);

    \filldraw[black] (0, 0) circle (2pt);
    \filldraw[black] (0, 1) circle (2pt);
    \filldraw[black] (0, -1) circle (2pt);
  \end{tikzpicture}
  \end{center}

Now that we have established the expressions and diagrams for one, two, and three-photon scattering processes, we can apply these and solve for the output state in photon transport problems.

\section{Applications}

In this Section, we apply the analytical framework developed in the previous section to specific photon transport scenarios in chiral waveguide QED systems. We begin by examining the two-photon transport problem, demonstrating how our formalism enables both exact solutions through iterative methods and intuitive understanding through diagrammatic representations. We then extend our analysis to the more complex case of three-photon transport, where we employ perturbation theory in powers of the coupling parameter $\beta$. Throughout both applications, we show how the connected $S$-matrix elements provide crucial insight into photon correlations and non-Gaussian quantum signatures that emerge from multi-photon interactions in atomic arrays.

\subsection{Two-Photon Problem} \label{subsec:two-photon}

We now focus on the scattering of two photons from an array of atoms chirally coupled to a waveguide. The exact expression for the two-photon scattering wavefunction has been computed in previous works~\cite{Mahmoodian2018,Ringel2014} by using an $M$-atom $S$-matrix. In this subsection, we show that this expression can also be derived by iterations of the solutions of single-atom scattering problem.

\subsubsection{Two-photon Scattering from a Single Atom}

Let us consider an arbitrary two-photon incoming state with momentum $k_1$ and $k_2$ at the $m$th atom in the array : $\psi_{m-1}(k_1,k_2)\hat{a}(k_1)^\dagger\hat{a}(k_2)^\dagger\ket{0}$. We wish to derive the $S$-matrices for the dissipative system:\,${}_{11}\hat{S}_{p,k}$, ${}_{22}\hat{S}_{p_1p_2,k_1k_2}$ and ${}_{21}\hat{S}_{\slashed{p}_1p_2,k_1k_2}$. Here, ${}_{11}\hat{S}_{p,k}$ and ${}_{22}\hat{S}_{p_1p_2,k_1k_2}$ describe single- and two-photon scattering without loss, i.e., all photons scatter off the atom back into the waveguide. Alternatively, ${}_{21}\hat{S}_{\slashed{p}_1p_2,k_1k_2}$ quantifies scattering of two incoming photons where $p_2$ is transmitted and $p_1$ is lost. The slash $\slashed{p}$ labels the momentum of the lost photon. The symbol ${}_{10}\hat{S}_{\slashed{p},k}$ describes one photon lost after scattering with an atom. ${}_{21}\hat{S}_{\slashed{p}_1p_2,k_1k_2}$ and ${}_{10}\hat{S}_{\slashed{p},k}$ are non-vanishing in a lossy system with $\beta < 1$.

The values of ${}_{11}\hat{S}_{p,k}$ and ${}_{10}\hat{S}_{\slashed{p},k}$  are equal to the well-known single-photon transmission coefficient $t_k=1-i\beta \Gamma_{\rm tot}/(k+i\Gamma_{\rm tot}/2)$, and reflection coefficient $r_k=-\sqrt{\beta(1-\beta)}i\Gamma_{\rm tot}/(k+i\Gamma_{\rm tot}/2)$, each multiplied by an energy conserving function $\delta(p-k)$~\cite{Mahmoodian2018, Yuds2008}. 

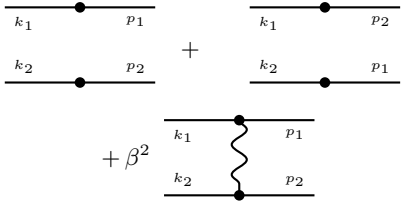
\begin{figure}[hbt]
  \centering
  % 第一行：两图加号
  \begin{minipage}{\columnwidth}
    \centering
    \begin{tabular}{@{}c@{\hspace{1em}}c@{\hspace{2em}}c@{}}
      % 第一个图形
      \begin{tikzpicture}[scale=1, baseline=(current bounding box.center)]
        \draw[thick] (-1,0) node[below right] {\tiny$k_1$} -- (1,0) node[below left] {\tiny$p_1$};
        \draw[thick] (-1,-1) node[above right] {\tiny$k_2$} -- (1,-1) node[above left] {\tiny$p_2$};
        \fill[black] (0,0) circle (2pt) (0,-1) circle (2pt);
      \end{tikzpicture}
      &
      \raisebox{-0.15cm}{$+$}
      &
      % 第二个图形
      \begin{tikzpicture}[scale=1, baseline=(current bounding box.center)]
        \draw[thick] (-1,0) node[below right] {\tiny$k_1$} -- (1,0) node[below left] {\tiny$p_2$};
        \draw[thick] (-1,-1) node[above right] {\tiny$k_2$} -- (1,-1) node[above left] {\tiny$p_1$};
        \fill[black] (0,0) circle (2pt) (0,-1) circle (2pt);
      \end{tikzpicture}
    \end{tabular}
  \end{minipage}

  \vspace{1em} % 行间距
  
  % 第二行：加号+图形
  \begin{minipage}{0.5\columnwidth}
    \centering
    \begin{tabular}{@{\hspace{0.5em}}c@{\hspace{0.5em}}c@{}}
      \raisebox{-0.1cm}{$+\,\beta^2$}
      &
      % 第三个图形
      \begin{tikzpicture}[scale=1, baseline=(current bounding box.center)]
        \draw[thick] (-1,0) node[below right] {\tiny$k_1$} -- (1,0) node[below left] {\tiny$p_1$};
        \draw[thick, decorate,
      decoration={
        snake,
        amplitude=1mm,
        segment length=4mm
      }] (0,0) -- (0,-1);
        \draw[thick] (-1,-1) node[above right] {\tiny$k_2$} -- (1,-1) node[above left] {\tiny$p_2$};
        \fill[black] (0,0) circle (2pt) (0,-1) circle (2pt);
      \end{tikzpicture}
    \end{tabular}
  \end{minipage}

  \caption{Diagrammatic representation of ${{}_{22}\hat{S}_{p_1p_2,k_1k_2}}$. The first two terms represent the disconnected parts ${}_{11}\hat{S}_{p_1,k_1}\,{}_{11}\hat{S}_{p_2,k_2}$ and ${}_{11}\hat{S}_{p_1,k_2}\,{}_{11}\hat{S}_{p_2,k_1}$, while the last term represents the connected part $\hat{S}_{p_1p_2,k_1k_2}^C$. We have explicitly included the $\beta$-dependence of the connected part.}
  \label{fig:diag for 22S }
\end{figure}

The computation of ${}_{22}\hat{S}_{p_1p_2,k_1k_2}$ is more involved. We start by decomposing the incoming state into its components in the even and odd subspace
\begin{eqnarray*}
   &&\int_{-\infty}^{\infty}dk_1dk_2 \, \psi_{m-1}(k_1,k_2)\Bigl[\beta\,\hat{a}_m^{\dagger(e)}(k_1)\hat{a}_m^{\dagger(e)}(k_2) \\
  +&&\sqrt{\beta(1-\beta)}\left(\hat{a}_m^{\dagger(e)}(k_1)\hat{a}_m^{\dagger(o)}(k_2)+\hat{a}_m^{\dagger(o)}(k_1)\hat{a}_m^{\dagger(e)}(k_2)\right)\\
  +&&(1-\beta)\hat{a}_m^{\dagger(o)}(k_1)\hat{a}_m^{\dagger(o)}(k_2)\Bigr]\ket{0}_e\ket{0}_o.
\end{eqnarray*}
Recall that the photons in the odd subspace do not interact so they just pick up delta functions for energy conservation after scattering. For the photons in the even subspace, $\hat{S}_{p_1p_2,k_1k_2}$ is multiplied in front of two-photon state, and $\hat{S}_{p,k}$ is multiplied in front of the one-photon state. The outgoing state then reads
\begin{eqnarray*}
  && \frac{1}{2}\int_{-\infty}^{\infty} dp_1dp_2dk_1dk_2\,\psi_{m-1}(k_1,k_2)\\
  &&\Big\{\beta \hat{S}_{p_1p_2,k_1k_2}\hat{a}_m^{\dagger(e)}(p_1)\hat{a}_m^{\dagger(e)}(p_2)\\
  &&+\sqrt{\beta(1-\beta)}\sum_{i,j\in \{1,2\},i\neq j}\hat{S}_{p_i,k_i} \delta(p_j-k_j)\hat{a}_m^{\dagger(e)}(p_i)\hat{a}_m^{\dagger(o)}(p_j)\\
  &&+(1-\beta)\delta(p_1-k_1)\delta(p_2-k_2)\hat{a}_m^{\dagger(o)}(p_1)\hat{a}_m^{\dagger(o)}(p_2)\\
  &&+p_1\leftrightarrow p_2\Big\}\ket{0}_e\ket{0}_o.
\end{eqnarray*}
 Transforming the above expression back to the original basis and using the definition (\ref{eq:def for S connected}), we obtain the $S$-matrix element for two-photon scattering on the $m$th atom, 
 \begin{eqnarray} \label{eq:two photon S matrix}
  {}_{22}\hat{S}_{p_1p_2,k_1k_2} =&& \frac{1}{2}({}_{11}\hat{S}_{p_1,k_1}\,{}_{11}\hat{S}_{p_2,k_2}+k_1\leftrightarrow k_2)+\beta^2 \hat{S}_{p_1p_2,k_1k_2}^C \nonumber,\\
  {}_{21}\hat{S}_{\slashed{p}_1p_2,k_1k_2} =&& \frac{1}{2}({}_{10}\hat{S}_{\slashed{p}_1,k_1}\,{}_{11}\hat{S}_{p_2,k_2}+k_1\leftrightarrow k_2) \nonumber\\
  &&+\beta^{3/2}\sqrt{(1-\beta)} \hat{S}_{p_1p_2,k_1k_2}^C,
 \end{eqnarray}
We obtain a key insight from the above equation:  the $\beta$ factors appearing in front of $\hat{S}_{p_1p_2,k_1k_2}^C$ reveal that the two-photon interaction scales as $\mathcal{O}(\beta^2)$. This was recently shown for two-photon scattering and used to approximate the two-photon output state~\cite{Schemmer2025}. In the next section, we use this insight to develop a perturbative series of leading-order diagrams for multi-atom, three-photon scattering. The $S$-matrix element for two-photon scattering can be visualized by the diagrams in Fig.~\ref{fig:diag for 22S } and \ref{fig:diag for 21S}. The connected part of $S$-matrix is represented by the connected diagram, while the disconnected parts are represented by the disconnected lines. A line ending at a cross represents the $-i\Gamma_{\rm tot}/(k+i\Gamma_{\rm tot}/2)$ term in $r_k$.

\begin{figure}[h]
  \centering
  % 第一行：两图加号
  \begin{minipage}{\columnwidth}
    \centering
    \begin{tabular}{@{}c@{\hspace{1em}}c@{\hspace{1em}}c@{}c@{}}
      % 第一个图形
    \raisebox{-0.15cm}{$\sqrt{\beta(1-\beta)}$}
    &
      \begin{tikzpicture}[scale=1, baseline=(current bounding box.center)]
        \draw[thick] (-1,0) node[below right] {\tiny$k_1$} -- (0,0) node[below right] {\tiny$p_1$};
        \draw[thick] (-1,-1) node[above right] {\tiny$k_2$} -- (1,-1) node[above left] {\tiny$k_2$};
        \fill[black] (0,0) circle (2pt) (0,-1) circle (2pt);
        \node[cross out, draw, line width=1pt] at (0,0) {};
      \end{tikzpicture}
      &
      \raisebox{-0.15cm}{$+\sqrt{\beta(1-\beta)}$}
      &
      % 第二个图形
      \begin{tikzpicture}[scale=1, baseline=(current bounding box.center)]
        \draw[thick] (-1,0) node[below right] {\tiny$k_1$} -- (1,0) node[below left] {\tiny$p_2$};
        \draw[thick] (-1,-1) node[above right] {\tiny$k_2$} -- (0,-1) node[above left] {\tiny$p_1$};
        \fill[black] (0,0) circle (2pt) (0,-1) circle (2pt);
        \node[cross out, draw, line width=1pt] at (0,-1) {};
      \end{tikzpicture}
    \end{tabular}
  \end{minipage}

  \vspace{1em} % 行间距
  
  % 第二行：加号+图形
  \begin{minipage}{0.5\columnwidth}
    \centering
    \begin{tabular}{@{}c@{\hspace{1em}}c@{}}
      \raisebox{-0.1cm}{$+\,\beta^{3/2}\sqrt{1-\beta}$}
      &
      % 第三个图形
      \begin{tikzpicture}[scale=1, baseline=(current bounding box.center)]
        \draw[thick] (-1,0) node[below right] {\tiny$k_1$} -- (0,0) node[below right] {\tiny$p_1$};
        \draw[thick, decorate,
        decoration={
          snake,
          amplitude=1mm,
          segment length=4mm
        }] (0,0) -- (0,-1);
        \draw[thick] (-1,-1) node[above right] {\tiny$k_2$} -- (1,-1) node[above left] {\tiny$p_2$};
        \fill[black] (0,0) circle (2pt) (0,-1) circle (2pt);
        \node[cross out, draw, line width=1pt] at (0,0) {};
      \end{tikzpicture}
    \end{tabular}
  \end{minipage}

  \caption{\label{fig:diag for 21S} Diagrammatic representation of ${}_{21}\hat{S}_{\slashed{p}_1p_2,k_1k_2}$. The first two terms represent the disconnected parts ${}_{10}\hat{S}_{\slashed{p}_1,k_1}\,{}_{11}\hat{S}_{p_2,k_2}$ and ${}_{10}\hat{S}_{\slashed{p}_1,k_2}\,{}_{11}\hat{S}_{p_2,k_1}$, while the last term represents the connected part $\hat{S}_{p_1p_2,k_1k_2}^C$. We have explicitly included the $\beta$-dependence of the connected part.}
\end{figure}

\subsubsection{Extension to Multiple Atoms} 

\label{subsec:two-photon multi-atom}
\begin{figure*}[t]
  \centering

  % ——— Top panel ———
  \subfloat[A concatenated two-photon diagram denotes the sum over all concatenations comprising a two-photon interaction diagram and \( M-1 \) individual scattering diagrams.\label{fig:concatenated diag top}]{
    \centering
    \begin{tabular}{@{}c@{\hspace{0.5em}}
                    c@{\hspace{0.5em}}
                    c@{\hspace{0.5em}}
                    c@{\hspace{0.5em}}
                    c@{\hspace{-2.5em}}
                    c@{\hspace{0.5em}}
                    c@{\hspace{-2.5em}}
                    c@{\hspace{0.5em}}
                    c@{\hspace{0.5em}}
                    c@{\hspace{0.5em}}
                    c@{\hspace{0.5em}}
                    c@{\hspace{0.5em}}}
      % First tikz
      \begin{tikzpicture}[scale=1,baseline=(current bounding box.center)]
        \draw[thick] (-2,1)--(2,1);
        \draw[thick] (-2,0)--(2,0);
        \draw[thick,decorate,decoration={snake,amplitude=1mm,segment length=4mm}]
             (0,1)--(0,0);
        \filldraw[black] (0,1) circle (2pt) (0,0) circle (2pt);
      \end{tikzpicture}
      &
      \raisebox{-0.05cm}{$=\beta^2\sum_{j=0}^{M-1}$}
      &
      % Second tikz
      \begin{tikzpicture}[scale=0.5,baseline=(current bounding box.center)]
        \draw[thick] (-1,0)--(1,0);
        \draw[thick] (-1,-1)--(1,-1);
        \fill (0,0) circle (2pt) (0,-1) circle (2pt);
      \end{tikzpicture}
      &
      $\dots$
      &
      % Third tikz
      \begin{tikzpicture}[scale=0.5,baseline=(current bounding box.center)]
        \draw[thick] (-1,0)--(1,0);
        \draw[thick] (-1,-1)--(1,-1);
        \fill (0,0) circle (2pt) (0,-1) circle (2pt);
      \end{tikzpicture}
      &
      \hspace{0.5em}
      &
      % j-th with brace
      \begin{tikzpicture}[scale=0.5,baseline=-8pt]
        \draw[thick] (-1,0)--(1,0);
        \draw[thick,decorate,decoration={snake,amplitude=1mm,segment length=4mm}]
             (0,0)--(0,-1);
        \draw[thick] (-1,-1)--(1,-1);
        \fill (0,0) circle (2pt) (0,-1) circle (2pt);
        \coordinate (sw) at (-1,-1);
        \coordinate (se) at (1,-1);
        \draw[decorate,decoration={brace,mirror,amplitude=6pt},
              shorten <=5pt,shorten >=5pt]
          (sw)--(se)
          node[midway,below=8pt]{the $(j+1)$th diagram};
      \end{tikzpicture}
      &
      \hspace{0.5em}
      &
      % Fourth tikz
      \begin{tikzpicture}[scale=0.5,baseline=(current bounding box.center)]
        \draw[thick] (-1,0)--(1,0);
        \draw[thick] (-1,-1)--(1,-1);
        \fill (0,0) circle (2pt) (0,-1) circle (2pt);
      \end{tikzpicture}
      &
      $\dots$
      &
      % Last tikz
      \begin{tikzpicture}[scale=0.5,baseline=(current bounding box.center)]
        \draw[thick] (-1,0)--(1,0);
        \draw[thick] (-1,-1)--(1,-1);
        \fill (0,0) circle (2pt) (0,-1) circle (2pt);
      \end{tikzpicture}
    \end{tabular}
  }

  \vspace{1em}

  % ——— Bottom panel ———
  \subfloat[A concatenated three-photon diagram denotes the sum of all concatenations comprising two two-photon interaction diagrams and \( M - 2 \) individual scattering diagrams.\label{fig:concatenated diag but}]{
    \centering
    \begin{tabular}{@{}c@{\hspace{0.5em}}
                    c@{\hspace{0.5em}}
                    c@{\hspace{0.5em}}
                    c@{\hspace{-2.5em}}
                    c@{\hspace{-2.5em}}
                    c@{\hspace{0.5em}}
                    c@{\hspace{0.5em}}
                    c@{\hspace{-4em}}
                    c@{\hspace{-3.5em}}
                    c@{\hspace{0.5em}}
                    c@{\hspace{0.5em}}
                    c@{\hspace{0.5em}}}
      % First tikz
      \begin{tikzpicture}[scale=1,baseline=(current bounding box.center)]
        \draw[thick] (-2,1)--(2,1);
        \draw[thick] (-2,0)--(2,0);
        \draw[thick] (-2,-1)--(2,-1);
        \draw[thick,decorate,decoration={snake,amplitude=1mm,segment length=4mm}]
             (-1,1)--(-1,0);
        \draw[thick,decorate,decoration={snake,amplitude=1mm,segment length=4mm}]
             (1,0)--(1,-1);
        \filldraw[black] (-1,1) circle (2pt)
                         (-1,0) circle (2pt)
                         (1,0) circle (2pt)
                         (1,-1) circle (2pt);
      \end{tikzpicture}
      &
      \raisebox{-0.05cm}{$=\beta^4\!\sum_{j=0}^{M-2}\sum_{m=0}^{M-j-2}$}
      &
      % 第一个图形
        \begin{tikzpicture}[scale=0.5, baseline=(current bounding box.center)]
        \draw[thick] (-1,0)-- (1,0) ;
        \draw[thick] (-1,-1)  -- (1,-1) ;
        \draw[thick] (-1,1)  -- (1,1) ;
        \fill[black] (0,0) circle (2pt) (0,-1) circle (2pt);
      \end{tikzpicture}
        &
        \raisebox{0.0cm}{$\dots$}
        &
        % 第二个图形
        \begin{tikzpicture}[scale=0.5,baseline=0pt]
        % 1) draw your usual diagram
        \draw[thick] (-1,1) -- (1,1);
        \draw[thick] (-1,0) -- (1,0);
        \draw[thick,decorate,decoration={snake,amplitude=1mm,segment length=4mm}]
    (0,1) -- (0,0);
        \draw[thick] (-1,-1) -- (1,-1);
        \fill (0,0) circle (2pt) (0,-1) circle (2pt);

  % 2) mark the two end‐points of the bottom line
        \coordinate (sw) at (-1,-1);
        \coordinate (se) at ( 1,-1);

  % 3) draw a mirrored brace between them
        \draw[decorate,decoration={brace,mirror,amplitude=6pt}]
        ([xshift=-5pt]sw) -- ([xshift=+5pt]se)
    node[midway,below=8pt]{the $(j+1)$th diagram};
        \end{tikzpicture}
        &
        \raisebox{0.0cm}{$\dots$}
        &
        \begin{tikzpicture}[scale=0.5, baseline=(current bounding box.center)]
        \draw[thick] (-1,0)  -- (1,0) ;
        \draw[thick] (-1,-1)  -- (1,-1) ;
        \draw[thick] (-1,1)  -- (1,1) ;
        \fill[black] (0,0) circle (2pt) (0,-1) circle (2pt);
        \end{tikzpicture}
        
        &
        \raisebox{0.0cm}{$\dots$}
        &
        \begin{tikzpicture}[scale=0.5,baseline=0pt]
        \draw[thick] (-1,1)  -- (1,1) ;
        % 1) draw your usual diagram
        \draw[thick] (-1,0) -- (1,0);
        \draw[thick,decorate,decoration={snake,amplitude=1mm,segment length=4mm}]
    (0,0) -- (0,-1);
        \draw[thick] (-1,-1) -- (1,-1);
        \fill (0,0) circle (2pt) (0,-1) circle (2pt);

  % 2) mark the two end‐points of the bottom line
        \coordinate (sw) at (-1,-1);
        \coordinate (se) at ( 1,-1);

  % 3) draw a mirrored brace between them
        \draw[decorate,decoration={brace,mirror,amplitude=6pt}]
        ([xshift=-5pt]sw) -- ([xshift=+5pt]se)
    node[midway,below=8pt]{the $(j+m+2)$th diagram};
        \end{tikzpicture}
        &
        \raisebox{0.0cm}{$\dots$}
        &
        \begin{tikzpicture}[scale=0.5, baseline=(current bounding box.center)]
        \draw[thick] (-1,0)  -- (1,0) ;
        \draw[thick] (-1,-1)-- (1,-1) ;
        \draw[thick] (-1,1)  -- (1,1) ;
        \fill[black] (0,0) circle (2pt) (0,-1) circle (2pt);
      \end{tikzpicture}
    \end{tabular}
  }

  \caption{\label{fig:concatenated diag} Two examples of the concatenated diagrams.}
  
\end{figure*}
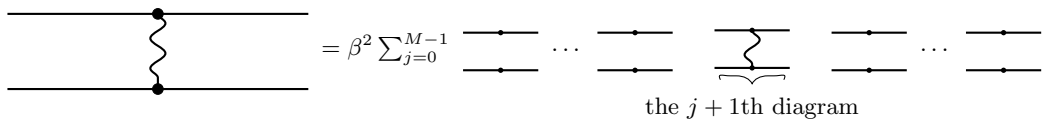
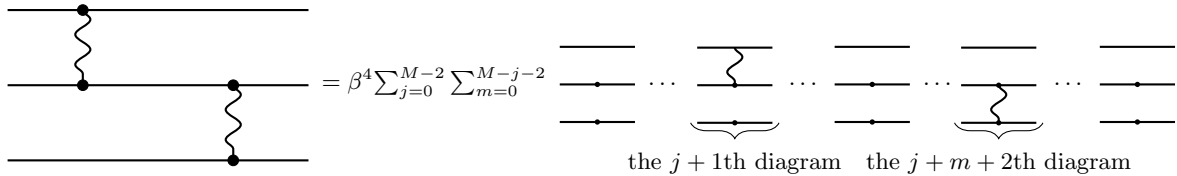

In an atomic array with cascaded interactions, each atom effectively sees only the forward-propagating input from its upstream neighbors. The total two-photon outgoing state from an array of $M$ atoms can be built up by iterating the single-atom scattering problem. If neither of the photons is scattered out of the array, the outgoing wavefunction after scattering off $d$ atoms can be obtained by $d$ iterations of the following equation:
\begin{eqnarray} \label{eq:two photon iteration}
\psi_{m+1}(p_1,p_2) &=& \int_{-\infty}^{\infty} dk_1\, dk_2 \, {}_{22}\hat{S}_{p_1 p_2, k_1 k_2} \psi_m(k_1, k_2) \nonumber,\\
& & \text{with}\quad m=0,1,2,\cdots,d-1,
\end{eqnarray}
where $\psi_0(k_1,k_2)=\delta(k_1)\delta(k_2)$ is the wavefunction of a resonant two-photon incoming state. Although performing the momentum integral
$d$ times appears challenging, this difficulty can be circumvented through a key observation: the frequencies of the two photons are always equally and oppositely shifted around the driving field frequency, and the pole in ${}_{22}\hat{S}_{p_1p_2,k_1k_2}$ is $ i\Gamma_{\rm tot}/2$. Based on this insight, we conjecture that the wavefunction coming out from $d$th atom $\psi_d(k_1, k_2)$ consists of a two-photon individual transmission term plus a multilinear polynomial of $1/(k_1+i\Gamma_{\rm tot}/2)$ and $1/(k_2+i\Gamma_{\rm tot}/2)$ with degree up to $d$. We write,
\begin{eqnarray} 
  \psi_d(k_1, k_2) =&& t_0^{2d}\delta(k_1)\delta{(k_2)} +\sum_{j,l=1}^{d} C_{j,l}^{(d)} \,\Gamma_{\rm tot}^{j+l-1} \nonumber\\
  && \times (k_1+i\Gamma_{\rm tot}/2)^{-j}(k_2+i\Gamma_{\rm tot}/2)^{-l},
  \label{eq:two photon conjecture}
\end{eqnarray}
 where $C_{j,l}^{(d)}$ denotes the polynomial coefficients. The conjectured form (\ref{eq:two photon conjecture}) is verified by mathematical induction in  Appendix~\ref{apx:two photon recursive}. As a byproduct of the induction, we obtain the recursive relation between $C_{j,l}^{(d)}$ and $C_{j,l}^{(d+1)}$. By iterating the recursion relations for the coefficients $C_{j,l}^{(d)}$, one can obtain the polynomial form of the outgoing wavefunction after any numbers of scattering events
\begin{eqnarray} \label{eq:two photon recursion}
    C_{1,1}^{(d+1)} &&= C_{1,1}^{(d)} + t_0^{2d} \left(\frac{2\beta^2}{\pi}\right)-\frac{\beta^2}{2\pi} \sum_{j,l=1}^{d} C_{j,l}^{(d)} F(j+1,l+1) \nonumber\\
    C_{j,l}^{(d+1)} &&= C_{j,l}^{(d)} -i\beta(C_{j-1,l}^{(d)}+C_{j,l-1}^{(d)})+(-i\beta)^2C_{j-1,l-1}^{(d)} \nonumber \\
    &&\text{for} \quad j+l\geq 3
  \end{eqnarray}
where $F(j,l) =  2\pi i^{-j-l} \frac{\prod_{k=1}^{l-1}(j+k-1)}{(l-1)!}$. Due to the permutation symmetry of the photon wavefunction, the coefficients $C_{j,l}^{(d)}$ are symmetric under the exchange of $j$ and $l$. The polynomial form of the two-photon wavefunction obtained by $d$ iterations of the recursion relation~(\ref{eq:two photon recursion}) agrees with the expression computed by eigenstate decomposition of the $S$-matrix~\cite{Mahmoodian2018}.

To compute the outgoing state with one photon lost on the $d+1$th atom, we simply perform $d$ iterations of the recursion relation (\ref{eq:two photon recursion}) and then multiply the result by the two-to-one scattering matrix ${}_{21}S_{\slashed{p}_1p_2,k_1k_2}$. The result is again a polynomial of $1/(k_1+i\Gamma_{\rm tot}/2)$ and $1/(k_2+i\Gamma_{\rm tot}/2)$ with degree up to $d+1$. The explicit calculation of the polynomial is given in Appendix~\ref{apx:two photon recursive}. The transmission (or loss) of the remaining photon through each of the following atoms is simply described by the multiplication of transmission and reflection coefficients ${}_{11}\hat{S}_{p,k}$ and ${}_{10}\hat{S}_{\slashed{p},k}$.

\subsubsection{Diagrammatic Method for Multiple Atoms}

Although the two-photon transport problem can be solved exactly using the iteration method presented above, we introduce here a complementary diagrammatic representation as preparation for addressing the more complex three-photon transport in the subsequent section. The single-atom scattering diagrams introduced in Sec.~\ref{subsec:connected S matrix} can be concatenated to build an \emph{concatenated} diagram that captures photon transport through the entire atomic array. In Fig.~\ref{fig:concatenated diag top}, the right-hand side shows this concatenation explicitly: for each number of atoms \( j \in \{0, \dots, M-1\} \) before the interaction, we insert the connected two-photon diagram \( \beta^2\hat{S}^{C}_{p_{1}p_{2},k_{1}k_{2}} \) at site \( j+1 \), while the remaining \( M-1 \) atoms contribute only their two-photon individual elastic scattering factors  $\frac{1}{2}({}_{11}\hat{S}_{p_1,k_1}\,{}_{11}\hat{S}_{p_2,k_2}+k_1\leftrightarrow k_2)$. Multiplying these \( M \) terms and summing over \( j \) therefore enumerates every possible location of the two-photon interaction along the chain. The concatenated diagrams on the left of the Fig. \ref{fig:concatenated diag}---with the extended horizontal lines---represents the entire superposition of \( M \) different concatenations on the right. The extended horizontal lines in the concatenated diagram represent photon propagation trajectories through the entire atomic array, distinguishing them from the short horizontal lines in single-atom scattering  diagrams. The three-photon concatenated diagram shown in Fig.~\ref{fig:concatenated diag but} is constructed in a similar way with the summation over $m$ enumerating all possible atom numbers between two interactions. The vertical wavy lines and filled black circles retain their earlier definitions, representing photon-photon interactions and scattering vertices, respectively. 
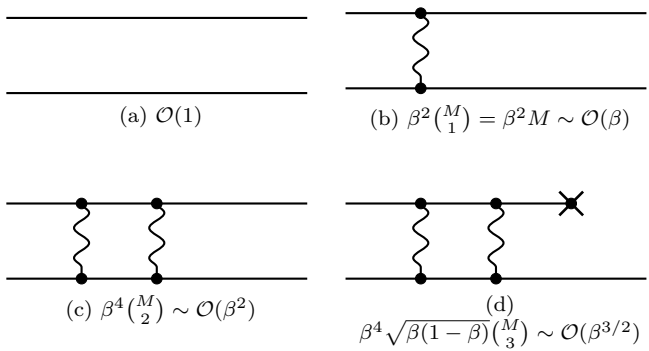
\begin{figure}[t]
  \centering
  % First row: two subfigures
  \subfloat[$\mathcal{O}(1)$\label{fig:two photon indv}]{%
    \begin{tikzpicture}
      % First tikzpicture content (upper left)
      \draw[thick] (-2, 1) -- (2, 1); % Top line
      \draw[thick] (-2, 0) -- (2, 0); % Middle line
    \end{tikzpicture}
  }
  \hfill
  \subfloat[$\beta^2\tbinom{M}{1}=\beta^2 M=\mathcal{O}(\beta)$\label{fig:two photon once}]{%
    \begin{tikzpicture}
      % Second tikzpicture content (upper right)
      \draw[thick] (-2, 1) -- (2, 1); % Top line
      \draw[thick] (-2, 0) -- (2, 0); % Middle line
      
      \draw[thick, decorate,
      decoration={
        snake,
        amplitude=1mm,
        segment length=4mm
      }] (-1, 1) -- (-1, 0); % Wavy line
      
      \filldraw[black] (-1, 1) circle (2pt); % Top dot
      \filldraw[black] (-1, 0) circle (2pt); % Middle dot
    \end{tikzpicture}
  }
  
  \vskip\baselineskip  % Forces a line break between rows
  
  % Second row: two subfigures
  \subfloat[$\beta^4\binom{M}{2}=\mathcal{O}(\beta^2)$\label{fig:two photon twice}]{%
    \begin{tikzpicture}
      % Third tikzpicture content (lower left)
      \draw[thick] (-2, 1) -- (2, 1);
      \draw[thick] (-2, 0) -- (2, 0);
      
      \draw[thick, decorate,
      decoration={
        snake,
        amplitude=1mm,
        segment length=4mm
      }] (-1, 1) -- (-1, 0);
      \draw[thick, decorate,
      decoration={
        snake,
        amplitude=1mm,
        segment length=4mm
      }] (0, 1) -- (0, 0);
      
      \filldraw[black] (-1, 1) circle (2pt);
      \filldraw[black] (-1, 0) circle (2pt);
      \filldraw[black] (0, 1) circle (2pt);
      \filldraw[black] (0, 0) circle (2pt);
    \end{tikzpicture}
  }
  \hfill
  \subfloat[$\beta^4\sqrt{\beta(1-\beta)}\binom{M}{3}=\mathcal{O}(\beta^{3/2})$\label{fig:two photon twice with one loss}]{%
    \begin{tikzpicture}
      % Fourth tikzpicture content (lower right)
      \draw[thick] (-2, 1) -- (1, 1);
      \draw[thick] (-2, 0) -- (2, 0);
      
      \draw[thick, decorate,
      decoration={
        snake,
        amplitude=1mm,
        segment length=4mm
      }] (-1, 1) -- (-1, 0);
      \draw[thick, decorate,
      decoration={
        snake,
        amplitude=1mm,
        segment length=4mm
      }] (0, 1) -- (0, 0);
      
      \filldraw[black] (-1, 1) circle (2pt);
      \filldraw[black] (-1, 0) circle (2pt);
      \filldraw[black] (0, 1) circle (2pt);
      \filldraw[black] (0, 0) circle (2pt);
      \filldraw[black] (1, 1) circle (2pt);

      \node[cross out, draw, line width=1pt,minimum size=8pt] at (1,1) {};
    \end{tikzpicture}
  }
  \caption{\label{fig:two photon diagrams} Concatenated diagrams of a few possible processes for unidirectional two-photon scattering. (a) two photon non-interacting transport through the waveguide. (b)-(d) each represents the sum of two-photon transport process where (b) two-photon interact exactly once at a single atom  (c) two-photon interaction happening twice at two different atoms (d) two photons interact twice at two atoms before one photon is lost at the third atom. The order estimates below each diagram show their respective scaling behavior in the large optical depth regime with $\beta M=\mathcal{O}(1)$.}
\end{figure}

Several illustrative processes of two-photon concatenated diagrams are shown in Figs.(\ref{fig:two photon indv})–(\ref{fig:two photon twice with one loss}). In these diagrams, we observe how photons may scatter elastically, interact once, or twice (without and with photon loss). Photons can generally interact up to $M$ times throughout the ensemble. While the recursive approach described earlier provides an exact solution to the two-photon transport problem, understanding the relative contributions of different interaction processes becomes increasingly important as we find the leading order terms in the perturbative treatment of three-photon transport with a small $\beta$. 

We now introduce a systematic method for estimating the relative amplitude of each concatenated diagram based on the coupling strength $\beta$ and combinatorial considerations depending on the atom number $M$. For the connected diagram without loss, the Weyl transformation (\ref{eq:Weyl_Transformation}) tells us that the creation operator of each photon gets multiplied by $\sqrt{\beta}$ when we transform from the propagating channel to the even channel, and multiplied by another $\sqrt{\beta}$ when we transform back, resulting the factor of $\beta$. For the diagram with loss, the creation operator of the lost photon gets multiplied by $\sqrt{\beta}$ when we transform from propagating channel to even channel, and multiplied by another $\sqrt{1-\beta}$ when we transform back, resulting the factor of $\sqrt{\beta(1-\beta)}$.
We summarize this estimation as follows:
\begin{enumerate}
    \item $\beta \ll 1$ and $\beta M =\mathcal{O}(1)$.
    \item Each vertex representing a `go through' photon (filled dot) contributes a factor of $\beta$.
    \item Each vertex representing a lost photon (crossed dot) contributes a factor of $\sqrt{\beta (1-\beta)}\approx \sqrt{\beta}$.
    \item The combinatorial multiplicity of the diagrams with $n$ photon-photon interactions (wavy lines) when $M$ atoms are in the array is given by $\binom{M}{n}$.
\end{enumerate}
As an explicit example, the diagram depicted in Fig.~\ref{fig:two photon twice} corresponds to an amplitude of order $\beta^4\binom{M}{2}=\beta^4 M(M-1)/2=\mathcal{O}(\beta^2)$.

It is worth clarifying that, based on the order estimates above, the leading-order diagrams at small to intermediate optical depth $\mathrm{OD}=4\beta M \sim o(1)$ (little-o notation is used here) form a subset of those at large optical depth ($\beta M =\mathcal{O}(1)$). This is because the power of $\beta$ associated with each diagram is fixed, while the overall magnitude of each diagram increases monotonically with $M$. As a result, all perturbative calculations carried out for the large-OD regime remain valid---and increase in accuracy---in the small and intermediate-OD regimes.

\subsection{Three-Photon Problem}\label{subsec:diag exp}

One of the main goals of this paper is to compute three-body observables, such as the third-order correlation function and the third-order quadrature cumulant. For three-photon transport, lost photons do not need to be considered when computing these quantities.  In the following analysis, we therefore focus on three-photon transport where all photons are transmitted through the waveguide. The outgoing wavefunction after the last atom in the array is challenging to obtain using the iteration method in the two-photon case. In principle, one could try to write down the expression of ${}_{33}\hat{S}_{p_1p_2p_3,k_1k_2k_3}$, obtaining the iteration equation similar to the Eq. (\ref{eq:two photon iteration}). Then, one would soon find that integrating over the incoming momenta is a formidable task because of the complicated pole structure. In addition, due to the complicated momentum exchange among three photons, it would be difficult to conjecture the multilinear polynomial form of the outgoing state. 

An alternative strategy is to write down the three-photon scattering matrix $\hat{T}_{p_1 p_2 p_3,k_1 k_2 k_3}$ for transport through the \textit{entire} atom array using a perturbative expansion. In the weak-coupling regime, the coupling constant $\beta \ll 1$ becomes an appropriate parameter for a perturbative expansion. As we have seen in the two-photon single-atom $S$-matrix (\ref{eq:two photon S matrix}), the contribution of the two-body connected part of the $S$-matrix is of order $\beta^2$.  The second item in order estimation at the end of the Sec.~\ref{subsec:two-photon multi-atom} suggests that, without a photon being scattered out of the waveguide, the $n$-body connected part of $S$-matrix is of order $\beta^n$. 

In the scattering matrix $\hat{T}_{p_1 p_2 p_3,k_1 k_2 k_3}$, the main contributions are from the scattering processes without three-body interactions because they are lower order in $\beta$. With the resonant input photons, the term describing the process of individual elastic photon transport (the transport without photon-photon interactions depicted in Fig.~\ref{fig:secB1a}) is just $t_0^{3M}$. Since $|t_0|<1$, this gives exponential decay with the atom number $M$.  The term describing the process with the two-photon interactions only happening once (depicted in Fig~\ref{fig:secB1b}) is $\mathcal{O}(\beta)$. Applying the similar estimations, we can see that the three-photon diagram with $n$ two-photon interactions is at least $\mathcal{O}(\beta^n)$. The order estimations motivate us to treat the influence of the three-body interaction as the perturbations at $\mathcal{O}(\beta^2)$ in addition to the processes with individual elastic photon transport and two-photon interactions. 

The scattering matrix $\hat{T}_{p_1 p_2 p_3,k_1 k_2 k_3}$ for the entire array can be expanded in the following form:
\begin{eqnarray} \label{eq:expansion of T}
  \hat{T} = \hat{T}^{(0)} + \hat{T}^{(1)} + \hat{T}^{(2)} + \mathcal{O}(\beta^4)
\end{eqnarray}
We henceforth suppress $T$'s momentum indices $p_1p_2p_3,k_1k_2k_3$ for notational 
simplicity. In the atomic array, $k_1=k_2=k_3=0$ due to the resonant 
driving field. The term $\hat{T}^{(0)}$ describes the individual elastic photon transport and two-photon 
interactions. Meanwhile, $\hat{T}^{(1)}$ and $\hat{T}^{(2)}$ describe
transport that includes three-photon interactions with leading 
order $\mathcal{O}(\beta^2)$ and sub-leading order $\mathcal{O}(\beta^3)$,
respectively.

\subsubsection{Three-Photon Transport with up to Two-Photon Interactions}

We begin our analysis of three-photon transport by considering the type of processes in which at most two photons interact throughout the ensemble, corresponding to the $\hat{T}^{(0)}$ contribution, which generalizes the two-photon scattering problem to include an individually propagating third photon.
\begin{figure}[t]
  \centering
  % First row: two subfigures
  \subfloat[$\mathcal{O}(1)$\label{fig:secB1a}]{%
    \begin{tikzpicture}
      % First tikzpicture content (upper left)
      \draw[thick] (-2, 1) -- (2, 1); % Top line
      \draw[thick] (-2, 0) -- (2, 0); % Middle line
      \draw[thick] (-2, -1) -- (2, -1); % Bottom line
    \end{tikzpicture}
  }
  \hfill
  \subfloat[$\beta^2\tbinom{M}{1}=\mathcal{O}(\beta)$\label{fig:secB1b}]{%
    \begin{tikzpicture}
      % Second tikzpicture content (upper right)
      \draw[thick] (-2, 1) -- (2, 1); % Top line
      \draw[thick] (-2, 0) -- (2, 0); % Middle line
      \draw[thick] (-2, -1) -- (2, -1); % Bottom line
      
      \draw[thick, decorate,
      decoration={
        snake,
        amplitude=1mm,
        segment length=4mm
      }] (-1, 1) -- (-1, 0); % Wavy line
      
      \filldraw[black] (-1, 1) circle (2pt); % Top dot
      \filldraw[black] (-1, 0) circle (2pt); % Middle dot
    \end{tikzpicture}
  }
  
  \vskip\baselineskip  % Forces a line break between rows
  
  % Second row: two subfigures
  \subfloat[$\binom{M}{2}\beta^4=\mathcal{O}(\beta^2)$\label{fig:secB1c}]{%
    \begin{tikzpicture}
      % Third tikzpicture content (lower left)
      \draw[thick] (-2, 1) -- (2, 1);
      \draw[thick] (-2, 0) -- (2, 0);
      \draw[thick] (-2, -1) -- (2, -1);
      
      \draw[thick, decorate,
      decoration={
        snake,
        amplitude=1mm,
        segment length=4mm
      }] (-1, 1) -- (-1, 0);
      \draw[thick, decorate,
      decoration={
        snake,
        amplitude=1mm,
        segment length=4mm
      }] (0, 1) -- (0, 0);
      
      \filldraw[black] (-1, 1) circle (2pt);
      \filldraw[black] (-1, 0) circle (2pt);
      \filldraw[black] (0, 1) circle (2pt);
      \filldraw[black] (0, 0) circle (2pt);
    \end{tikzpicture}
  }
  \hfill
  \subfloat[$\binom{M}{3}\beta^6=\mathcal{O}(\beta^3)$\label{fig:secB1d}]{%
    \begin{tikzpicture}
      % Fourth tikzpicture content (lower right)
      \draw[thick] (-2, 1) -- (2, 1);
      \draw[thick] (-2, 0) -- (2, 0);
      \draw[thick] (-2, -1) -- (2, -1);
      
      \draw[thick, decorate,
      decoration={
        snake,
        amplitude=1mm,
        segment length=4mm
      }] (-1, 1) -- (-1, 0);
      \draw[thick, decorate,
      decoration={
        snake,
        amplitude=1mm,
        segment length=4mm
      }] (0, 1) -- (0, 0);
      \draw[thick, decorate,
      decoration={
        snake,
        amplitude=1mm,
        segment length=4mm
      }] (1, 1) -- (1, 0);
      
      \filldraw[black] (-1, 1) circle (2pt);
      \filldraw[black] (-1, 0) circle (2pt);
      \filldraw[black] (0, 1) circle (2pt);
      \filldraw[black] (0, 0) circle (2pt);
      \filldraw[black] (1, 1) circle (2pt);
      \filldraw[black] (1, 0) circle (2pt);
    \end{tikzpicture}
  }
  \caption{\label{fig:secB1} Concatenated diagrams of the first a few terms in $\hat{T}^{(0)}$. In these diagrams, we do not explicitly label the incoming and outgoing momenta. All incoming momenta on the left side of the diagram are zero (resonant photons), while the outgoing momenta on the right side correspond to $p_1$, $p_2$, and $p_3$, ordered from top to bottom. All the diagrams in the rest of this article follows this convention of labeling.}
\end{figure}
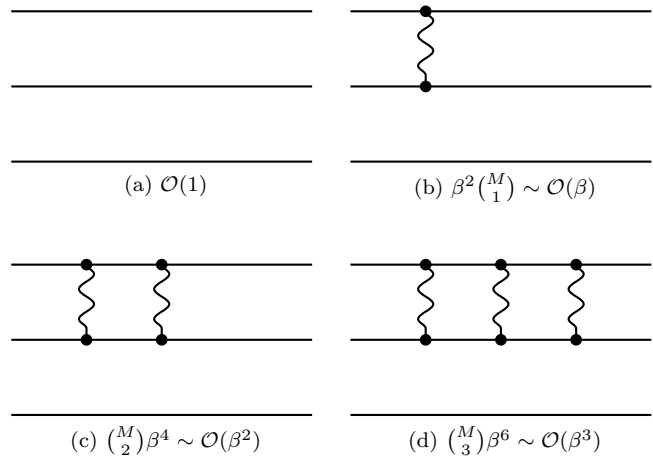

The $\hat{T}^{(0)}$ term includes two distinct transport scenarios: non-interacting elastic three-photon transport and two-photon interactions with one individual elastically scattered photon. As illustrated in the concatenated diagrams in Fig.~\ref{fig:secB1a}-\ref{fig:secB1d}, which are built in analogy with Fig.~\ref{fig:concatenated diag but}, these processes represent straightforward extensions of the two-photon scattering problem addressed in Sec.~\ref{subsec:two-photon}. The third photon, represented by the bottom line in these diagrams, scatters individually, with its scattering term factoring out from the two-photon interaction terms at each atomic site. Consequently, the scattering matrix $\hat{T}^{(0)}$ for these transport processes can be expressed as the product of the two-photon scattering matrices multiplied by the individual scattering term of the third photon. In addition, this expression inherently includes the three-photon individual scattering contribution in Fig.~\ref{fig:secB1a}, as ${}_{22}\hat{S}_{p_1p_2,k_1k_2}$ is the full two-photon scattering matrix including both the connected and factorisable parts and thus already accounts for individual elastic scattering of two photons at a single atomic site. The $\hat{T}^{(0)}$ term can therefore be concisely written as,
\begin{equation*}
 \hat{T}^{(0)}=({}_{22}\hat{S}_{p_1p_2,k_1k_2})^M (_{11}\hat{S}_{p_3,k_3})^M+\text{permutations}.
\end{equation*}

\subsubsection{Three-Photon Transport with Tree-Level Three-Body Interactions}

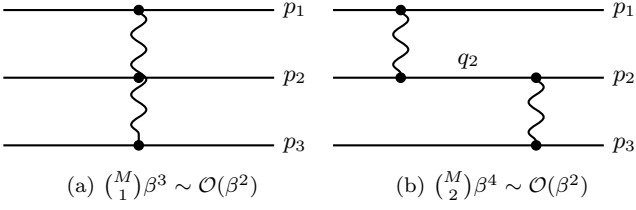
\begin{figure}[t]
  \centering
  % First subfigure
  \subfloat[$\binom{M}{1}\beta^3= \mathcal{O}(\beta^2)$\label{fig:fork}]{%
    \begin{tikzpicture}[baseline=(current bounding box.center),scale=0.9]
      % Horizontal lines
      \draw[thick] (-2, 1) -- (2, 1) node[right]{$p_1$}; % Top line
      \draw[thick] (-2, 0) -- (2, 0) node[right]{$p_2$}; % Middle line
      \draw[thick] (-2, -1) -- (2, -1) node[right]{$p_3$}; % Bottom line
      
      % Wavy line
      \draw[thick, decorate,
        decoration={
          snake,
          amplitude=1mm,
          segment length=4mm
        }] (0, 1) -- (0, -1);
      
      % Dots
      \filldraw[black] (0, 0) circle (2pt); % Middle dot
      \filldraw[black] (0, 1) circle (2pt); % Top dot
      \filldraw[black] (0, -1) circle (2pt); % Bottom dot
    \end{tikzpicture}
  }
  \hfill
  % Second subfigure
  \subfloat[$\binom{M}{2}\beta^4= \mathcal{O}(\beta^2)$\label{fig:crank}]{%
    \begin{tikzpicture}[baseline=(current bounding box.center),scale=0.9]
      % Horizontal lines
      \draw[thick] (-2, 1) -- (2, 1) node[right]{$p_1$}; % Top line
      \draw[thick] (-2, 0) -- (2, 0) node[right]{$p_2$}; % Middle line
      \draw[thick] (-2, -1) -- (2, -1) node[right]{$p_3$}; % Bottom line
      
      % Wavy lines
      \draw[thick, decorate,
        decoration={
          snake,
          amplitude=1mm,
          segment length=4mm
        }] (-1, 1) -- (-1, 0);
      \draw[thick, decorate,
        decoration={
          snake,
          amplitude=1mm,
          segment length=4mm
        }] (1, 0) -- (1, -1);
      
      % Dots
      \filldraw[black] (-1, 0) circle (2pt); % Middle dot
      \filldraw[black] (1, 0) circle (2pt);
      \filldraw[black] (-1, 1) circle (2pt); % Top dot
      \filldraw[black] (1, -1) circle (2pt); % Bottom dot
      \node at (0,0.25) {$q_2$};
    \end{tikzpicture}
  }
  \caption{\label{fig:tree-level diagrams} Concatenated diagrams of the terms in $\hat{T}^{(1)}$. The order estimates below each diagram show their respective scaling behavior in the large optical depth regime.}
  
\end{figure}
  We now examine the $\hat{T}^{(1)}$ contribution, which captures three-photon transport processes involving the connected tree diagrams, which generate nontrivial three-photon correlations. The $\hat{T}^{(1)}$ term in the Eq. (\ref{eq:expansion of T}) includes all transport processes with the occurrence of either one three-body interaction or two connected two-body interactions. Throughout the remainder of the atomic array, photons then undergo individual elastic scattering. These two processes are represented as the \textit{connected} concatenated diagrams in Figs.~\ref{fig:fork} and \ref{fig:crank}. Both diagrams exhibit comparable orders of magnitude $\sim\mathcal{O}(\beta^2)$ at larger OD. It is worth noting that the diagram in Fig.~\ref{fig:crank} exhibits asymmetry with respect to photon exchanges, necessitating the summation of all six permutations when calculating the outgoing state.
  
  The connectedness of the diagrams signifies that these processes generate three-photon correlations. As we will see in Secs.~\ref{subsec:g3g3c} and \ref{subsec:squz fluc obs}, the outgoing wavefunction after these interactions will be the dominant contribution of the third-order quadrature cumulant and the connected three-point correlation function $g_{c}^3$. The $\hat{T}^{(1)}$ term can be written as,
  \begin{eqnarray} \label{eq:three photon T1}
    \hat{T}^{(1)}&=& \hat{T}^{3v}+(\hat{T}^{4v}+\text{permutations}), \nonumber\\
    \text{with} \nonumber\\
    \hat{T}^{3v}&=& \sum_{j=0}^{M-1} (t_{p_1}t_{p_2}t_{p_3})^{M-j-1}\, \beta^3 \hat{S}_{p_1p_2p_3,000}^C\, t_0^{3j}, \nonumber\\
    \text{and}  \\
    \hat{T}^{4v}&=& \sum_{j=0}^{M-2} \sum_{m=0}^{M-j-2} t_{p_1}^{M-j-1} (t_{p_2}t_{p_3})^{M-j-m-2} \beta^2 \hat{S}_{p_2p_3,q_2 0}^C\; t_{q_2}^m \nonumber\\
    &&\beta^2 \hat{S}_{p_1q_2,00}^C  t_0^{3j+m+1},  \nonumber
  \end{eqnarray}
  In this expression, the variable $j$ in the summation for $T_{3v}$ counts the number of atoms before the first three-photon interaction occurs. Similarly, in $T_{4v}$, $j$ in the first summation denotes the number of atoms before the first two-photon interaction takes place, while $m$ in the second summation represents the number of atoms between the first and second two-photon interactions.

\subsubsection{Three-Photon Transport with Loop-Level Three-Body Interactions}

\begin{figure}[t]
  \centering
  % --- Row 1 ---
  \subfloat[$\binom{M}{2}\beta^5=\mathcal{O}(\beta^3)$\label{fig: three two diag}]{%
    \begin{tikzpicture}[baseline=(current bounding box.center)]
      % Horizontal lines
      \draw[thick] (-2, 1) -- (2, 1); % Top line
      \draw[thick] (-2, 0) -- (2, 0); % Middle line
      \draw[thick] (-2, -1) -- (2, -1); % Bottom line
      
      % Wavy lines
      \draw[thick, decorate,
        decoration={snake, amplitude=1mm, segment length=4mm}]
        (-1, 1) -- (-1, -1);
      \draw[thick, decorate,
        decoration={snake, amplitude=1mm, segment length=4mm}]
        (1, 1) -- (1, 0);
      
      % Dots
      \filldraw[black] (-1, -1) circle (2pt);
      \filldraw[black] (-1, 0) circle (2pt); % Middle dot
      \filldraw[black] (1, 0) circle (2pt);
      \filldraw[black] (-1, 1) circle (2pt); % Top dot
      \filldraw[black] (1, 1) circle (2pt);  % (Label comment unchanged)
    \end{tikzpicture}
  }
  \hfill
  \subfloat[$\binom{M}{2}\beta^5=\mathcal{O}(\beta^3)$\label{fig:two three diag}]{%
    \begin{tikzpicture}[baseline=(current bounding box.center)]
      % Horizontal lines
      \draw[thick] (-2, 1) -- (2, 1);
      \draw[thick] (-2, 0) -- (2, 0);
      \draw[thick] (-2, -1) -- (2, -1);
      
      % Wavy lines
      \draw[thick, decorate,
        decoration={snake, amplitude=1mm, segment length=4mm}]
        (-1, 1) -- (-1, 0);
      \draw[thick, decorate,
        decoration={snake, amplitude=1mm, segment length=4mm}]
        (1, 1) -- (1, -1);
      
      % Dots
      \filldraw[black] (1, 1) circle (2pt);
      \filldraw[black] (-1, 0) circle (2pt); % Middle dot
      \filldraw[black] (1, 0) circle (2pt);
      \filldraw[black] (-1, 1) circle (2pt); % Top dot
      \filldraw[black] (1, -1) circle (2pt); % Bottom dot
    \end{tikzpicture}
  }
  
  \vspace{1em} % Vertical space between rows
  
  % --- Row 2 ---
  \subfloat[$\binom{M}{3}\beta^6=\mathcal{O}(\beta^3)$\label{fig:twotwotwo non pla diag}]{%
    \begin{tikzpicture}[baseline=(current bounding box.center)]
      % Horizontal lines
      \draw[thick] (-2, 1) -- (2, 1);
      \draw[thick] (-2, 0) -- (2, 0);
      \draw[thick] (-2, -1) -- (2, -1);
      
      % Wavy lines
      \draw[thick, decorate,
        decoration={snake, amplitude=1mm, segment length=4mm}]
        (-1, 1) -- (-1, 0);
      \draw[thick, decorate,
        decoration={snake, amplitude=1mm, segment length=4mm}]
        (0,0) -- (0,-1);
      \draw[thick, decorate,
        decoration={snake, amplitude=1mm, segment length=4mm}]
        (1,1) .. controls (2,1) and (2,-1) .. (1,-1);
      
      % Dots
      \filldraw[black] (-1, 0) circle (2pt); % Middle dot
      \filldraw[black] (-1, 1) circle (2pt); % Top dot
      \filldraw[black] (0, -1) circle (2pt);
      \filldraw[black] (0, 0) circle (2pt);
      \filldraw[black] (1, -1) circle (2pt);
      \filldraw[black] (1, 1) circle (2pt); % (Label comment unchanged)
    \end{tikzpicture}
  }
  \hfill
  \subfloat[$\binom{M}{3}\beta^6=\mathcal{O}(\beta^3)$\label{fig:twotwotwo last diag}]{%
    \begin{tikzpicture}[baseline=(current bounding box.center)]
      % Horizontal lines
      \draw[thick] (-2, 1) -- (2, 1);
      \draw[thick] (-2, 0) -- (2, 0);
      \draw[thick] (-2, -1) -- (2, -1);
      
      % Wavy lines
      \draw[thick, decorate,
        decoration={snake, amplitude=1mm, segment length=4mm}]
        (-1, 1) -- (-1, 0);
      \draw[thick, decorate,
        decoration={snake, amplitude=1mm, segment length=4mm}]
        (0,1) -- (0,0);
      \draw[thick, decorate,
        decoration={snake, amplitude=1mm, segment length=4mm}]
        (1, 0) -- (1, -1);
      
      % Dots
      \filldraw[black] (0, 1) circle (2pt);
      \filldraw[black] (0, 0) circle (2pt);
      \filldraw[black] (-1, 0) circle (2pt); % Middle dot
      \filldraw[black] (1, 0) circle (2pt);
      \filldraw[black] (-1, 1) circle (2pt); % Top dot
      \filldraw[black] (1, -1) circle (2pt); % Bottom dot
    \end{tikzpicture}
  }
  
  \vspace{1em} % Vertical space between rows
  
  % --- Row 3 ---
  \subfloat[$\binom{M}{3}\beta^6=\mathcal{O}(\beta^3)$\label{fig:twotwotwo mid diag}]{%
    \begin{tikzpicture}[baseline=(current bounding box.center)]
      % Horizontal lines
      \draw[thick] (-2, 1) -- (2, 1);
      \draw[thick] (-2, 0) -- (2, 0);
      \draw[thick] (-2, -1) -- (2, -1);
      
      % Wavy lines
      \draw[thick, decorate,
        decoration={snake, amplitude=1mm, segment length=4mm}]
        (-1, 1) -- (-1, 0);
      \draw[thick, decorate,
        decoration={snake, amplitude=1mm, segment length=4mm}]
        (1,1) -- (1,0);
      \draw[thick, decorate,
        decoration={snake, amplitude=1mm, segment length=4mm}]
        (0, 0) -- (0, -1);
      
      % Dots
      \filldraw[black] (-1, 0) circle (2pt); % Middle dot
      \filldraw[black] (-1, 1) circle (2pt); % Top dot
      \filldraw[black] (0, -1) circle (2pt);
      \filldraw[black] (0, 0) circle (2pt);
      \filldraw[black] (1, 0) circle (2pt);
      \filldraw[black] (1, 1) circle (2pt); % (Label comment unchanged)
    \end{tikzpicture}
  }
  \hfill
  \subfloat[$\binom{M}{3}\beta^6=\mathcal{O}(\beta^3)$\label{fig:twotwotwo first diag}]{%
    \begin{tikzpicture}[baseline=(current bounding box.center)]
      % Horizontal lines
      \draw[thick] (-2, 1) -- (2, 1);
      \draw[thick] (-2, 0) -- (2, 0);
      \draw[thick] (-2, -1) -- (2, -1);
      
      % Wavy lines
      \draw[thick, decorate,
        decoration={snake, amplitude=1mm, segment length=4mm}]
        (-1, 0) -- (-1, -1);
      \draw[thick, decorate,
        decoration={snake, amplitude=1mm, segment length=4mm}]
        (1,1) -- (1,0);
      \draw[thick, decorate,
        decoration={snake, amplitude=1mm, segment length=4mm}]
        (0, 1) -- (0,0);
      
      % Dots
      \filldraw[black] (-1, 0) circle (2pt); % Middle dot
      \filldraw[black] (-1, -1) circle (2pt); % (Label comment unchanged)
      \filldraw[black] (0, 1) circle (2pt);
      \filldraw[black] (0, 0) circle (2pt);
      \filldraw[black] (1, 0) circle (2pt);
      \filldraw[black] (1, 1) circle (2pt);
    \end{tikzpicture}
  }
  
  \caption{\label{fig:loop-level diagrams} Concatenated diagrams of the terms in $\hat{T}^{(2)}$. Each diagram contains one loop structure, which introduce a loop momentum which we have to integrate over. The order estimation below each diagram show their respective scaling behavior in the large optical depth regime.}
\end{figure}
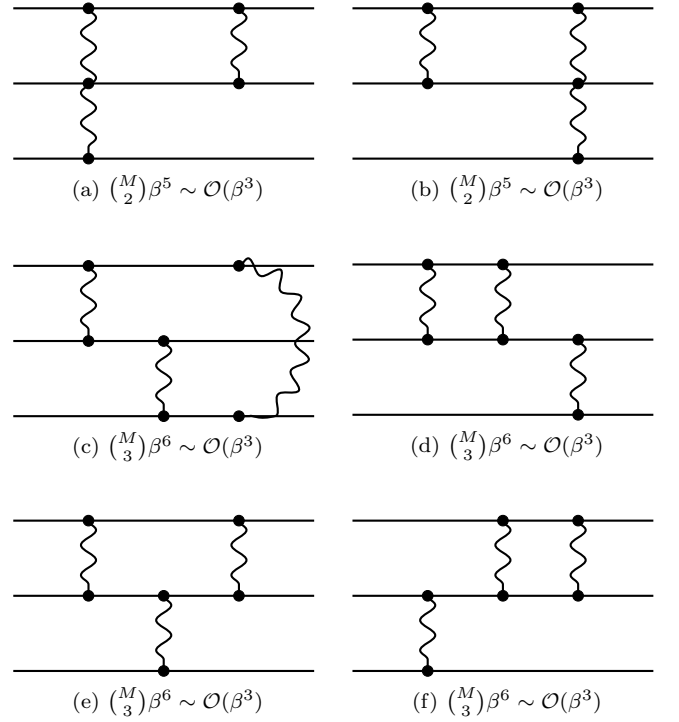

In this subsection, we analyze the $\hat{T}^{(2)}$ contribution to the three-photon scattering amplitude, highlighting the structure and scaling of loop-level diagrams. The $\hat{T}^{(2)}$ term in Eq. (\ref{eq:expansion of T}) encompasses two distinct types of processes: (1) those containing one three-body interaction combined with one two-body interaction (Fig.~\ref{fig: three two diag} and~\ref{fig:two three diag}), and (2) those containing three separate two-body interactions (Fig.~\ref{fig:twotwotwo non pla diag}-\ref{fig:twotwotwo first diag}). As with Fig.~\ref{fig:crank}, all concatenated diagrams presented here are asymmetric with respect to photon exchange. Each diagram in Fig.~(\ref{fig:loop-level diagrams}) represents a class of diagrams that can be generated by permuting the photon labels on the right side. When calculating $\hat{T}^{(2)}$, we must sum over all possible permutations for each diagram. All diagrams in Fig.~\ref{fig:loop-level diagrams} are of order $\mathcal{O}(\beta^3)$ and feature loop structures, which necessitate loop momentum integration when computing the scattering amplitudes. Due to the complexity of the resulting expressions, we present the explicit form of $\hat{T}^{(2)}$ in Appendix~\ref{apx:loop order}. To conclude this section, we formally write down the two-photon and three-photon outgoing wavefunction below, in terms of the connectedness of the concatenated diagrams,
\begin{eqnarray} \label{eq:expanded psi}
  \psi_2(x_1,x_2) &=& t_0^{2M} + \phi_2(x_1,x_2), \nonumber\\
  \psi_3(x_1,x_2,x_3) &=& t_0^{3M} + t_0^{M}\bigl[\phi_2(x_1,x_2) + \phi_2(x_2,x_3) \nonumber\\
  &+& \phi_2(x_1,x_3)\bigr] + \phi_3(x_1,x_2,x_3),
\end{eqnarray}
where $\phi_2(x_1,x_2)$ is the two-photon entangled wavefunction, including all two-photon connected diagrams with both photons transmitted through the entire ensemble, as shown in  Figs.~\ref{fig:two photon once} and \ref{fig:two photon twice}. Here, $\phi_3(x_1,x_2,x_3)$ is the three-photon entangled wavefunction, corresponding to the leading-order three-photon diagrams shown in in Figs.~\ref{fig:tree-level diagrams} and \ref{fig:loop-level diagrams}, noting that all three photons are transmitted through the entire ensemble . Each $t_0^M$ describes an individual elastically scattered photon that is disconnected from the other photons.

\section{Results} \label{sec:results}

In this Section, we apply the analytical framework developed in the previous sections to calculate experimentally observable quantities using parameters that can be realised in state-of-the-art experiments with atomic ensembles. Accordingly, we examine the special case of a coherent input state. A key advantage of our diagrammatic approach is its adaptability to various input states.

\subsection{Connected Third-Order Correlations Function} \label{subsec:g3g3c}

One observable that directly identifies correlations induced by three-photon interactions is the connected three-point correlation function $g_c^{(3)}$~\cite{Stiesdal2018,Jachymski2016},
\begin{equation} \label{eq:def of g3c}
  g_c^{(3)}(x_1, x_2, x_3) = 2 + g^{(3)}(x_1, x_2, x_3) - \sum_{i < j} g^{(2)}(x_i, x_j),
\end{equation}
where $g^{(2)}$ and $g^{(3)}$ are the normalized second- and third-order correlation function.  This definition of $g_c^{(3)}$, arises by considering a normalised third-order intensity cumulant ~\cite{Kubo1962} of the field. This observable is significant in experiments where the atomic ensemble is driven by a coherent laser field. This is because the laser field is classical and has Gaussian correlations in its intensity and electric field quadratures. Non-zero values of $g^{(3)}_c$ therefore indicate that photon transport through the atomic ensemble induces genuine three-body intensity correlations that are not present in the input light.

\subsubsection{Coherent input states}

Our aim in this section is to calculate the connected third-order intensity correlation function for a resonant coherent state propagating through the ensemble. To do this,  we first expand the input coherent state in the Fock basis,
\begin{equation} \label{eq:coherent exp}
\ket{\alpha}= e^{-\frac{\alpha^2}{2}}\sum_{n=0}^{\infty} \frac{P_{\rm in}^{\frac{n}{2}}}{n!} \int_{-L/2}^{L/2} \prod_{j=1}^n dx_j \hat{a}^\dagger (x_j) \ket{0}.
\end{equation}
The input power is $P_{\rm in}=|\alpha|^2/L$, where $L$ is the quantization length. For practical calculations, we employ the simplifying assumption that \emph{in each $n$-photon sector, genuine three-photon interactions depicted in concatenated diagrams Fig.~\ref{fig:secB1}-\ref{fig:loop-level diagrams} occur at most once among three photons, while the remaining $n-3$ photons undergo either linear transmission or scattering followed by loss without participating in the interaction.} This approximation is well-justified when $\beta \ll 1$ and allows us to compute correlation functions that directly reveal the non-Gaussian signatures in the output light field when the ensemble is driven by a coherent laser. Under this assumption, the outgoing \(n\)-photon amplitude $\Psi_n$ (before symmetrizing among $n$-photons) factorizes into
\begin{equation*}
    \Psi_n(x_1,\dots,x_n)\;=\;\psi_{3}(x_1,x_2,x_3)\sum_{\text{configurations}}\;\prod_{j\neq 1,2,3}^n s(x_j),
\end{equation*}
where \(\psi_{3}\) is the symmetrized three-photon wavefunction calculated from the concatenated diagrams in Figs.~\ref{fig:secB1}--\ref{fig:loop-level diagrams}, and \(s(x)\) is a single-photon tranmission/lost amplitude (which could be \(t_0^{M}\), \(t_0^{\,m}r_0\), etc.). The sum is over all the configurations of $n-3$ single-photon scatterings through the array. The full outgoing state is a sum of $n$-photon amplitudes weighted by Poisson factors
\begin{equation} \label{eq:out state fock expansion}
    \ket{\rm out} = e^{-\frac{\alpha^2}{2}}\sum_{n=0}^{\infty} \frac{P_{\rm in}^{\frac{n}{2}}}{n!} \int_{-L/2}^{L/2}  \Psi_n^S(x_1,\dots,x_n)\prod_{j=1}^n dx_j \hat{a}^\dagger (x_j) \ket{0},
\end{equation}
where $\Psi_n^S(x_1,\dots,x_n)$ stands for the symmetrized $\Psi_n(x_1,\dots,x_n)$. When we evaluate observables---say the third-order correlator 
$G_3(x_1,x_2,x_3)=\bra{\rm out}\hat a^\dagger(x_1)\hat a^\dagger(x_2)\hat a^\dagger(x_3)\hat a(x_3)\hat a(x_2)\hat a(x_1)\ket{\rm out}$---we sum over all $n$-photon Fock sectors of the outgoing state (\ref{eq:out state fock expansion}). As shown in Appendix~\ref{apx:Coherent state convertion}, \(|\psi_{3}(x_1,x_2,x_3)|^{2}\) and a factor of \(P_{\text{in}}^{3}\) factor out, while the sum of all products of individual-scattering terms $\prod_{j\neq 1,2,3}^n s(x_j)$ cancel the overall normalization \(\exp(-|\alpha|^{2})\).

\subsubsection{Perturbation expansion with  $\beta \ll 1$ and low power}

In Sec.~\ref{subsec:diag exp} we introduced our perturbative expansion in $\beta$ for calculating three-photon transport, which we argued works well for atomic ensembles as typically $\beta \ll 1$. In this Section we have thus far used a Fock state expansion of the coherent state input and have restricted photon transport to considering only up to three-photon-connected interactions. This restricts the power of the input drive field $P_{\rm in}$ relative to the total decay rate of the atoms, i.e., the ratio $P_{\rm in}/\Gamma_{\rm tot}$. Since we are now using perturbation theory in both the relative input power $P_{\rm in}/\Gamma_{\rm tot}$ and the coupling magnitude $\beta$ we need to specify the relative magnitude of these two quantities.

To do this, we realise that our assumption of considering only up to three-photon-connected processes in the coherent state is reliable provided we can neglect the contribution of four-photon concatenated diagrams to $g_c^{(3)}$. Specifically, the contribution from the four-photon concatenated diagrams is calculated similarly by assuming that only four photons undergo the transport processes depicted by the leading-order four-photon concatenated diagrams, while the remaining $n-4$ photons either scatter or are lost individually. The contribution to $g_c^{(3)}$ of such concatenated diagrams has one more factor of $P_{\rm in}/\Gamma_{\rm tot}$ than three-photon concatenated diagrams. This means that, quantitatively, the weak drive condition,
\begin{equation} \label{eq:drive cond1}
\mathcal{O}\left(\frac{P_{\text{in}}}{\Gamma_{\text{tot}}}\right)=\mathcal{O}(\beta^{2}),
\end{equation}
guarantees that the largest four-photon correction, of order \(\beta\,P_{\text{in}}/\Gamma_{\text{tot}}\) (see Appendix~\ref{apx:power correc}), remains negligible compared with the leading three-photon contribution depicted in Fig.~\ref{fig:tree-level diagrams}, which scales as \(\beta^{2}\) in \(g^{(3)}_{c}\). We can then approximate $g^{(3)}$ and $g^{(2)}$ as,
\begin{eqnarray} \label{eq:g3ng2}
g^{(3)}(x_1, x_2, x_3)&=&\frac{\expval{\hat{a}^\dagger(x_1)\hat{a}^\dagger(x_2)\hat{a}^\dagger(x_3)\hat{a}(x_3)\hat{a}(x_2)\hat{a}(x_1)}}{\prod_{i=1}^3\expval{\hat{a}^\dagger(x_i) \hat{a}(x_i)}} \nonumber\\
&=&\frac{|\psi_3(x_1, x_2, x_3)|^2}{t_0^{6M}}+\mathcal{O}\left(\beta\frac{P_{\rm in}}{\Gamma_{\rm tot}}\right), \nonumber\\
g^{(2)}(x_1,x_2)&=&\frac{|\psi_2(x_1,x_2)|^2}{t_0^{4M}}+\mathcal{O}\left(\beta\frac{P_{\rm in}}{\Gamma_{\rm tot}}\right),
\end{eqnarray}
where $\psi_2$ and $\psi_3$ are the two- and three-photon outgoing wavefunctions.

The condition (\ref{eq:drive cond1}) is so strict that it may lead to and experimentally undetectable output signal from the system. However, as we calculate in Appendix \ref{apx:power correc}, the leading-order three- and four-photon concatenated diagrams merely add a position-independent shift to the value of $G^{(2)}$ and $G^{(3)}$, respectively. Since $g^{(3)}_c(x_1,x_2,x_3)$ must decay to zero when two of its arguments are far apart, this constant shift would cancel with the leading-order two-photon constant correction in $\expval{\hat{a}^\dagger(x)\hat{a}(x)}$ at $\mathcal{O}\left(\beta\frac{P_{\rm in}}{\Gamma_{\rm tot}}\right)$ in $g^{(2)}$ and $g^{(3)}$. We may therefore relax the constraint on the drive strength to, 
\begin{equation} \label{eq:drive cond2}
\mathcal{O}\left(\frac{P_{\text{in}}}{\Gamma_{\text{tot}}}\right)=\mathcal{O}(\beta),
\end{equation}
while still using the approximation in (\ref{eq:g3ng2}) to calculate $g_c^{(3)}$. The condition (\ref{eq:drive cond2}) guarantees that the leading-order five-photon concatenated diagram, which is expected to be at order $\beta P_{\rm in}^2/\Gamma_{\rm tot}^2$, is negligible compared with the term at order $\beta^2$ in $g_c^{(3)}$. Numerical simulations in Sec. \ref{subsec:num ver} confirm that, within this stronger drive regime, our perturbative calculation of $g_c^{(3)}$ agrees closely with those obtained from the full cascaded-master-equation treatment. 

We substitute Eq. (\ref{eq:expanded psi}) and (\ref{eq:approx G3c})
into the definition of $g_c^{(3)}$ (\ref{eq:def of g3c}), and use $\expval{\hat{a}^\dagger (x)\hat{a}(x)}=P_{\rm in}\left(t_0^{2M}+\mathcal{O}(\beta\frac{P_{\rm in}}{\Gamma_{\rm tot}})\right)$. We obtain,
\begin{equation} \label{eq:approx g3c}
    g_c^{(3)} = \frac{G_c^{(3)}(x_1,x_2,x_3)}{P_{\rm in}^3 t_0^{6M}}+\mathcal{O}\left(\beta\frac{P_{\rm in}}{\Gamma_{\rm tot}}\right),
\end{equation}
where
\begin{eqnarray} \label{eq:approx G3c}
    & &G_c^{(3)}(x_1,x_2,x_3)=\expval{\hat{a}^\dagger(x_1)\hat{a}^\dagger(x_2)\hat{a}^\dagger(x_3)\hat{a}(x_3)\hat{a}(x_2)\hat{a}(x_1)} \nonumber\\
    &-&\expval{\hat{a}^\dagger(x_1)\hat{a} (x_1)}\expval{\hat{a}^\dagger(x_2)\hat{a}^\dagger(x_3)
    \hat{a}(x_3)\hat{a}(x_2)} \nonumber\\
    &-&\expval{\hat{a}^\dagger(x_2)\hat{a}(x_2)}\expval{\hat{a}^\dagger(x_1)\hat{a}^\dagger(x_3)
    \hat{a}(x_3)\hat{a}(x_1)} \nonumber\\
    &-&\expval{\hat{a}^\dagger(x_3)\hat{a}(x_3)}\expval{\hat{a}^\dagger(x_1)\hat{a}^\dagger(x_2)
    \hat{a}(x_2)\hat{a}(x_1)} \nonumber\\
    &+&2\expval{\hat{a}^\dagger(x_1)\hat{a}(x_1)}\expval{\hat{a}^\dagger(x_2)\hat{a}(x_2)}\expval{\hat{a}^\dagger(x_3)\hat{a}(x_3)} \nonumber\\
    &\approx& P_{\rm in}^3\Big\{2t_0^{3M}\phi_3(x_1,x_2,x_3)\\
    &+&2t_0^{2M} \big[\phi_2(x_1,x_2)\phi_2(x_2,x_3)
    +\phi_2(x_1,x_2)\phi_2(x_1,x_3) \nonumber\\
    &+&\phi_2(x_1,x_3)\phi_2(x_2,x_3)\big]+2t_0^M\phi_3(x_1,x_2,x_3)\big[\phi_2(x_1,x_2) \nonumber\\
    &+&\phi_2(x_1,x_3)+\phi_2(x_2,x_3)\big] + \phi_3(x_1,x_2,x_3)^2\Big\}. \nonumber
\end{eqnarray}
Here, we highlight a feature that not only $\phi_3$ but also $\phi_2$ contributes to $g_c^{(3)}$. As seen from the terms of the form $\phi_2(x_i,x_j)\phi_2(x_j,x_k)$, $\phi_2$ can connect three photons via contributions from the bra and ket when evaluating the correlation function. This result also highlights the distinction between non-Gaussian intensity correlations and non-Gaussian field correlations. The connected third-order intensity correlation can be non-zero even if the field correlations are Gaussian as  products of  $\phi_2$ lead to non-zero $G_c^{(3)}$. As we discuss at the end of this Section and in Sec.~\ref{subsec:squz fluc obs}, leading-order non-Gaussian field correlations require $\phi_3 \neq 0$.

\begin{figure*}[ht] \label{g3c}
  \includegraphics[width=\textwidth]{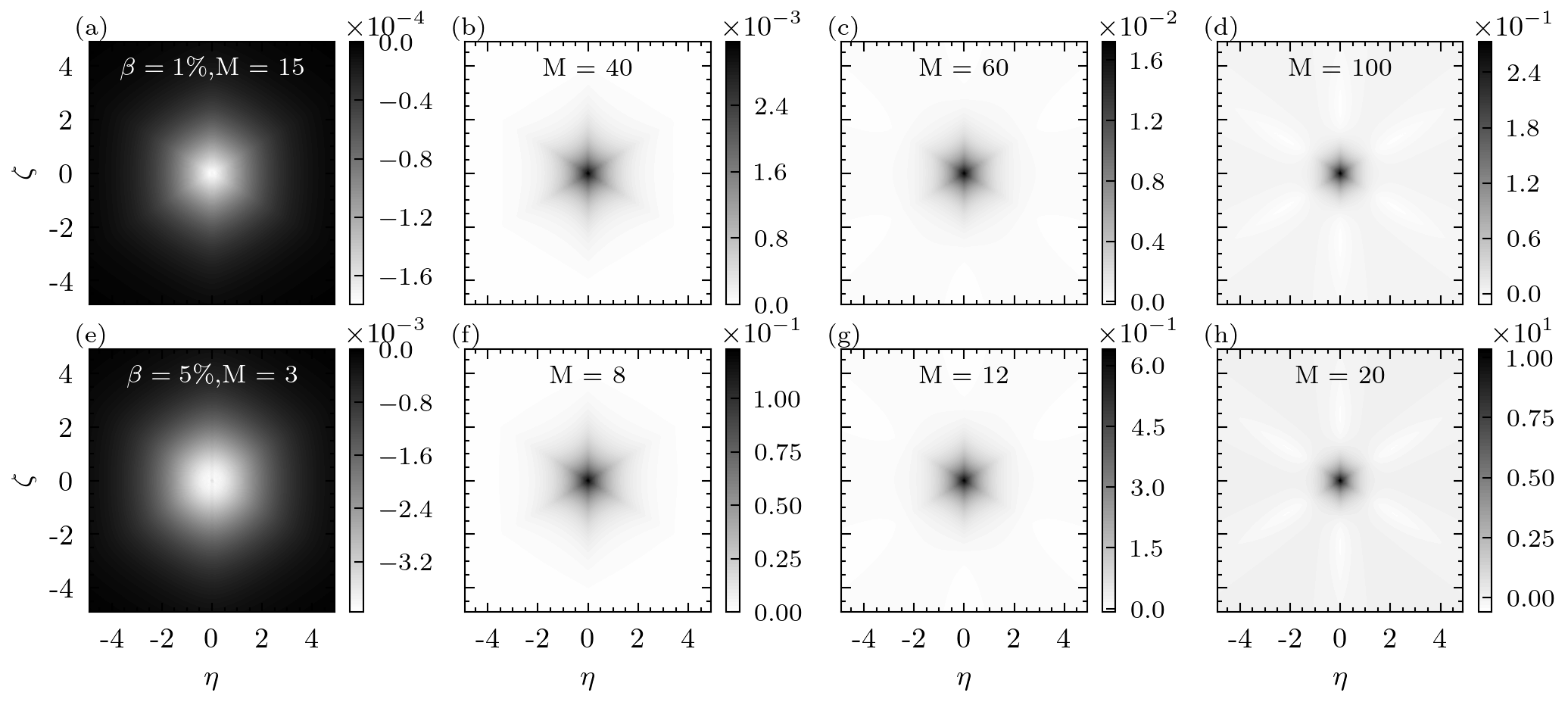}
  \caption{\label{fig:g3c} Connected third-order correlation function $g_c^{(3)}(R,\eta,\zeta)$ in Jacobi coordinates with various OD from 0.8 to 4. The center of mass $R=0$. The coupling strength $\beta$ is 1\% in  the first row (a)-(d), while $\beta$ is 5\% in the second row (e)-(h). $g_c^{(3)}$ is computed with tree-level diagram for $\beta=1\%$ while the loop order correction is added for $\beta= 5\%$. The six-fold symmetry reflects three symmetry axes corresponding to two-photon coincidences. }
\end{figure*}

In Fig.~\ref{fig:g3c}, we plot $g_c^{(3)}$ in Jacobi coordinates: $R=(x_1+x_2+x_3)/\sqrt{3}$, $\eta=(x_1-x_2)/\sqrt{2}$ and $\zeta = \sqrt{2/3}[(x_1+x_2)/2-x_3]$, where the center of mass $R$ is set to zero due to the translational invariance of the steady state. For our results we henceforth consider two different values of the coupling efficiency $\beta = 0.05$ and $\beta=0.01$. The connected third-order correlation function $g_c^{(3)}$ exhibits six-fold symmetry arising from two distinct sources. First, the permutation symmetry of the three-body wavefunction inherently generates three-fold symmetry. This symmetry is then doubled because $g_c^{(3)}$ does not distinguish between events where a photon pair arrives at the detector earlier than a single photon or vice versa.  In the $\eta$-$\zeta$ plane, three axes $\eta=0$, $\zeta=\sqrt{3}/3 \eta$ and $\zeta=-\sqrt{3}/3 \eta$ are the symmetry axes of the $g_c^{(3)}$, corresponding to two-photon coincidence events. The center $\eta=\zeta=0$ corresponds to three-photon coincidence events. The pattern of $g_c^{(3)}$ at various ODs in Fig.~\ref{fig:g3c} can be understood by a semi-quantitative analysis of the first two terms in Eq. (\ref{eq:approx G3c}).
\begin{enumerate}
    \item At low enough OD (Fig.~\ref{fig:g3c}a and ~\ref{fig:g3c}e), \(g_c^{(3)}\) is negative everywhere in the $\eta$-$\zeta$ plane.  
      Equations~\eqref{eq:approx g3c}–\eqref{eq:approx G3c} show that the leading term is the three-photon wavefunction \(\phi_{3}\propto\beta^{3}\);  
      the products of two-photon amplitudes \(\phi_{2}\phi_{2}\propto\beta^{4}\) are one order higher in \(\beta\) and are negligible.  
      Within \(\phi_{3}\) the 3-vertex diagram (\( \sim\beta^{3}\)) dominates over the 4-vertex diagram (\(\sim\beta^{4}\)), and it carries a negative sign inherited from the connected three-body \(S\)-matrix \(S^{C}_{p_1p_2p_3,k_1k_2k_3}\) in \(\hat T^{3v}\).  

    \item As OD increases (Fig.~\ref{fig:g3c}b,\ref{fig:g3c}c,\ref{fig:g3c}f,\ref{fig:g3c}g), the 4-vertex diagram becomes comparable to the 3-vertex term. Because it contains a product of two negative two-body matrices \(S^{C}_{p_ip_j,k_ik_j}\), its overall contribution is positive. Meanwhile, \(\phi_{2}\) is negative at these ODs, so the product \(\phi_{2}\phi_{2}\) adds another positive term. The negative component from the 3-vertex diagram is concealed by these positive terms.     
    
    \item At large OD (Fig.~\ref{fig:g3c}d and \ref{fig:g3c}h), the products of two-photon amplitudes dominate because the contribution of $\phi_3$ is weighed by one more factor of $t_0^M$ in Eq.~(\ref{eq:approx G3c}).  Since \(\phi_{2}\) is negative when two photons are nearly coincident and positive at moderate separations, \(g_c^{(3)}\) develops a positive peak at the origin (three-photon coincidence) surrounded by six negative ``legs'' along the coincidence axes. Physically, this indicates that three photons are most likely to arrive nearly simultaneously, whereas events in which a photon pair is followed by a third are suppressed.
\end{enumerate}

\begin{figure*}[ht]
  \centering
    %\hspace*{-0.5cm} % Adjust this value to move the figure to the left
    \includegraphics[width=\textwidth]{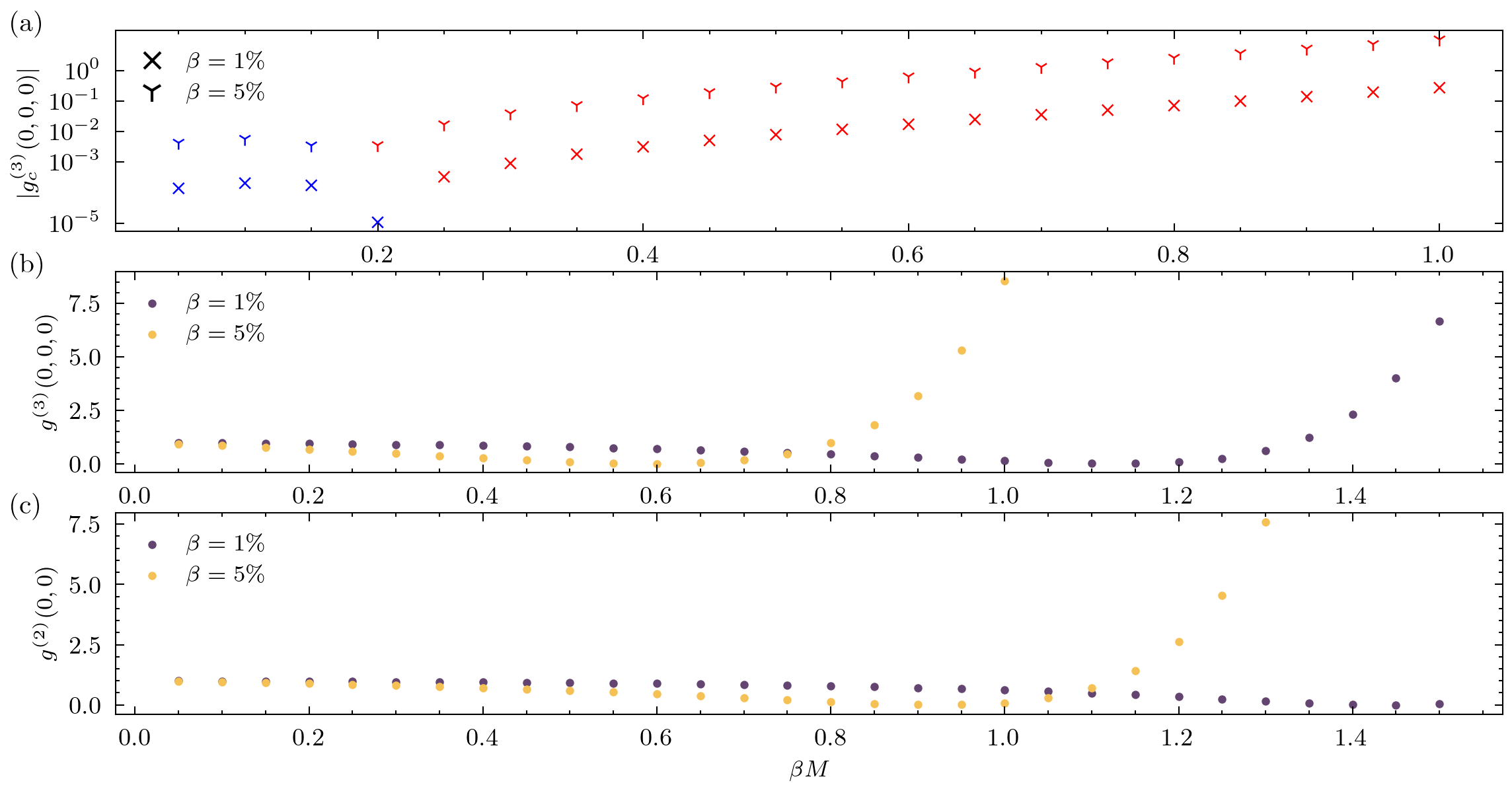}
    \caption{\label{fig:g3g3c} (a) Connected three-point correlation function $g_c^{(3)}$ evaluated at the origin $x_1 = x_2 = x_3 = 0$. The colors indicate the sign of $g_c^{(3)}(0,0,0)$, with red marks corresponding to positive values and blue marks corresponding to negative values. The cusp is an artefact of the sign change of $g_c^{(3)}$ inside the absolute value. (b) Third- and (c) second-order correlation function evaluated at the origin.}
\end{figure*}

In Fig.~\ref{fig:g3g3c}a, we present the absolute value of the connected third-order correlation at the origin as a function of optical depth. At low optical depths, $g_c^{(3)}(0,0,0)$ displays negative values, indicating the presence of non-Gaussian anti-correlations. As OD increases to approximately $4\beta M \approx 0.6$, the wavefunction begins to receive significant positive contributions from multiple photon-photon scattering processes as we explained in item 2 above, which counteract and ultimately reverse the initial decreasing trend in $g_c^{(3)}(0,0,0)$. When OD exceeds 2 ($\beta M > 0.5$), these positive contributions become dominant and increase. Meanwhile, the output power $\expval{\hat{a}^\dagger (x)\hat{a}(x)}$ in the denominator exponentially decays at large OD, causing $g_c^{(3)}(0,0,0)$ to transition into a regime of exponential growth.

In Fig.~\ref{fig:g3g3c}b and \ref{fig:g3g3c}c, we observe that both correlation functions $g^{(2)}(0,0)$ and $g^{(3)}(0,0,0)$ vanish at specific OD values. Notably, $g^{(3)}(0,0,0)$ reaches zero at a lower OD compared to $g^{(2)}(0,0)$. This phenomenon can be explained by examining the outgoing wavefunctions in Eq.~\ref{eq:expanded psi}. In both \(\psi_{2}\) and \(\psi_{3}\), the overall behavior is set by two competing terms: the two-photon entangled amplitude \(\phi_{2}\) and the product of single-photon scattering coefficients $t_0^M$. The three-photon amplitude \(\phi_{3}\) is higher order in \(\beta\) and therefore sub-dominant. Importantly, \(\phi_{2}\) is negative and decays only sub-exponentially with optical depth, whereas the individual-scattering factors \(t_{0}^{2M}\) and \(t_{0}^{3M}\) are positive and decay exponentially. Their opposite signs mean that each correlation function crosses zero at a particular OD. In \(\psi_{3}(0,0,0)\), the positive single-photon term is enhanced by a combinatorial factor of three relative to \(\psi_{2}(0,0)\), so the complete cancellation---and thus the zero of \(g^{(3)}(0,0,0)\)---occurs at a lower OD than the corresponding zero of \(g^{(2)}(0,0)\).

We now discuss the importance of non-Gaussian field correlations in intensity correlation function $g_c^{(3)}$. For each creation/annihilation operator denoted as $\hat{A}_i=\hat{a}(t)$ or $\hat{a}^\dagger(x)$, we define its fluctuation operator $\delta \hat{A}_i=\hat{A}_i-\expval{\hat{A}_i}$ whose mean is shifted to zero. Isserlis's theorem states that for a Gaussian state, $N$th order moment of $\delta\hat{A}_i$ equal to zero when $N$ is odd, and equal to the sum of all the distinct products of all the second order moments whose random variables are the pairing among $\{1,2,\cdots,N\}$ when $N$ is even~\cite{Isserlis1918,Shi2018,Hackl2021}. Mathematically, the expression reads,
\begin{align} \label{eq:isserlis}
    &\expval{:\delta \hat{A}_1\delta\hat{A}_2\cdots \delta \hat{A}_N:} \\
    &=\begin{cases}\sum_{s\in\mathcal{P}_2} \prod_{\{i,j\}\in s} \expval{:\delta\hat{A}_i \delta\hat{A}_j:} & N\text{ is even,} \\
    0 & N \text{ is odd,} \nonumber
    \end{cases}
\end{align}
where $\mathcal{P}_2$ denotes the partition of $\{1,2,\cdots,N\}$ into the subset of 2 elements. If the outgoing state $\ket{\mathrm{out}}$ has Gaussian field correlations, Isserlis's theorem holds for any observable which is linear in creation/annihilation operators. This means that we can express $g_c^{(3)}$ solely in terms of its mean and second order moments. Substituting Eq.(\ref{eq:isserlis}) into Eq.(\ref{eq:approx G3c}), combined with the approximation of for the mean $\expval{\hat{a}(x)} \approx \sqrt{P}_{\mathrm{in}} t_0^M$, we obtain the expression,
\begin{eqnarray} \label{eq:gaus G3c}
    &\widetilde{G}_c^{(3)}(x_1,x_2,x_3) = \nonumber
    2 \operatorname{Re}\Big\{N_{12}N_{23}N_{31}\\&+M_{12}^* M_{32} N_{31}+ M_{12}^* M_{31} N_{32} +M_{13}^* M_{23} N_{21} \nonumber\\
    &+ P_{\mathrm{in}}t_0^{2M}\big[M_{12}^* (N_{31}+N_{32})+M_{13}^* (N_{21}+N_{23}) \nonumber\\
    &+M_{23}^* (N_{12}+N_{13}) +N_{21}N_{13}+N_{12}N_{23}+N_{13}N_{32} \nonumber\\
    &+M_{21}^*M_{13}+M_{12}^*M_{23}+M_{23}^* M_{31}\big].
    \Big\}
\end{eqnarray}
Here, we used the shorthanded notation for the moments $M_{ij}=\expval{\delta\hat{a}(x_i)\delta\hat{a}(x_j)}$ and $N_{ij}=\expval{\delta\hat{a}^\dagger(x_i)\delta\hat{a}(x_j)}$. Equation (\ref{eq:gaus G3c}) agrees with the cumulant of the intensity correlation derived in~\cite{Cardin2024}. Furthermore, since our theory applies to the low-power limit, keeping terms up to $\mathcal{O}(\beta P_{\mathrm{in}}/\Gamma_{\mathrm{tot}})$ in $g_c^{(3)}$, we can approximate $M_{ij}\approx P_{\mathrm{in}}\phi_2(x_1,x_2)$ and $N_{ij}\approx 0$ in Eq.(\ref{eq:gaus G3c}). This gives a simple form
\begin{align} \label{eq:approx gaus G3c}
    &\widetilde{G}_c^{(3)}(x_1,x_2,x_3)\approx 2 P_{\mathrm{in}}^3 t_0^{2M} \big[\phi_2(x_1,x_2)\phi_2(x_1,x_3) \nonumber\\&+\phi_2(x_1,x_2)\phi_2(x_2,x_3)+\phi_2(x_2,x_3)\phi_2(x_1,x_3)\big].
\end{align}
Here, we notice that after applying the same perturbative approximation to $\widetilde{G}_c^{(3)}$, as we did to $G_c^{(3)}$. $\widetilde{G}_c^{(3)}$ in Eq.(\ref{eq:approx gaus G3c}) equals to the second term in Eq.(\ref{eq:approx G3c}. Moreover, their difference equal to 
\begin{eqnarray*}
    &G_c^{(3)}(x_1,x_2,x_3) - \widetilde{G}_c^{(3)} (x_1,x_2,x_3) \\
    &\approx P_{\rm{in}}^3 \big[ 2 t_0^{3M} \phi_3(x_1,x_2,x_3) + \phi_3(x_1,x_2,x_3)^2 \big] \nonumber
\end{eqnarray*}
which vanishes when there is no three photon connected scattering. Therefore, we conclude that \textit{at the order of interest in our perturbative theory, non-Gaussian field correlations in the outgoing state arise from the wavefuntion of three-photon interactions $\phi_3$}.

\subsection{Outgoing Wavefunction and Quadrature Cumulant Operator} \label{subsec:squz fluc obs}

\hspace{-0.6cm}
\begin{figure*}[hbt]
  \includegraphics[width=\textwidth]{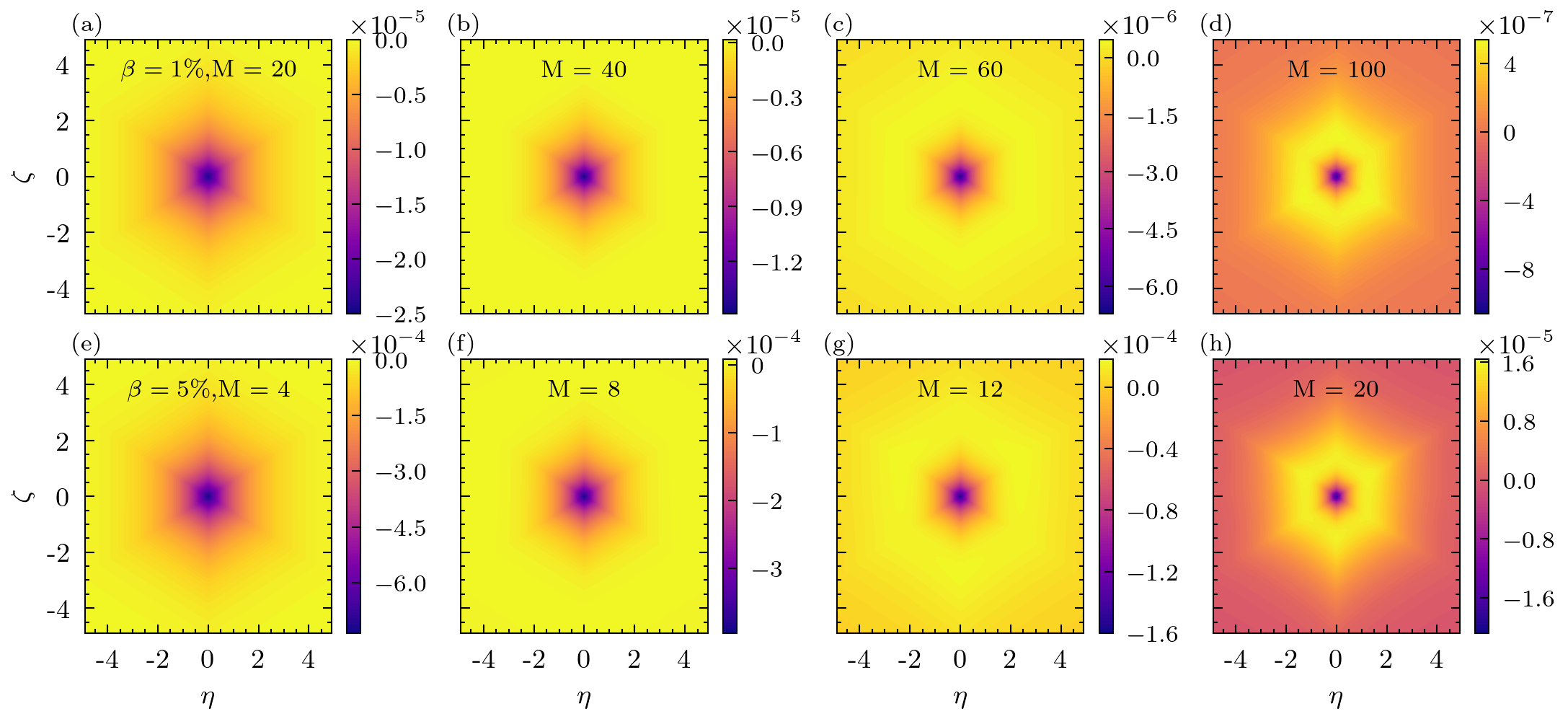}
  \caption{\label{fig:DXDXDX}  Quadrature cumulant $\expval{:\Delta  \hat{X}_\theta(x_1)\Delta  \hat{X}_\theta(x_2)\Delta  \hat{X}_\theta(x_3):}/P_{\rm in}^{3/2}=\frac{1}{4}\phi_3(x_1,x_2,x_3)$ with $\theta=0$ plotted in Jacobi coordinates with various OD from 0.8-4. The center of mass $R=0$. The coupling strength $\beta$ is 1\% in  the first row (a)-(d), while $\beta$ is 5\% in the second row (e)-(h). $\expval{:\Delta  \hat{X}_\theta(x_1)\Delta  \hat{X}_\theta(x_2)\Delta  \hat{X}_\theta(x_3):}$ is computed with tree-level diagram for $\beta=1\%$ while the loop order correction is added for $\beta= 5\%$.}
\end{figure*}
\begin{figure*}
  \centering
  \begin{tikzpicture}
    % (1) Place the full‐width graphic in a node called “main”
    \node[inner sep=0] (main) {\includegraphics[width=\textwidth]{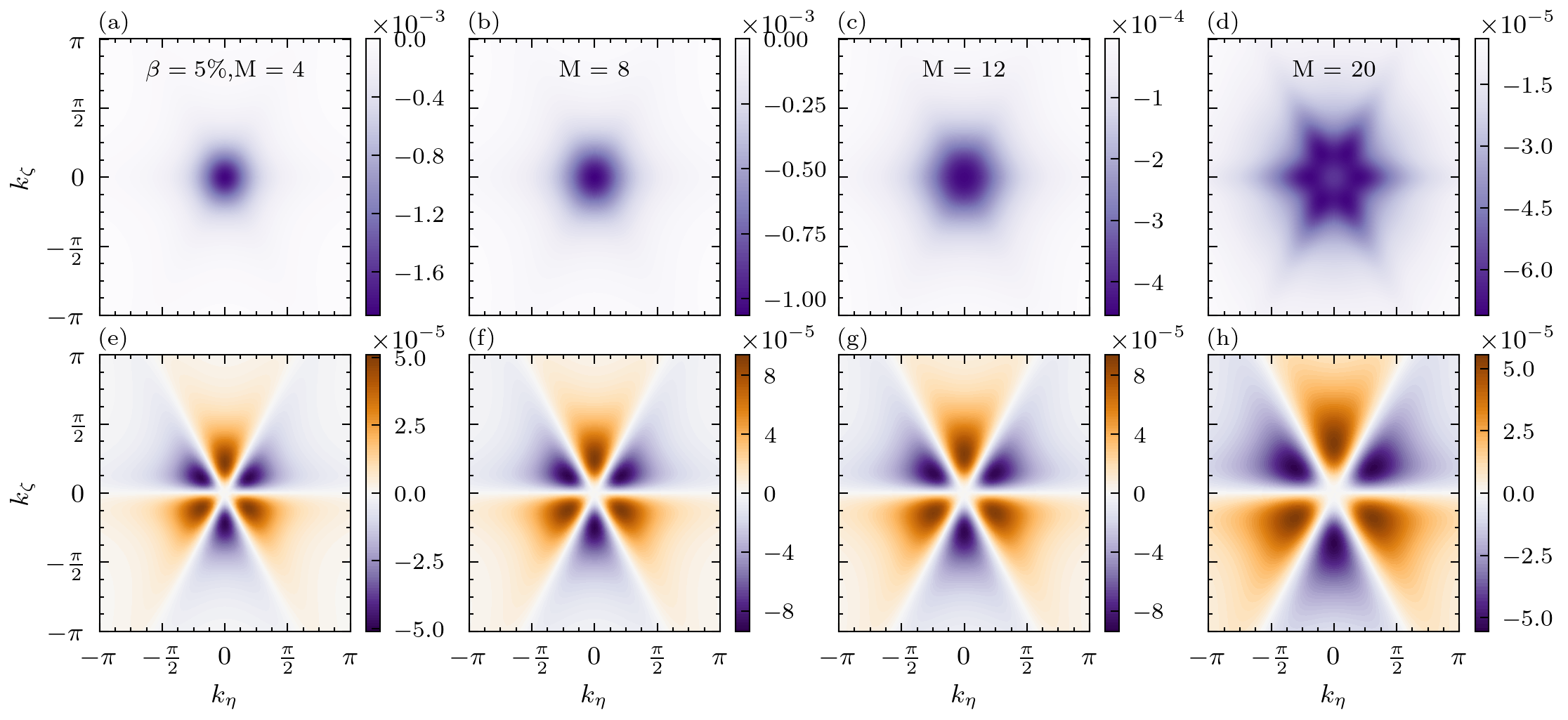}};
    % (2) Now overlay a second node at the top‐right of “main”
    \node[anchor=north east, xshift=2.7cm, yshift=-2.7cm] at (main.north west) {%
      \begin{tikzpicture}[scale=0.3]
        %— same inset code as above —
        \draw[thick] (-2, 1) -- (2, 1) ; % Top line
      \draw[thick] (-2, 0) -- (2, 0) ; % Middle line
      \draw[thick] (-2, -1) -- (2, -1) ; % Bottom line
      
      % Wavy line
      \draw[thick, decorate,
        decoration={
          snake,
          amplitude=1mm,
          segment length=4mm
        }] (0, 1) -- (0, -1);
      
      % Dots
      \filldraw[black] (0, 0) circle (2pt); % Middle dot
      \filldraw[black] (0, 1) circle (2pt); % Top dot
      \filldraw[black] (0, -1) circle (2pt); % Bottom dot
      \end{tikzpicture}
    };
  \end{tikzpicture}
  \caption{\label{fig:Forkbeta5XXX} The momentum space outgoing steady state wavefunction generated by the 3-vertex diagram (depicted in the inset at the lower-left corner of (a)) scattering at various OD with $\beta=5\%$. The first row (a)-(d) are the real part of the wavefunction, while the second row (e)-(h) are the imaginary part of the wavefunction.}
\end{figure*}

\begin{figure*}
  \centering
  \begin{tikzpicture}
    % (1) Place the full‐width graphic in a node called “main”
    \node[inner sep=0] (main) {\includegraphics[width=\textwidth]{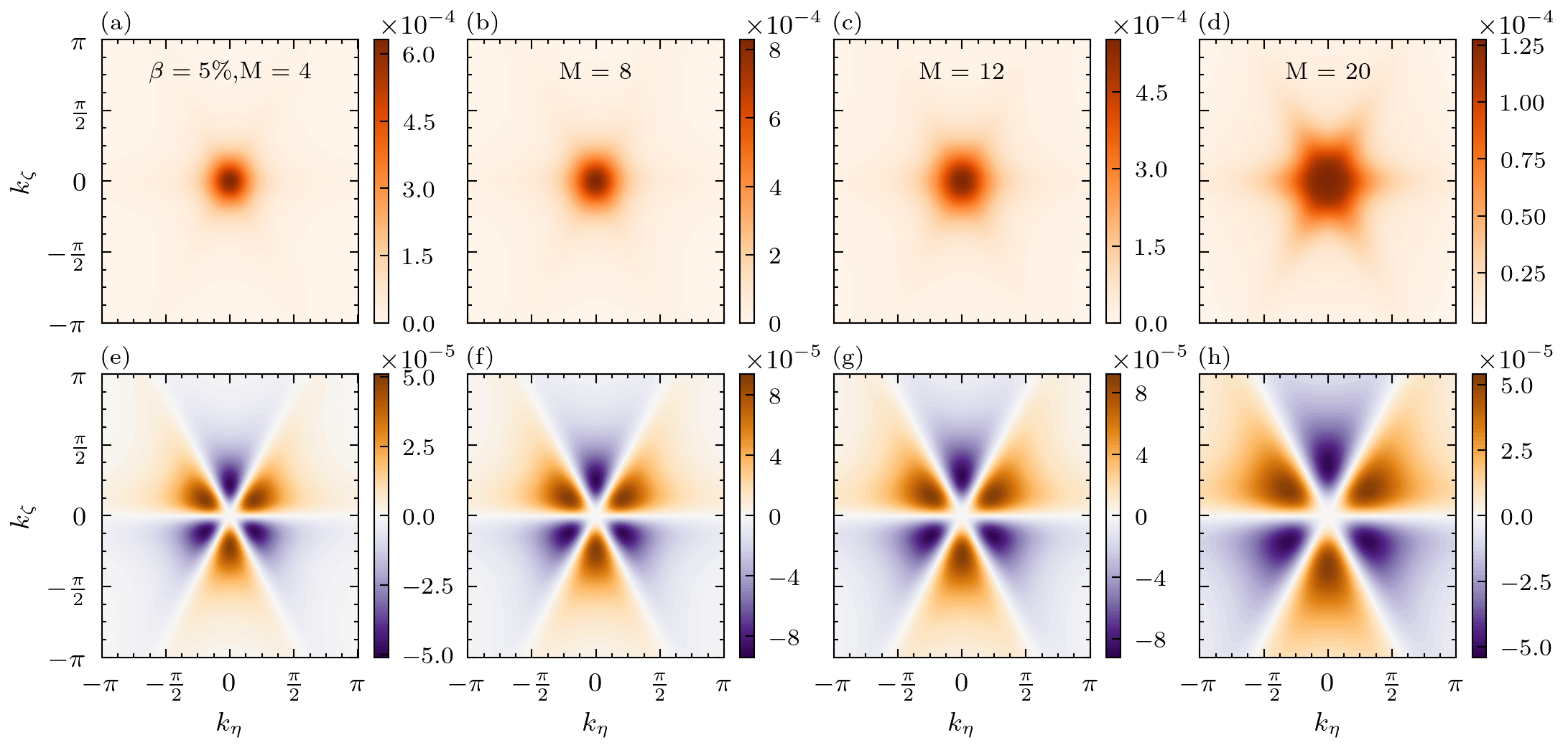}};
    % (2) Now overlay a second node at the top‐right of “main”
    \node[anchor=north east, xshift=2.7cm, yshift=-2.7cm] at (main.north west) {%
      \begin{tikzpicture}[scale=0.3]
       % Horizontal lines
      \draw[thick] (-2, 1) -- (2, 1) ; % Top line
      \draw[thick] (-2, 0) -- (2, 0) ; % Middle line
      \draw[thick] (-2, -1) -- (2, -1) ; % Bottom line
      
      % Wavy lines
      \draw[thick, decorate,
        decoration={
          snake,
          amplitude=1mm,
          segment length=4mm
        }] (-1, 1) -- (-1, 0);
      \draw[thick, decorate,
        decoration={
          snake,
          amplitude=1mm,
          segment length=4mm
        }] (1, 0) -- (1, -1);
      
      % Dots
      \filldraw[black] (-1, 0) circle (2pt); % Middle dot
      \filldraw[black] (1, 0) circle (2pt);
      \filldraw[black] (-1, 1) circle (2pt); % Top dot
      \filldraw[black] (1, -1) circle (2pt); % Bottom dot
      \end{tikzpicture}
    };
  \end{tikzpicture}
  \caption{\label{fig:Crankbeta5XXX} The momentum space outgoing steady-state wavefunction generated by the 4-vertex diagram (depicted in the inset at the lower-left corner of (a)) scattering at various OD with $\beta=5\%$. The first row (a)-(d) are the real part of the wavefunction, while the second row (e)-(h) are the imaginary part of the wavefunction.}
\end{figure*}

So far we have considered intensity correlation functions and cumulants. It is also useful to consider electric-field correlation functions of the output state. These can be measured using a balanced homodyne setup, where the output state is interfered with a local oscillator and a differential photocurrent is measured using photodetectors \cite{Mahmoodian2021}. Just like the photon intensity, one can also quantify correlations and cumulants of the scattered field and relate these to the scattered photon wavefunction. Previous work\,\cite{Mahmoodian2021} has found that, the second-order quadrature cumulant operator is proportional to the two-photon entangled part of the outgoing wavefunction in the weak driving regime.
\begin{equation} \label{eq:DXDX relation}
\begin{split}
    \expval{:\Delta \hat{X}_\theta (x_1) \Delta \hat{X}_\theta (x_2):} =& -\frac{P_{\rm in}}{2}  \biggl\{\operatorname{Re} \left[e^{2i\theta}\phi_2(x_1,x_2)\right]\\
    & +\mathcal{O}\left(\beta^2\frac{P_{\rm in}}{\Gamma_{\rm tot}}\right)\biggl\}
    \end{split}
\end{equation}
where $\hat{X}_\theta(t)=\frac{1}{2}[\hat{a}(t)e^{i\theta}+\hat{a}^\dagger (t)e^{-i\theta}]$ is the quadrature operator. One can control the phase $\theta$ of the measured field, by changing the phase of the local oscillator. This second-order field correlator is a Fourier-Transform pair with the photon squeezing spectrum. In this work, we extend this result by showing that the third-order quadrature cumulant operator is approximately proportional to the three-photon entangled part of the wavefunction,
\begin{equation}  \label{eq:DXDXDX relation}
\begin{split}
  \expval{:\prod_{i=1}^3 \Delta \hat{X}_\theta (x_i) :}=&\frac{P_{\rm in}^{\frac{3}{2}}}{4}\biggl\{\operatorname{Re}[e^{i3\theta} \phi_3(x_1,x_2,x_3)]\\
  &+\mathcal{O}\left(\beta^3\frac{P_{\rm in}}{\Gamma_{\rm tot}}\right)\biggl\}.
\end{split}
\end{equation}
The derivation of this equation is presented in Appendix~\ref{apx:DXDXDX}. Note that in this work we consider a resonant driving field and the output wavefunction is real valued. We therefore set $\theta=0$ and can determine $\phi_3(x_1,x_2,x_3)$ using this single quadrature measurement.  This provides a direct way to measure the non-Gaussian field correlations of the output state which lies in the three-photon entangled part of the wavefunction.

In Fig.~\ref{fig:DXDXDX}, we use Jacobi coordinates to show $\expval{:\prod_{i=1}^3 \Delta \hat{X}_\theta (x_i):}$ of the transmitted photons  for different atom numbers. Similar to $g_c^{(3)}$ in Fig.~\ref{fig:g3c}, we here set the center of mass coordinate $R=0$ due to the translational invariance of the steady state. In the low optical depth limit (Fig. \ref{fig:DXDXDX}a and \ref{fig:DXDXDX}e), $\phi_3$ are negative, as we explained in item 1 in the semi-quantitative analysis in Sec.\ref{subsec:g3g3c}. For larger optical depths, $\phi_3$ shows a remarkable structure of \textit{a ring of positive values around the negative center}. The numerical values of $\phi_3$ are small in the plots due to the weak atom-waveguide coupling. This small value should be compared with $\expval{X_\theta(x)}^3/P_{\rm in}^{3/2}$ and $\expval{:\prod_{i=1}^3 \hat{X}_\theta (x_i) :}/P_{\rm in}^{3/2}$, both of which are  $\sim10^{-1}$ for small OD, and $\sim10^{-3}$ at large OD.  The ring-like structure can be analyzed by examining the momentum-space outgoing wavefunctions generated by the leading-order diagrams (tree-level), which are displayed in the Fig.~\ref{fig:Forkbeta5XXX} and \ref{fig:Crankbeta5XXX}. Since the center of mass Jacobi coordinate $R=0$, the center of mass momentum $k_R=0$ as well. The relation between the original outgoing momentum $p_i$ and Jacobi momentum $k_j$ reads: $k_R=(p_1+p_2+p_3)/\sqrt{3}$, $k_\eta= (p_1-p_2)/\sqrt{2}$ and $k_\zeta=\sqrt{2/3}[(p_1+p_2)/2-p_3]$. The three axes in Fig.~\ref{fig:DXDXDX} correspond to the state with one resonant photon and two photons with opposite momenta with respect to the resonance frequency. There are a few points about Fig.~\ref{fig:DXDXDX} worth noting: 
\begin{enumerate}
    \item The momentum-space amplitude of 3-vertex diagram in Fig.~\ref{fig:Forkbeta5XXX} and 4-vertex diagram in Fig.~\ref{fig:Crankbeta5XXX} have opposite signs in their real and imaginary parts, respectively.
    
    \item The amplitude of both 3 and 4-vertex are symmetric under transformation $\psi(\vec{p})\rightarrow \psi^*(-\vec{p})$. This is the direct consequence of the reality of the steady-state wavefunction in the position space~\cite{MandelandWolf}.

    \item At a sufficiently large optical depth $OD \gtrsim 4$, the amplitude of the momentum-space wavefunction near the origin ($k_\zeta = k_\eta = 0$) decreases, eventually forming a hole. This occurs because the on-resonance photons are more likely to be scattered out of the waveguide than the off-resonance photons.
       
    \item By comparing the real part of the amplitude Fig.~\ref{fig:Forkbeta5XXX}a-d and ~\ref{fig:Crankbeta5XXX}a-d, we see the amplitude from 3-vertex diagram spreads more broadly around the center, while the 4-vertex diagram amplitude is more concentrated at the center. After inverse Fourier transforming the amplitudes to position space, this behavior reverses: in position space, the negative amplitude of 3-vertex diagram becomes concentrated at the center, whereas the positive amplitude of 4-vertex diagram spreads out. Adding these two amplitude components together in position space gives the ``negative center and positive ring'' pattern observed in Fig.~\ref{fig:DXDXDX}c d and Fig.~\ref{fig:DXDXDX}g h.
 \end{enumerate}
    
%The probability integral of three-photon entangled wavefunctions $\int_{-\infty}^{\infty} dx_1 dx_2 dx_3 |\phi_3(x_1,x_2,x_3)|^2$ first  increases with atom number and then eventually decreases. This is because of the competition between the combinatorics of diagrams and photon loss. When the number of atoms is small, the value of the integral remains small because the first few atoms rarely interact with the light for the small $\beta$. However, the magnitude of the entangled three-photon wavefunction increases linearly with atom number at the start of the chain as the amplitude of the wavefunction dominated by 3-vertex diagram is proportional to the number of atoms in the array. The contribution of 4-vertex diagram is negligible at the for small $M$ because it is one order higher in $\beta$. When the OD is sufficiently large, the photons have a greater likelihood of being scattered out of the waveguide, and thus the amplitude decays at larger OD. 

\subsection{Numerical Verification} \label{subsec:num ver}

To assess the precision of our diagrammatic expansion developed and its application to resonant coherent inputs, we benchmark its predictions against numerical solutions of the cascaded master equation with the quantum-regression theorem (QRT). The comparison focuses on third-order quadrature cumulants and $g_c^{(3)}$ for small atom numbers, where a full numerical simulation is tractable.

For numerical calculation of third-order quadrature cumulants, the left-hand side of Eq.~(\ref{eq:DXDXDX relation}) can be expressed as a summation of correlators involving creation and annihilation operators, each of which can be numerically evaluated using the QRT applied to the cascaded master equation derived in \cite{Mahmoodian2023}. For $g_c^{(3)}$, we perform similar simulations by calculating $g^{(2)}$ and $g^{(3)}$ using the QRT.

We evaluated $\expval{:\Delta \hat{X}_\theta (x_1) \Delta \hat{X}_\theta (x_2)\Delta \hat{X}_\theta (0):}$ and $g_c^{(3)}(x_1,x_2,0)$ across a $50 \times 50$ discrete grid with $x_1, x_2 \in [0,5]$, resulting in a matrix $\mathcal{M}$. Meanwhile, we analytically calculated these two observables by using our perturbative approach, which generated a matrix $\mathcal{M}^{\prime}$.

The relative error between these matrices was quantified using the formula:
\begin{equation}
\text{Relative Error} = \frac{\|\mathcal{M}-\mathcal{M}^{\prime}\|_F}{\|\mathcal{M}\|_F},
\end{equation}
where $\| \cdot \|_F$ is Frobenius norm defined for any $m\times n$ matrix $A$ as $\|A\|_F=\sqrt{\sum_{i=1}^m \sum_{j=1}^n |a_{ij}|^2}$. 
Our analysis shows that the relative error between the third-order quadrature cumulant values obtained through loop-order calculation and master equation simulation remains below 1.2\% for system parameters $P_{\rm in}=0.02\Gamma_{\rm tot}$, $\beta=5\%$, and $M\leq8$. The error remained below 2.8\% with a stronger input power of $P_{\rm in}=0.06\Gamma_{\rm tot}$ for $\beta=5\%$ and $M\leq 8$. 

Similarly, we compared the analytically computed $g_c^{(3)}$ (using the approximations in Eq. (\ref{eq:g3ng2}) for $g^{(3)}$ and $g^{(2)}$) with numerical QRT calculations, finding a relative error of less than 2.0\% for $\beta=5\%$, $M\leq 8$, $P_{\rm in}=0.02\Gamma_{\rm tot}$ and less than 6.2\% for $P_{\rm in}=0.06\Gamma_{\rm tot}$ for $\beta=5\%$, $M\leq 8$. The slightly higher error in $g_c^{(3)}$ can be attributed to its greater sensitivity to higher-order diagrams in $P_{\rm in}/\Gamma_{\rm tot}$ that were omitted in our truncated perturbative expansion. These results confirm that our perturbative approach carried to the loop level provides a highly accurate calculation of the system's quantum correlations at small to moderate optical depth $OD\leq 1.6$.

\subsection{Non-Gaussian Intensity Correlation Signal} \label{sec:expCountRate}

\begin{figure*}
    \hspace*{-0.8cm}
    \centering
    \includegraphics[width=0.8\textwidth]{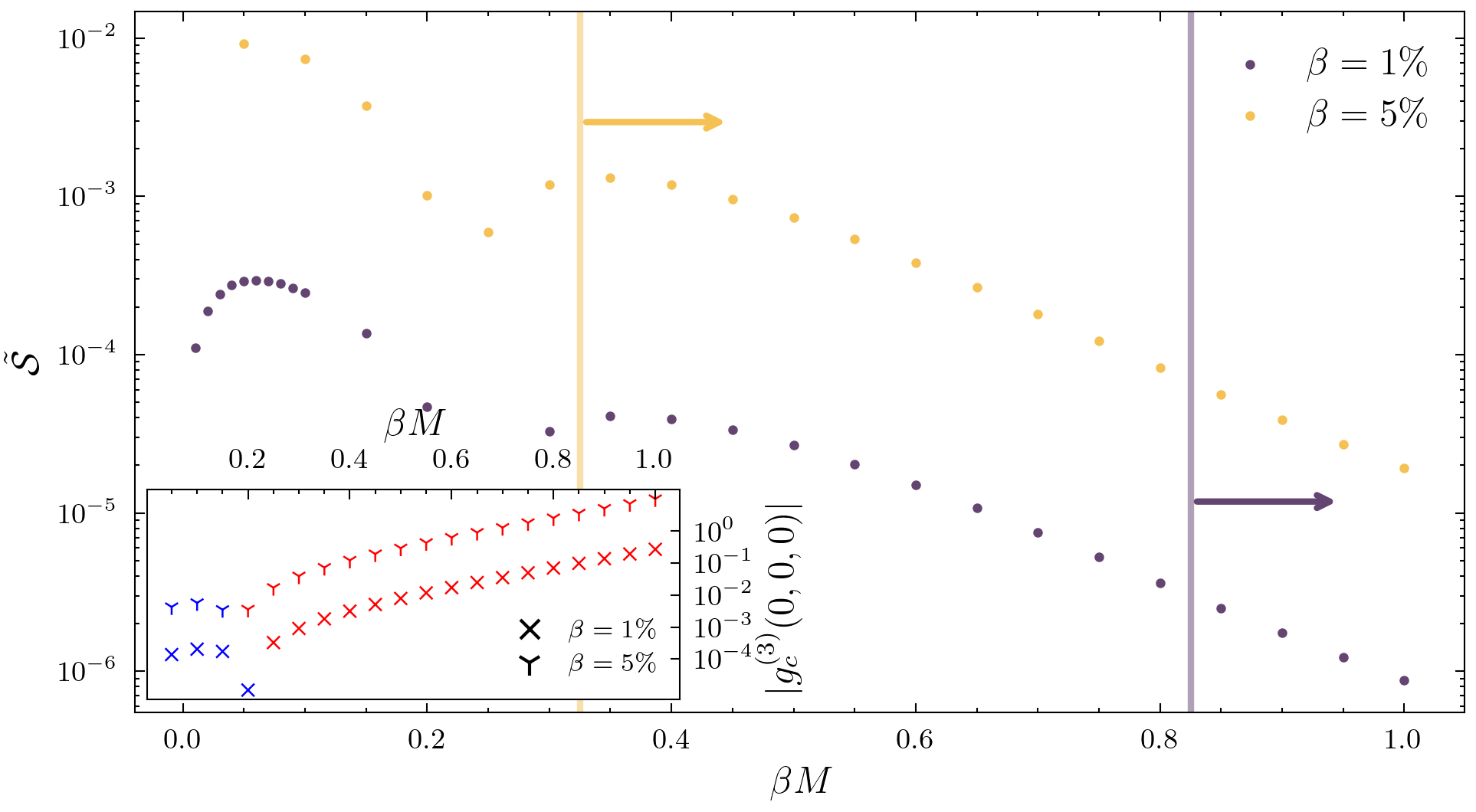}
    \caption{\label{fig:count rate} 
Dimensionless triple-coincidence signal  
\(
  \tilde{\mathcal{S}}
  = \int_{0}^{3}\!d\tau_{1}\,d\tau_{2}\;
    \bigl|G^{(3)}_{c}(t_{1},t_{2},0)\bigr|/P_{\mathrm{in}}^{3}
\) 
as a function of \(\beta M\).  
Yellow (\(\beta=5\%\)) and purple (\(\beta=1\%\)) points are evaluated with
tree-level transport processes only.  
Vertical lines mark the smallest \(\beta M\) for which
\(\lvert g^{(3)}_{c}(0,0,0)\rvert\ge 0.1\);
the arrows indicate the region to the right where the
non-Gaussian signal should be experimentally accessible.  
Inset: Fig.~\ref{fig:g3g3c}a logarithmic plot of \(\lvert g^{(3)}_{c}(0,0,0)\rvert\) versus \(\beta M\)
for the same coupling strengths.}
\end{figure*}

In this subsection, we analyze the experimentally observable signal strength of non-Gaussian photon intensity correlation to identify regimes where three-photon intensity correlations are detectable under experimental conditions. This quantity is interesting because it is directly related to the measurability of $g_c^{(3)}$ which reflects three-photon physics.

To experimentally isolate the physics arising from three-photon interactions, we quantify the non-Gaussian photon intensity correlation signal $\mathcal{S}$ within a window time of $3/\Gamma_{\rm tot}$ as,
\begin{equation}
\mathcal{S}=\int_0^{3/\Gamma_{\text{tot}}}dt_1 dt_2\,|G_c^{(3)}(t_1,t_2,0)|.
\end{equation}
The choice of a window time of $3/\Gamma_{\rm tot}$ is arbitrary, as $G_c^{(3)}(t_1,t_2,0)$ decays at large times and is square integrable, but simplifies numerical convergence. Here, the absolute value is employed because any non-vanishing $G_c^{(3)}$ indicates a deviation of the outgoing state from Gaussian statistics. Without it, the positive and negative components of $G_c^{(3)}$ could cancel during integration, even in the presence of true non-Gaussian Intensity correlation. The absolute value ensures all departures from Gaussian correlation contribute to $\mathcal{S}$. The signal $\mathcal{S}$ can be factorized into the product of a dimensionful part and a dimensionless part $\tilde{\mathcal{S}}$. The former part solely depends on the driving power $P_{\rm in}$, while the second part $\tilde{\mathcal{S}}$ is estimated with full two-photon and three-photon tree-level wavefunctions. Therefore, $\tilde{\mathcal{S}}$ reflects the fraction of the signal among all the input photons to the ensemble,
\begin{eqnarray*}
    \mathcal{S}&=& P_{\rm in}\left(\frac{P_{\rm in}}{\Gamma_{\rm tot}}\right)^2\tilde{S}, \\
    \tilde{\mathcal{S}}&=&\int_{0}^{3} d\tau_1 d\tau_2 \frac{|G_c^{(3)}(\tau_1,\tau_2,0)|}{P_{\rm in}^3},\quad \tau_i=\Gamma_{\rm tot} t_i.
\end{eqnarray*}
Because \(\mathcal{S} \propto P_{\rm in}^3\), increasing the drive power significantly enhances detectability of correlated photon triples. However, our perturbative theory remains accurate under the condition \(\mathcal{O}(P_{\rm in}/\Gamma_{\rm tot}) =\mathcal{O}(\beta)\). For quantitative agreement with our perturbation theory at all values of OD, we therefore require \(P_{\rm in} \lesssim 4\beta \Gamma_{\rm tot}\). Numerically, for a system with \(\beta = 5\%\) and \(M \leq 8\), we observe that raising the input power to \(P_{\rm in} = 10\beta \Gamma_{\rm tot}\) moderately reduces the magnitude of \(g_c^{(3)}(t_1, t_2, 0)\) and $\tilde{\mathcal{S}}$, while preserving its overall structure~\cite{WM2025}, indicating qualitative agreement with our low-power perturbative theory. Due to the cubic dependence of $\mathcal{S}$ on $P_{\rm in}$ and slow decay of $g_c^{(3)}$ and $\tilde{\mathcal{S}}$ with increasing power, experimental measurements can easily balance count rates with the magnitude of $g_c^{(3)}$.

As observed with $|g_c^{(3)}(0,0,0)|$, $\tilde{\mathcal{S}}$ in Fig.~\ref{fig:count rate} for $\beta=1\%$ and $5\%$ exhibits a dip at OD$\sim 1$, which indicates transition from three-photon non-Gaussian intensity anti-correlations to correlations. We also observe distinct patterns in the signal curves: for $\beta=1\%$, $\tilde{\mathcal{S}}$ initially increases with atom number before decreasing, while for $\beta=5\%$, $\tilde{\mathcal{S}}$ reaches its maximum with just a single atom. The decrease of $\tilde{\mathcal{S}}$ with OD is because the dominating terms in $G_c^{(3)}$ are weighed by $t_0^M$, as we see in Eq.~(\ref{eq:approx G3c}). The maxima at low optical depths shown in Fig.~\ref{fig:count rate}, can be explained through semi-quantitative analysis.

We now examine $G^{(3)}_c$ in the low optical depth regime $M\beta\ll 1$. This allows us to make three key simplifications to Eq. (\ref{eq:approx G3c}) that yields a simple analytic expression for $G^{(3)}_c$(0,0,0). These are:
\begin{enumerate}
    \item The dominant transport processes contributing to $\phi_3$ and $\phi_2$ are represented by Fig.~\ref{fig:fork} and Fig.~\ref{fig:two photon once}, respectively.

    \item Within the summand of each concatenated diagram,  the wavefunction amplitudes remain approximately equal regardless of which specific atoms host the interactions. This equivalence occurs because all summand of a concatenated diagram share the same connected $S$-matrix element, which dominates the amplitude calculation, while the effects of individual photon scattering before/after interaction sites become negligible in the low optical depth regime, i.e. when $M\beta \ll 1$. This allows us to apply approximation $\phi_3(0,0,0)\approx M\chi_3(0,0,0)$ and $\phi_2(0,0)\approx M\chi_2(0,0)$, where $\chi_2$ and $\chi_3$ are the entangled two- and three-photon position-space wavefunction for single-atom scattering.

    \item Single-atom scattering calculations reveal that $\chi_3(0,0,0)/\beta^3=-(\chi_2(0,0)/\beta^2)^2$
\end{enumerate}
These simplifications give,
\begin{eqnarray} \label{eq:approx G3czero}
    G_c^{(3)}(0,0,0)&\approx& 2P_{\rm in}^3[t_0^{3M}M\chi_3(0,0,0)+3t_0^{2M}(M\chi_2(0,0))^2] \nonumber\\
    &=&2P_{\rm in}^{3}\chi_3(0,0,0)t_0^{2M}(t_0^M M-3\beta M^2).
\end{eqnarray}
Since $\tilde{\mathcal{S}}$ within a short time interval $3/\Gamma_{\rm tot}$ can be approximated as $\tilde{\mathcal{S}}\approx 9 |G_c^{(3)}(0,0,0)|/P_{\rm in}^3$, the quadratic equation~(\ref{eq:approx G3czero}) predicts local maxima at $M\approx 1.6$ for $\beta = 5\%$ and $M \approx 8.3$ for $\beta = 1\%$, consistent with the more rigorous calculation using tree-level scattering wavefunctions.

\section{Conclusion}

In conclusion, we have developed an analytical and diagrammatic framework to unravel the complexities of three-photon interactions in atomic ensembles. By employing a Bethe Ansatz method combined with Yudson’s representation, we derived explicit expressions for the multi-photon $S$-matrix elements that not only capture individual transmission but also isolate genuine photon–photon interactions through their connected parts. Our recursive approach to the two-photon scattering problem led to a conjectured polynomial form for the outgoing wavefunction laid down the foundation for extending the analysis to three-photon processes. Future work could focus on developing the integral technique to address the integral over incoming photon momenta when $N$-photon connected part of $S$-matrix ($N\geq 3$) is involved.

The diagrammatic expansion, leaning on the perturbative parameter  $\beta$, allowed us to distinguish between tree level, one-loop level and higher-order contributions. The connectedness of the diagrams reflects how successive photon–photon interactions build up non-trivial correlations, as reflected in the behavior of both the connected third-order correlation function  $g_c^{(3)}$  and the third-order quadrature cumulant. These two observables indicate that the photons have non-Gaussian correlations. Notably, our analytical predictions are supported by numerical simulations using a cascaded master equation, which confirm that even in the weak-coupling regime, the emergent multi-photon effects are robust and experimentally observable.

Our diagrammatic framework could be extended to address many-photon input cases, paving the way to study richer non-Gaussian correlations and emergent many-body effects in atomic ensembles. Moreover, by applying similar calculations to a variety of input states---such as squeezed, or engineered superposition states---we can explore the generation of novel non-Gaussian light fields with potential applications in quantum information processing. These directions promise to deepen our understanding of complex photon–photon interactions and to advance the design of innovative photonic quantum devices.

\begin{acknowledgments}
Y.M. W. acknowledges support through Sydney Quantum Academy. S.M. acknowledges support from the Australian Research Council (ARC) via the Future Fellowship, ‘Emergent many-body phenomena in engineered
quantum optical systems’, project no. FT200100844. Noé Demazure thanks the École Normale Supérieure Paris-Saclay for its financial support.
\end{acknowledgments}

\newpage
% The \nocite command causes all entries in a bibliography to be printed out
% whether or not they are actually referenced in the text. This is appropriate
% for the sample file to show the different styles of references, but authors
% most likely will not want to use it.
\nocite{*}

\bibliography{apssamp}% Produces the bibliography via BibTeX.

\newpage
\appendix
\onecolumngrid
\section{\label{apx:n photon S matrix} Derivation of the $S$-matrix element for $n$-photon scattering in momentum space}

The Fourier Transformation of the position space $S$-matrix element (\ref{eq:real Sm}) is given by the integral:
\begin{eqnarray} \label{eq:fourier int}
  &&\int_{-\infty}^\infty d^N x\, e^{i(k_1 x_1+\dots+k_N x_N)}S_{y_1\dots y_N,x_1\dots x_N} \nonumber\\
  =&&\frac{1}{N!} \sum_{P}\int_{y_1-t}^0 e^{ik_1 x_1}f(y_{P_1},\xi_1)dx_1 \prod_{j=2}^N \int_{y_j-t}^{y_{j-1}-t} e^{ik_j x_j}f(y_{P_j},\xi_j)dx_j \nonumber\\
\end{eqnarray}
where $f(y,\xi)$ is the propagator for a single photon scattering in even subspace.
\begin{equation*}
    f(y,\xi)=\delta(y-\xi)-\Gamma_{\rm tot}\theta(y<\xi)e^{\frac{\Gamma_{\rm tot}}{2}(y-\xi)}
\end{equation*}
The sum $\sum_{P}$ is not evaluated over all the permutations of ${y_1,...,y_N}$ but part of the permutations, which abide by the condition:
\begin{equation} \label{cond}
    P_j \geq j-1,\,j=2,...,N.
\end{equation}
To perform the integral over $x_j$'s, we first observe that the $N$-fold integral over $x_j$ in (\ref{eq:fourier int}) are factorizable and can be rewritten below as a product of two types of integrals $K_1$ and $K_2$ over each $x_j$, whose domain is restricted by theta function $\theta(y_N \leq \xi_N  \leq ... y_1 \leq \xi_1)$ and the condition $y_1 > \dots > y_N$. The expression for these two types of integrals reads (after taking the asymptotic limit $t\rightarrow \infty$),
\begin{eqnarray} \label{eq:K1}
    K_1(y_{P_1},y_1)=&&\int_{y_1-t}^0 e^{ik_1 x_1}f(y_{P_1},\xi_1)dx_1 \nonumber\\
    =&&\begin{cases}
        e^{-ik_1 t+ik_1y_1}\left(1-\frac{i\Gamma_{\rm tot}}{k_1+i\frac{\Gamma_{\rm tot}}{2}}\right),&{P_1}=1 \\
        \frac{-i\Gamma_{\rm tot}}{k_1+i\frac{\Gamma_{\rm tot}}{2}}e^{-ik_1 t+ik_1 y_1-\frac{\Gamma_{\rm tot}}{2}(y_1-y_{P_1})}, & {P_1}>1 
    \end{cases} \nonumber\\
\end{eqnarray}
as well as,
\begin{eqnarray} \label{eq:K2}
    && K_2(y_{P_j},y_j)=\int_{y_j-t}^{y_{j-1}-t} e^{ik_j x_j} f(y_{P_j},\xi_j)dx_j \nonumber\\
    =&& \begin{cases}
        e^{-ik_jt + i k_j y_{P_j}} ,& P_j=j-1\\
        e^{-ik_jt} [\frac{k_j-i\frac{\Gamma_{\rm tot}}{2}}{k_j+i\frac{\Gamma_{\rm tot}}{2}}e^{ik_j y_j} + \frac{i\Gamma_{\rm tot}}{k_j+i\frac{\Gamma_{\rm tot}}{2}}e^{ik_j y_{j-1}-\frac{\Gamma_{\rm tot}}{2}(y_{j-1}-y_j)}] ,& P_j=j\\
        e^{-ik_jt} \frac{i\Gamma_{\rm tot}}{k_{j}+i\frac{\Gamma_{\rm tot}}{2}}[e^{ik_{j}y_{j-1}-\frac{\Gamma_{\rm tot}}{2}(y_{j-1}-y_{P_j})}
        - e^{ik_{j}y_{j}-\frac{\Gamma_{\rm tot}}{2}(y_{j}-y_{P_j})}] ,&P_j>j
    \end{cases} \nonumber\\
    &&
\end{eqnarray}
for $j=2,...,N$. Here, (\ref{eq:fourier int}) is just the sum of the product of $ K_1(y_{P_1},y_1)$ and all the $K_2(y_{P_j},y_j)$ with the various permutations in the set $P$, 
\begin{equation} \label{eq:K prod}
  I_N =\frac{1}{N!}\sum_{P} K_1(y_{P_1},y_1) \prod_{j=2}^N K_2(y_{P_j},y_j).
\end{equation}
One can show that $P$ includes $2^{N-1}$ elements. The above formulae also imply that the some terms in the sum associated with different permutations may share a common factor.

Next, we wish to further simplify the sum in the Eq. (\ref{eq:fourier int}). Intuitively, we could understand the condition (\ref{cond}) as: Any index $j$ are allowed to move towards right by up to 1 position in a legitimate permutation. Such that the index $j$ can become $P_{j+1}$ but not $P_{j+2}$ in a legitimate permutation. This condition motivates us to partition the set $P$ into the subsets in terms of the position of the largest index $N$, the position of the second largest index $N-1\dots$. In the following, $P(N)$ labels the permutation set for $N$ photons. For instance, the permutation set $P(2)$ is partitioned into two subsets:
\begin{equation*}
  \text{subset with $P_1 =2$},\quad \{(2,1)\}\quad \text{and}\quad P_2=2\quad \{(1,2)\}.
\end{equation*}
$P(3)$ is partitioned into three subsets.\\
\begin{eqnarray*}
  \text{subsets with $P_1=3$},\quad & \{(3,1,2)\}\\
  P_2=3,\quad & \{(1,3,2)\}\\
  P_3=3,\quad & \{(2,1,3),(1,2,3)\}.
\end{eqnarray*}
$P(4)$ is partitioned into four subsets:
\begin{eqnarray*}
  \text{subsets with $P_1=4$},\quad & \{(4,1,2,3)\} \\
  P_2=4,\quad & \{(1,4,2,3)\}\\
  P_3=4,\quad & \{(2,1,4,3),(1,2,4,3)\}\\
  P_4=4,\quad & \{(3,1,2,4),(1,3,2,4),\\
  & (2,1,3,4),(1,2,3,4)\}.
\end{eqnarray*}

Here, we notice an important feature that there is an obvious one-to-one correspondence between the elements in $P(2)$ and the elements in the subset of $P(3)$ with $P_3=3$. In the meantime, the subsets of $P(4)$ with $P_4=4$ is isomorphic to $P(3)$, while its subsets with $P_3=4$ is also isomorphic to $P(2)$. This is the natural consequence of the condition (\ref{cond}). Due to its constraint, when the largest index $N$ is placed at $P_{m}$, the position of the indices on its right hand side, such as $P_{m+1}$, $P_{m+2}$,...$P_N$ are fixed and must equal to $m$, $m+1$,...,$N-1$, respectively. In contrast, all the indices on the left hand side of $P_m$, which can only be the integers from 1 to $m-1$, are allowed to permute among themselves, just as the permutations in $P(m-1)$.

Based on the above analysis, the terms in the sum $\sum_{P}$ associated with the permutations within the same class can be grouped together. In the exemplary 4-photon case, the Eq. (\ref{eq:K prod}) simply reads
  \begin{eqnarray*}
      I_4=&&\frac{1}{4!}\Bigl\{K_1(y_4,y_1) K_2(y_1,y_2) K_2(y_2,y_3) K_2(y_3,y_4)\quad \text{subset $P_1=4$}\\
      +&&K_1(y_1,y_1) K_2(y_4,y_2) K_2(y_2,y_3) K_2(y_3,y_4)\quad \text{subset $P_2=4$ }\\
      +&&\Bigl[K_1(y_2,y_1) K_2(y_1,y_2)+K_1(y_1,y_1) K_2(y_2,y_2)\Bigr]K_2(y_4,y_3) K_2(y_3,y_4)\quad \text{subset $P_3=4$}\\
      +&&\Big[K_1(y_3,y_1)K_2(y_1,y_2)K_2(y_2,y_3)+K_1(y_1,y_1)K_2(y_3,y_2)K_2(y_2,y_3)\quad \text{subset $P_4=4$}\\
      +&&K_1(y_2,y_1)K_2(y_1,y_2)K_2(y_3,y_3)+K_1(y_1,y_1)K_2(y_2,y_2)K_2(y_3,y_3)
      \Big]K_2(y_4,y_4)
      \Bigr\}.
  \end{eqnarray*}

We recognize that the terms in the two square brackets are nothing but (\ref{eq:K prod}) with $2!I_2$ and $3!I_3$. Let us define $\widetilde{I}_N  \equiv N! I_N $ and rewrite the above equation.

\begin{eqnarray*} \label{eq:expr I4}
    \widetilde{I}_4 (y_1,...,y_4)=&&\Bigl[K_1(y_4,y_1) K_2(y_1,y_2) K_2(y_2,y_3) K_2(y_3,y_4)
    +\widetilde{I}_1(y_1) K_2(y_4,y_2) K_2(y_2,y_3) K_2(y_3,y_4) \\
     +&&\widetilde{I}_2(y_1,y_2)K_2(y_4,y_3) K_2(y_3,y_4)+\widetilde{I}_3(y_1,y_2,y_3) K_2(y_4,y_4) \Bigr].
\end{eqnarray*}

This expression can easily be generalized to the $N$-photon case with,
\begin{eqnarray} \label{eq:expr IN}
  &&\widetilde{I}_N (y_1,...,y_N)= \nonumber\\
  &&K_1(y_N,y_1) K_2(y_1,y_2) K_2(y_2,y_3) \cdots K_2(y_{N-1},y_N)  \nonumber
  + \sum_{j=1}^{N-1} \widetilde{I}_j(y_1,...,y_j) K_2(y_N,y_{j+1}) K_2(y_{j+1},y_{j+2}) \cdots K_2(y_{N-1},y_N) \nonumber\\
  &&
\end{eqnarray}

We substitute $K_1(\cdot,\cdot)$ and $K_2(\cdot,\cdot)$ by their expression (\ref{eq:K1}) and (\ref{eq:K2}), obtaining $\widetilde{I}_N$ in terms of the $\widetilde{I}_{N-1}$'s, $\widetilde{I}_{N-2}$'s, ...$\widetilde{I}_1$'s. We drop the arguments of $\widetilde{I}_{j}$ in the following expressions for notational simplicity,
\begin{eqnarray*}
  \widetilde{I}_N  =&&\frac{-i\Gamma_{\rm tot}}{k_1+i\frac{\Gamma_{\rm tot}}{2}}e^{ik_1y_1+i\sum_{j=2}^N k_j y_{j-1}-\frac{\Gamma_{\rm tot}}{2}(y_1-y_N)} 
  + \sum_{j=1}^{N-2} \widetilde{I}_j \frac{i\Gamma_{\rm tot}}{k_{j+1}+i\frac{\Gamma_{\rm tot}}{2}}[e^{ik_{j+1}y_j-\frac{\Gamma_{\rm tot}}{2}(y_j-y_{N})}\\
  -&& e^{ik_{j+1}y_{j+1}-\frac{\Gamma_{\rm tot}}{2}(y_{j+1}-y_{N})}]e^{\sum_{m=j+2}^{N+1}i k_m y_{m-1}} 
  + \widetilde{I}_{N-1}\Bigl[\frac{k_N-i\frac{\Gamma_{\rm tot}}{2}}{k_N+i\frac{\Gamma_{\rm tot}}{2}}e^{ik_N y_N}\\
  +&&  \frac{i\Gamma_{\rm tot}}{k_N+i\frac{\Gamma_{\rm tot}}{2}}e^{ik_N y_{N-1}-\frac{\Gamma_{\rm tot}}{2}(y_{N-1}-y_N)}\Bigr]. 
\end{eqnarray*}
This recursive relation can be simplified by the elementary trick of subtracting $\widetilde{I}_N$ from $\widetilde{I}_{N+1}$. We obtain the following expression for $\widetilde{I}_{N+1}$ in terms of $\widetilde{I}_{N}$'s,
\begin{eqnarray}
  \widetilde{I}_{N+1} =&&\widetilde{I}_N \Bigl[\frac{k_{N+1}-i\frac{\Gamma_{\rm tot}}{2}}{k_{N+1}+i\frac{\Gamma_{\rm tot}}{2}} e^{i k_{N+1}y_{N+1}}
  +\frac{k_{N+1}+\frac{3i\Gamma_{\rm tot}}{2}}{k_{N+1} +\frac{i\Gamma_{\rm tot}}{2}} e^{\frac{\Gamma_{\rm tot}}{2}(y_{N+1}-y_N)+i k_{N+1}y_N}\Bigr] \nonumber\\
  -&&\widetilde{I}_{N-1} e^{i(k_{N+1}+k_N)y_N+\frac{\Gamma_{\rm tot}}{2}(y_{N+1}-y_N)}.
\end{eqnarray}
The solution of the above recursive relation is a bit complicated. We define the following notation in order to concisely write down the solution.
\begin{equation*}
  [N]= \{ x\in \mathbb{N}|1\leq x \leq N \},
\end{equation*}
\begin{equation*}
  \mathcal{P}_{2p}([N]) = \{S\in \mathcal{P}([N])\Big | \quad|S|=2p\}.
\end{equation*}
where $\mathcal{P}([N])$ denotes the power set of $[N]$ with total order, and $\mathcal{P}_{2p}([N])$ denotes the lexicographically-ordered set of all the subsets of $[N]$ with the cardinality of $2p$. We use $\mathcal{P}_{2p}^{(m)}([N])$ denotes the $m$th element in the set $\mathcal{P}_{2p}([N])$, and $\Theta_{pmk}$ denotes the $k$th element in the set $\mathcal{P}_{2p}^{(m)}([N])$. For each set $\mathcal{P}_{2p}^{(m)}([N])$, we define
\begin{equation*}
  \mathcal{C}(\mathcal{P}_{2p}^{(m)}([N])) = [N]\backslash (\cup_{q=1}^{p} [\Theta_{pm(2q-1)},\Theta_{pm(2q)}]).
\end{equation*}
where $\backslash$ is the set difference $A\backslash B=\{x\in A|x\notin B\}$. 

The general solution of $\widetilde{I}_N$ reads

  \begin{eqnarray} \label{eq:half transformed IN}
    \widetilde{I}_N =&&\left(\prod_{j=1}^{N} \frac{k_j-i\frac{\Gamma_{\rm tot}}{2}}{k_j+i\frac{\Gamma_{\rm tot}}{2}} e^{i k_j y_j}\right) + \left(\prod_{j=1}^{N} \frac{1}{k_j+i\frac{\Gamma_{\rm tot}}{2}}\right) \sum_{p=1}^{\lfloor N/2\rfloor} (i\Gamma_{\rm tot})^p \sum_{m=1}^{\binom{N}{2p}} \left(\prod_{c\in \mathcal{C}(\mathcal{P}_{2p}^{(m)}([N]))} e^{i k_c y_c} (k_c-\frac{i\Gamma_{\rm tot}}{2})\right) \nonumber\\
    \times && \prod_{q=1}^{p} e^{i k_{\Theta_{pm(2q-1)}} y_{\Theta_{pm(2q-1)}}+i\sum_{v=\Theta_{pm(2q-1)}+1}^{\Theta_{pm(2q)}} k_v y_{v-1}}(k_{\Theta_{pm(2q-1)}}-k_{\Theta_{pm(2q-1)}+1}-i\Gamma_{\rm tot}) A(\Theta_{pm(2q-1)},\Theta_{pm(2q)}) \nonumber\\
    \times &&  e^{-\frac{\Gamma_{\rm tot}}{2}(y_{\Theta_{pm(2q-1)}}-y_{\Theta_{pm(2q)}})},
  \end{eqnarray}
where, 
\begin{equation*}
  A(i,j) =
\begin{cases}
1, & \text{if } j = i+1, \\[1em]
\displaystyle \prod_{r=i+2}^{\,j} \left( k_r + \frac{3i\,\Gamma_{\mathrm{tot}}}{2} \right), & \text{if } j \ge i+2.
\end{cases}
\end{equation*}
Taking $N=3$ and $k_j=0$ for all $j=1,2,3$ reproduce Eq. (6) in the previous work~\cite{Stiesdal2018}.

Directly Fourier Transforming the expression (\ref{eq:half transformed IN}) with respect to the outgoing coordinates $y_1,...,y_N$ to obtain the full $N$-photon $S$-matrix is not straightforward. For the practical purpose of this work, we only need to work out $S_{p_1 p_2,k_1 k_2}$ and $S_{p_1 p_2 p_3, k_1 k_2, k_3}$ or, equivalently, $S_{p_1 p_2,k_1 k_2}^C$ and $S_{p_1 p_2 p_3, k_1 k_2, k_3}^C$.

Recall the definition (\ref{eq:def for S connected}),
\begin{eqnarray} \label{eq:derivation for S2C}
  S_{p_1 p_2,k_1 k_2}^C =&& S_{p_1 p_2,k_1 k_2} - \frac{1}{2}(S_{p_1,k_1}S_{p_2,k_2}+S_{p_2,k_1}S_{p_1,k_2}) \nonumber\\
  = &&\frac{1}{2!} \sum_{\hat{\sigma}(\{p_j\})}\sum_{\hat{\sigma}(\{k_j\})}\int_{-\infty}^{\infty} dy_1 dy_2\theta(y_1>y_2) e^{-i p_1 y_1-i p_2 y_2}[\widetilde{I}_2(k_1,k_2, y_1,y_2)-I_1(k_1,y1)I_1(k_2,y2) ] \nonumber\\
  = && 2\pi i\Gamma_{\rm tot}^2\frac{(p_1+p_2+i\Gamma_{\rm tot})\delta(p_1+p_2-k_1-k_2)}{(p_1+i\frac{\Gamma_{\rm tot}}{2})(p_2+i\frac{\Gamma_{\rm tot}}{2})(k_1+i\frac{\Gamma_{\rm tot}}{2})(k_2+i\frac{\Gamma_{\rm tot}}{2})},
\end{eqnarray}
where
\begin{eqnarray*}
  I_1 (k,y)=&& \frac{k-i\frac{\Gamma_{\rm tot}}{2}}{k+i\frac{\Gamma_{\rm tot}}{2}}e^{i k y},\\
  \widetilde{I}_2(k_1,k_2, y_1,y_2) =&& \frac{k_1-i\frac{\Gamma_{\rm tot}}{2}}{k_1+i\frac{\Gamma_{\rm tot}}{2}} \frac{k_2-i\frac{\Gamma_{\rm tot}}{2}}{k_2+i\frac{\Gamma_{\rm tot}}{2}}e^{i k_1 y_1+ik_2 y_2}+ e^{i(k_1+k_2)y_1 -\frac{\Gamma_{\rm tot}}{2} (y_1-y_2)}\frac{\Gamma_{\rm tot}(k_1-k_2-i\Gamma_{\rm tot})}{({k_1+i\frac{\Gamma_{\rm tot}}{2}})({k_2+i\frac{\Gamma_{\rm tot}}{2}})},
\end{eqnarray*}
In the second equality, since $I_N$ is derived under the sector $x_1>x_2>\dots>x_N$ and $y_1>y_2>\dots>y_N$, the full $S$-matrix should be obtained by symmetrizing both the incoming and outgoing momenta. Specifically:
\begin{itemize}
  \item The contribution of the other sectors corresponding to various permutation among $x_j$'s can be obtained by symmetrizing the incoming momenta.
  \item The contribution of the other sectors corresponding to various permutation among $y_j$'s can be obtained by symmetrizing the outgoing momenta.
\end{itemize}
In the equations of $S_{p_1 \dots p_n,k_1\dots k_n}^C$ in the main text, a factor of $1/(2\pi)^n$ factor is absorbed into $S$-matrix. It arises from the normalization of Fourier transformation $\hat{a}^\dagger(x)=\int \frac{dp}{2\pi}\, \exp(-ipx)\hat{a}^\dagger(p)$.
\begin{eqnarray*}
  \hat{S}_{p_1p_2p_3,k_1k_2k_3}^C =&& \hat{S}_{p_1p_2p_3,k_1k_2k_3} 
- \frac{1}{3!}(S_{p_1,k_1}S_{p_2,p_3,k_2,k_3}^C+\text{permutations}) 
- \frac{1}{3!}(S_{p_1,k_1}S_{p_2,k_2}S_{p_3,k_3}+\text{permutations}) \\
=&& \frac{1}{3!}\sum_{\hat{\sigma}(\{p_j\})}\sum_{\hat{\sigma}(\{k_j\})}\int_{-\infty}^{\infty} dy_1dy_2dy_3\,\theta(y_1>y_2>y_3) e^{-ip_1 y_1-ip_2 y_2-ip_3 y_3}\Bigl\{\widetilde{I}_3(k_1,k_2,k_3,y_1,y_2,y_3)\\
-&&I_1(k_1,y_1)[\widetilde{I}_2(k_2,k_3,y_2,y_3)-I_1(k_2,y_2)I_1(k_3,y_3)]-I_1(k_2,y_2)[\widetilde{I}_2(k_1,k_3,y_1,y_3)-I_1(k_1,y_1)I_1(k_3,y_3)] \\
-&&I_1(k_3,y_3)[\widetilde{I}_2(k_1,k_2,y_1,y_2)-I_1(k_1,y_1)I_1(k_2,y_2)]-I_1(k_1,y_1)I_1(k_2,y_2)I_1(k_3,y_3)]]\Bigr\}\\
 =&& -\frac{4i\pi\Gamma_{\rm tot}^3}{3}\sum_{\hat{\sigma}(\{p_j\})}\sum_{\hat{\sigma}(\{k_j\})}\frac{\delta(p_1+p_2+p_3-k_1-k_2-k_3)}{(p_1+p_2-k_1+i \frac{\Gamma_{\rm tot}}{2})(p_1+i\frac{\Gamma_{\rm tot}}{2})(k_1+i\frac{\Gamma_{\rm tot}}{2})(k_2+i\frac{\Gamma_{\rm tot}}{2})(k_3+i\frac{\Gamma_{\rm tot}}{2})}.
\end{eqnarray*}

\section{\label{apx:two photon recursive} Proof for the conjectured polynomial form (\ref{eq:two photon conjecture})}

The first step of the induction is to show that the outgoing wavefunction after the first atom $\psi_1(k_1,k_2)$ satisfies the polynomial form (\ref{eq:two photon conjecture}), 
\begin{equation*}
  \psi_1(p_1,p_2) = \int_{-\infty}^{\infty} dk_1 dk_2 \, {}_{22}S_{p_1 p_2, k_1 k_2} \psi_0(k_1, k_2), 
\end{equation*}
where $\psi_0(k_1,k_2)=\delta(k_1)\delta(k_2)$ is  on-resonant two-photon state (we here temporarily dropped the normalization factor $e^{-|\alpha|^2}\alpha^2/(2L)$ from the coherent state). The explicit form of the $S$-matrix ${}_{22}S$ is 
\begin{align*}
  &{}_{22}S_{p_1 p_2, k_1 k_2} = \frac{1}{2}t_{p_1}t_{p_2}\left(\delta(p_1-k_1)\delta(p_2-k_2)+\delta(p_1-k_2)\delta(p_2-k_1)\right)+\frac{i\beta^2 \Gamma_{\rm tot}^2}{2\pi
  }\frac{(p_1+p_2+i\Gamma_{\rm tot})\delta(p_1+p_2-k_1-k_2)}{(p_1+i\frac{\Gamma_{\rm tot}}{2})(p_2+i\frac{\Gamma_{\rm tot}}{2})(k_1+i\frac{\Gamma_{\rm tot}}{2})(k_2+i\frac{\Gamma_{\rm tot}}{2})}.
\end{align*}
After the momentum integral, we obtain:
\begin{equation*}
  \psi_1(k_1,k_2) = t_0^2\delta(k_1)\delta(k_2) + \frac{2\beta^2}{\pi}\frac{\Gamma_{\rm tot}\,\delta(k_1+k_2)}{(k_1+i\frac{\Gamma_{\rm tot}}{2})(k_2+i\frac{\Gamma_{\rm tot}}{2})}.
\end{equation*}
We see that $\psi_1(k_1,k_2)$ satisfies the polynomial form (\ref{eq:two photon conjecture}) with $d=1$.

Next, we assume that the outgoing wavefunction after the $d$th atom $\psi_d(k_1,k_2)$ satisfies the polynomial form (\ref{eq:two photon conjecture}) with degree upto $d$. We wish to show that the outgoing wavefunction after the $(d+1)$th atom $\psi_{d+1}(k_1,k_2)$ also satisfies the polynomial form (\ref{eq:two photon conjecture}) with degree upto $d+1$. Let us look at the following integral:
\begin{align*}
  &\psi_{d+1}(p_1,p_2) = \int_{-\infty}^{\infty} dk_1 dk_2 \, {}_{22}S_{p_1 p_2, k_1 k_2} \psi_d(k_1, k_2), \\
  &\text{with} \\
  &\psi_d(k_1,k_2) = t_0^{2d}\delta(k_1)\delta{(k_2)}
  +\sum_{j,l=1}^{d} C_{j,l}^{(d)}\,\Gamma_{\rm tot}^{j+l-1} \left(k_1+i\frac{\Gamma_{\rm tot}}{2}\right)^{-j}\left(k_2+i\frac{\Gamma_{\rm tot}}{2}\right)^{-l}.
\end{align*}
We here notice that the above integral is nothing but a linear combination of the following dimensionless integral which we define as $F(j,l)$.
\begin{equation*}
  F(j,l)=\int_{-\infty}^{\infty} dk_1 dk_2\, \delta(k_1+k_2) \frac{\Gamma_{\rm tot}^{j+l-1}}{(k_1+i\frac{\Gamma_{\rm tot}}{2})^{j}(k_2+i\frac{\Gamma_{\rm tot}}{2})^{l}}.
\end{equation*}
Here, $F(j,l)$ can be solved either by residue theorem or by the IBP technique for Feynman integral. The result is:
\begin{equation*}
  F(j,l) =  2\pi i^{-j-l} \frac{\prod_{k=1}^{l-1}(j+k-1)}{(l-1)!}.
\end{equation*}
The result of the above momentum integral is,
\begin{align*}
  &\psi_{d+1}(k_1,k_2) =  t_0^{2(d+1)}\delta(k_1)\delta(k_2) +\delta(k_1+k_2)\Biggl\{\Gamma_{\rm tot} (k_1+i\frac{\Gamma_{\rm tot}}{2})^{-1}(k_2+i\frac{\Gamma_{\rm tot}}{2})^{-1} \\
  &\Bigl[C_{1,1}+t_0^{2d}\left(\frac{2\beta^2}{\pi}\right)-\frac{\beta^2}{2\pi}\sum_{j,l=1}^{d}C_{j,l}F(j+1,l+1)\Bigr] +\sum_{j+l\geq 3}^{d+1} \Gamma_{\rm tot}^{j+l-1}\left(k_1+i\frac{\Gamma_{\rm tot}}{2}\right)^{-j}\left(k_2+i\frac{\Gamma_{\rm tot}}{2}\right)^{-l} \\
  &\Bigl[C_{j,l}^{(d)}-i\beta(C_{j-1,l}^{(d)}+C_{j,l-1}^{(d)})+(-i\beta)^2C_{j-1,l-1}^{(d)}\Bigr]\Biggr\}.
\end{align*}
We see that $\psi_{d+1}(k_1,k_2)$ also satisfies the polynomial form (\ref{eq:two photon conjecture}) with degree upto $d+1$. This completes the induction. In addition, if we compare the coefficient of the polynomial term with the one in the conjectured form, we obtain the recursive relation (\ref{eq:two photon recursion}) in the main text.

Including photon loss in the transmission model can be done by similar calculation. To consider the process where one photon is lost at the $d+1$th atom, we simply   multiply the two-photon outgoing wavefunction after the $d$th atom $\psi_d(k_1,k_2)$ with ${}_{21}S_{\slashed{p}_1p_2,k_1k_2}$, 
\begin{equation*}
  \psi_{d+1}(\slashed{p}_1,p_2) = \int_{-\infty}^{\infty} dk_1 dk_2 \, {}_{21}S_{\slashed{p}_1p_2, k_1 k_2}  \psi_d(k_1, k_2). 
\end{equation*}

Performing the momentum integral, we obtain:
\begin{align*}
  &\psi_{d+1}(\slashed{k}_1,k_2) = r_0t_0^{2d+1}\delta(k_1)\delta(k_2) + \frac{\beta^{3/2} \sqrt{1-\beta}}{2\pi} \Bigl[4 t_0^{2d}  - \sum_{j,l=1}^{d} C_{j,l}^{(d)} F(j+1,l+1)\Bigr] \frac{\Gamma_{\rm tot} \delta(k_1+k_2)}{(k_1+i\Gamma_{\rm tot}/2)(k_2+i\Gamma_{\rm tot}/2)} \\
  &-i\sqrt{\beta(1-\beta)}\sum_{j+l\geq 3}^{d+1} \left(C_{j-1,l}^{(d)}-i\beta C_{j-1,l-1}^{(d)}\right)\frac{\Gamma_{\rm tot}^{j+l-1}\delta(k_1+k_2)}{(k_1+i\Gamma_{\rm tot}/2)^j (k_2+i\Gamma_{\rm tot}/2)^l}
\end{align*}

The outgoing state after one photon loss is 
\begin{equation*}
  \int_{-\infty}^{\infty} dk_1dk_2 \,\psi_{d+1}(\slashed{k}_1,k_2)\hat{b}_{d+1}^\dagger (k_1)\hat{a}^\dagger(k_2)+\psi_{d+1}(k_1,\slashed{k}_2)\hat{a}^\dagger(k_1)\hat{b}_{d+1}^\dagger (k_2)\ket{0}
\end{equation*}

\section{\label{apx:loop order}The Calculation of Loop-Level Three-Photon Transport}

The loop-level scattering matrix \(\hat{T}^{(2)}\) is defined as the sum of the unsymmetrized expressions \ref{eq:C1}--\ref{eq:C6} below, each of which should be symmetrized over the outgoing momenta \(p_1, p_2, p_3\)  before being added together. To perform the integral over loop momentum in the diagrams at order $\mathcal{O}(\beta^3)$, we first observe that the integrals in all six diagrams can be reduced to several fundamental forms. Each fundamental form can be calculated by using residual theorem, and the residual is computed by Laurent expansion near the pole.

For the reader's convenience, we provide below a list of these fundamental integrals. Unless otherwise specified, the parameters $a$, $b$, and $c$ are treated as positive integers throughout.
\begin{eqnarray*}
    \int_{-\infty}^{\infty} \frac{1}{(l + i\,\Gamma_{\text{tot}}/2)^a \cdot (- l + i\,\Gamma_{\text{tot}}/2)^b} \, dl = 2\pi i^{-a-b} \Gamma_{\text{tot}}^{1-a-b} \frac{(a+b-2)!}{(a-1)!(b-1)!},
\end{eqnarray*}
\begin{eqnarray*}
    \int_{-\infty}^{\infty}&& \frac{1}{(l + i\,\Gamma_\text{tot}/2)^a \cdot (-l + i\,\Gamma_\text{tot}/2)^b \cdot (l + p_1 + p_2 + i\,\Gamma_\text{tot}/2)^c} \, dl = \\
    -&&2\pi i^{1-a} (p_1+p_2+i\,\Gamma_\text{tot})^{1-b-c} \cdot \Gamma_\text{tot}^{-a} \cdot \binom{b+c-2}{b-1} \cdot {}_2F_1\left(a,1-b,2-b-c,\frac{-i(p_1+p_2+i\,\Gamma_\text{tot})}{\Gamma_\text{tot}}\right),
\end{eqnarray*}
with $c\geq 2$, and ${}_2F_1(a,b,c,z)$ is the ordinary hypergeometric function. We also have,
\begin{eqnarray*}
\int_{-\infty}^{\infty}&& \frac{1}{(l + i\,\Gamma_{\text{tot}}/2)^a \cdot (-l + i\,\Gamma_{\text{tot}}/2)^b \cdot (l + p_1 + p_2 + i\,\Gamma_{\text{tot}}/2)} \, dl = \\
-&&2i(-p_1-p_2)^{-a} \pi (p_1+p_2+i\,\Gamma_{\text{tot}})^{-b} \left[1-b\,\text{B}\left(1-\frac{i(p_1+p_2)}{\Gamma_{\text{tot}}};b,a\right) \binom{a+b-1}{b}\right],
\end{eqnarray*}
where $\text{B}(z;a,b)$ is incomplete Beta function, and,
\begin{eqnarray*}
    \int_{-\infty}^{\infty} \frac{1}{(l + i\,\Gamma_{\text{tot}}/2)^a \cdot (-l - p_1 + i\,\Gamma_{\text{tot}}/2)^b} \, dl = -2\pi i \binom{a+b-2}{b-1} \cdot (-p_1+i\,\Gamma_{\text{tot}})^{(-a-b+1)}.
\end{eqnarray*}

We give expressions for the contribution of the diagrams to the outgoing wavefunction in the subsections below. We dedicate one subsection for the evaluation of each diagram. The diagrams below show photon transport from the left to the right. The sum index $j$ labels the possible number of atoms before the first photon-photon interaction happen. The index $m$ labels the number of atoms between two interactions. The convention of labeling also applies to the sum in the next subsection.

\subsection{The expression for the Fig.~\ref{fig: three two diag}}

\begin{figure}[hbt]
\centering
    \begin{tikzpicture}
      % Horizontal lines
      \draw[thick] (-2, 1) -- (2, 1) node[right]{$p_1$}; % Top line
      \draw[thick] (-2, 0) -- (2, 0) node[right]{$p_2$}; % Middle line
      \draw[thick] (-2, -1) -- (2, -1)node[right]{$p_3=-p1-p2$}; % Bottom line
      % Wavy lines
      \draw[thick, decorate,
        decoration={snake, amplitude=1mm, segment length=4mm}]
        (-1, 1) -- (-1, -1);
      \draw[thick, decorate,
        decoration={snake, amplitude=1mm, segment length=4mm}]
        (1, 0)-- (1, -1);
      
      % Dots
      \filldraw[black] (-1, -1) circle (2pt);
      \filldraw[black] (-1, 0) circle (2pt); % Middle dot
      \filldraw[black] (1, 0) circle (2pt);
      \filldraw[black] (-1, 1) circle (2pt); % Top dot
      \filldraw[black] (1, -1) circle (2pt);  % (Label comment unchanged)
      \node at (0, 0.2) {$l$};
      \node at (0,-1.25) {$-p_1-l$};
    \end{tikzpicture}
\end{figure}
The contribution to the outgoing wavefunction of this diagram is,
\begin{eqnarray} \label{eq:C1}
  &&\int_{-\infty}^{\infty} dl\, S_{p_1 l\,(-p_1-l),000}^C S_{p_2(-p_1-p_2),l(-p_1-l)}^C \times \sum_{j=0}^{M-2}\sum_{m=0}^{M-j-2} t_0^{3j}(t_l t_{-p_1-l})^m t_{p_1}^{M-j-1} (t_{p_2}t_{-p_1-p_2})^{M-j-m-2}\nonumber\\
  =&& \frac{-2 i \beta^5 \Gamma_{\text{tot}}^3 (-p_1 + i \Gamma_{\text{tot}})}
  {\pi^2 p_1 (-p_1 - p_2 + i \Gamma_{\text{tot}}/2) (p_2 + i \Gamma_{\text{tot}}/2)}\sum_{j=0}^{M-2} \sum_{m=0}^{M-j-2} 
  t_0^{3j} t_{p_1}^{M - j - 1} (t_{p_2} t_{-p_1 - p_2})^{M - j - m - 2} \nonumber\\
  &&  \times \sum_{r=0}^{m} \sum_{s=0}^{m} 
  \binom{m}{r} \binom{m}{s} (-i \beta \Gamma_{\text{tot}})^{r+s} \times \Bigg\{ 
  \frac{1}{p_1 - i \Gamma_{\text{tot}}/2} \binom{r+s+2}{r+1} (-p_1 + i \Gamma_{\text{tot}})^{-r-s-2} \nonumber\\
  && + \frac{1}{p_1 + i \Gamma_{\text{tot}}/2} 
  \left( (-p_1)^{-r-1}(i \Gamma_{\text{tot}})^{-s-1} \right. \times \Biggl[ 1 - (s+1)\binom{r+s+1}{s+1} 
  \text{B}\left(\frac{-i \Gamma_{\text{tot}}}{p_1 - i \Gamma_{\text{tot}}}; s+1, r+1\right) \Biggr]  \nonumber\\
  &&  + (-p_1)^{-s-1} (i \Gamma_{\text{tot}})^{-r-1} \left.\times \Biggl[1 - (r+1) \binom{r+s+1}{r+1} 
  \text{B}\left(\frac{-i \Gamma_{\text{tot}}}{p_1 - i \Gamma_{\text{tot}}}; r+1, s+1\right) \right) \Biggr]\Bigg\}.
  \end{eqnarray}

\subsection{The expression for the Fig.~\ref{fig:two three diag}}

\begin{figure*}[hbt]
    \centering
    \begin{tikzpicture}
      % Horizontal lines
      \draw[thick] (-2, 1) -- (2, 1) node[right]{$p_1$};
      \draw[thick] (-2, 0) -- (2, 0) node[right]{$p_2$};
      \draw[thick] (-2, -1) -- (2, -1) node[right]{$p_3=-p_1-p_2$};
      
      % Wavy lines
      \draw[thick, decorate,
        decoration={snake, amplitude=1mm, segment length=4mm}]
        (-1, 1) -- (-1, 0);
      \draw[thick, decorate,
        decoration={snake, amplitude=1mm, segment length=4mm}]
        (1, 1) -- (1, -1);
      
      % Dots
      \filldraw[black] (1, 1) circle (2pt);
      \filldraw[black] (-1, 0) circle (2pt); % Middle dot
      \filldraw[black] (1, 0) circle (2pt);
      \filldraw[black] (-1, 1) circle (2pt); % Top dot
      \filldraw[black] (1, -1) circle (2pt); % Bottom dot
      \node at (0,1.3) {$l$};
      \node at (0,-0.2) {$-l$};
    \end{tikzpicture}
\end{figure*}

The contribution to the outgoing wavefunction of this diagram is,
\begin{eqnarray}
&&\int_{-\infty}^{\infty}dl\,S_{l(-l),00}^CS_{p_1 p_2 (-p_1-p_2),l(-l)0}^C \sum_{j=0}^{M-2}\sum_{m=0}^{M-j-2} t_0^{3j+m+1} (t_l t_{-l})^m (t_{p_1}t_{p_2}t_{-p_1-p_2})^{M-j-m-2} \nonumber\\
&&= -\frac{2 \beta^5 \Gamma_{\text{tot}}^3}{3 \pi^3} 
\sum_{j=0}^{M-2} 
\sum_{m=0}^{M-j-2} 
t_0^{3j + m + 1} 
(t_{p_1} t_{p_2} t_{p_3})^{M - j - m - 2} \sum_{r=0}^{m} 
\sum_{s=0}^{m} 
\binom{m}{r} \binom{m}{s} 
(-i \beta \Gamma_{\text{tot}})^{r+s} \nonumber\\
&&\times \sum_{\sigma(\{p_k\})} 
\frac{1}{p_1 + i \frac{\Gamma_{\text{tot}}}{2}} 
\left( 
\frac{1}{p_1 + p_2 + i \frac{\Gamma_{\text{tot}}}{2}} 
\frac{2 i^{-r-s} \pi \Gamma_{\text{tot}}^{-3-r-s} (2 + r + s)!}{(1 + r)! (1 + s)!} 
\right) \nonumber\\
&& - 2i (-p_1 - p_2)^{-2 - r} \pi (p_1 + p_2 + i \Gamma_{\text{tot}})^{-2 - s} \nonumber\\
&&\times \left( 
1 - (2 + s) \, \text{B} \left(1 - \frac{i (p_1 + p_2)}{\Gamma_{\text{tot}}}; 2 + s, 2 + r \right) 
\binom{3 + r + s}{2 + s} \right) \nonumber\\
&&-2i(-p_1-p_2)^{-2-s}\pi(p_1+p_2+i\Gamma_{\text{tot}})^{-2 - r} \nonumber\\
&&\times \left(1 - (2 + r) \, \text{B} \left(1 - \frac{i (p_1 + p_2)}{\Gamma_{\text{tot}}}; 2 + r, 2 + s \right) 
\binom{3 + r + s}{2 + r} 
\right).
\end{eqnarray}

\subsection{The expression for the Fig.~\ref{fig:twotwotwo mid diag}}

The sum index $j$ here again labels the number of atoms before the first photon-photon interaction from the left to the right. $q$ labels the number of atoms between the first and the second interaction, while $m$ labels the number of atoms between the second and the third interaction. This convention also applies to the diagrams in the next three subsections.

\begin{figure*}[hbt]
     \begin{tikzpicture}
      % Horizontal lines
      \draw[thick] (-2, 1) -- (2, 1) node[right]{$p_1$};
      \draw[thick] (-2, 0) -- (2, 0) node[right]{$p_2$};
      \draw[thick] (-2, -1) -- (2, -1) node[right]{$p_3=-p_1-p_2$};
      
      % Wavy lines
      \draw[thick, decorate,
        decoration={snake, amplitude=1mm, segment length=4mm}]
        (-1, 1) -- (-1, 0);
      \draw[thick, decorate,
        decoration={snake, amplitude=1mm, segment length=4mm}]
        (1,1) -- (1,0);
      \draw[thick, decorate,
        decoration={snake, amplitude=1mm, segment length=4mm}]
        (0, 0) -- (0, -1);
      
      % Dots
      \filldraw[black] (-1, 0) circle (2pt); % Middle dot
      \filldraw[black] (-1, 1) circle (2pt); % Top dot
      \filldraw[black] (0, -1) circle (2pt);
      \filldraw[black] (0, 0) circle (2pt);
      \filldraw[black] (1, 0) circle (2pt);
      \filldraw[black] (1, 1) circle (2pt); % (Label comment unchanged)
      \node at (0,1.3) {$-l$};
      \node at (-0.5,-0.3) {$l$};
      \node at (1.0,-0.3) {$l+p_1+p_2$};
    \end{tikzpicture}
\end{figure*}

The contribution to the outgoing wavefunction of this diagram is,

\begin{eqnarray}
  &&\sum_{j=0}^{M-3}\sum_{m=0}^{M-j-3}\sum_{q=0}^{M-j-m-3}\int_{-\infty}^{\infty}dl\,S_{(-l)l,00}^C S_{(l+p_1+p_2)(-p_1-p_2),l0}^C S_{p_1 p_2,(-l)(l+p_1+p_2)}^C   \nonumber\\
  &&\times t_0^{3j+1+m} t_{-l}^{m+q+1} t_l^m t_{l+p_1+p_2 }^q  (t_{p_1}t_{p_2})^{M-j-m-q-3} t_{-p_1-p_2}^{M-j-m-2} \nonumber\\
  &&= \frac{i \beta^6 (p_1 + p_2 + i \Gamma_{\text{tot}}) \Gamma_{\text{tot}}^4}
  {\pi^2 (p_1 + i \Gamma_{\text{tot}}/2) (p_2 + i \Gamma_{\text{tot}}/2) (-p_1 - p_2 + i \Gamma_{\text{tot}}/2)}   \times \sum_{j=0}^{M-3} \sum_{m=0}^{M-j-3} \sum_{q=0}^{M-j-m-3} 
  t_0^{3j + m + 1} (t_{p_1} t_{p_2})^{M - j - m - q - 3} t_{-p_1 - p_2}^{M - j - m - 2} \nonumber\\
  &&  \times \sum_{r=0}^{m+q+1} \sum_{s=0}^{m} \sum_{t=0}^{q} 
  \binom{m+q+1}{r} \binom{m}{s} \binom{q}{t} \beta^{r+s+t} \Gamma_{\text{tot}}^{r+t-1} (-1)^{s+1} i^{-r-t}  \times (p_1 + p_2 + i \Gamma_{\text{tot}})^{-r-t-3} \binom{r+t+2}{r+1} \nonumber\\
  &&  \times \left( 
  2 \, {}_2F_1\left(-r-1, s+1, -r-t-2, 1 - \frac{i (p_1 + p_2)}{\Gamma_{\text{tot}}} \right)   +  {}_2F_1\left(-r-1, s+2, -r-t-2, 1 - \frac{i (p_1 + p_2)}{\Gamma_{\text{tot}}} \right) 
  \right).
  \end{eqnarray}

\subsection{The expression for the Fig.~\ref{fig:twotwotwo first diag}}

\begin{figure*}[hbt]
\centering
    \begin{tikzpicture}
      % Horizontal lines
      \draw[thick] (-2, 1) -- (2, 1) node[right]{$p_1$};
      \draw[thick] (-2, 0) -- (2, 0) node[right]{$p_2$};
      \draw[thick] (-2, -1) -- (2, -1) node[right]{$p_3=-p_1-p_2$};
      
      % Wavy lines
      \draw[thick, decorate,
        decoration={snake, amplitude=1mm, segment length=4mm}]
        (-1, 0) -- (-1, -1);
      \draw[thick, decorate,
        decoration={snake, amplitude=1mm, segment length=4mm}]
        (1,1) -- (1,0);
      \draw[thick, decorate,
        decoration={snake, amplitude=1mm, segment length=4mm}]
        (0, 1) -- (0,0);
      
      % Dots
      \filldraw[black] (-1, 0) circle (2pt); % Middle dot
      \filldraw[black] (-1, -1) circle (2pt); % (Label comment unchanged)
      \filldraw[black] (0, 1) circle (2pt);
      \filldraw[black] (0, 0) circle (2pt);
      \filldraw[black] (1, 0) circle (2pt);
      \filldraw[black] (1, 1) circle (2pt);
      \node at (-0.4,-0.3){$p_1+p_2$};
      \node at (0.5,1.3){$-l$};
      \node at (0.5,0.3){$l$};
    \end{tikzpicture}
\end{figure*}

The contribution to the outgoing wavefunction of this diagram is,

\begin{eqnarray}
  && \sum_{j=0}^{M-3}\sum_{m=0}^{M-j-3}\sum_{q=0}^{M-j-m-3} \int_{-\infty}^{\infty} dl\,S_{(p_1+p_2)(-p_1-p_2),00}^C S_{(-l)l,0(p_1+p_2)}^C S_{p_1p_2,(-l)l}^C \nonumber\\
  &&\times t_0^{3j+m+1} t_{p_1+p_2}^m (t_{-l}t_l)^q (t_{p_1}t_{p_2})^{M-j-m-q-3} t_{-p_1-p_2}^{M-j-1} \nonumber\\
  &&= -\frac{\beta^6 (p_1 + p_2 + i \Gamma_{\text{tot}}) \Gamma_{\text{tot}}^5}
  {\pi^3 (p_1 + i \Gamma_{\text{tot}}/2) (-p_1 - p_2 + i \Gamma_{\text{tot}}/2) 
  (p_2 + i \Gamma_{\text{tot}}/2) } \times \frac{1}{(p_1 + p_2 + i \Gamma_{\text{tot}}/2)^2} \nonumber\\
  &&  \times \sum_{j=0}^{M-3} \sum_{m=0}^{M-j-3} \sum_{q=0}^{M-j-m-3} 
  t_0^{3j + m + 1} t_{p_1 + p_2}^m (t_{p_1} t_{p_2})^{M - j - m - q - 3} 
  t_{-p_1 - p_2}^{M - j - 1} \nonumber\\
  &&  \times \sum_{r=0}^{q} \sum_{s=0}^{q} 
  \binom{q}{r} \binom{q}{s} (-i \beta \Gamma_{\text{tot}})^{r+s} 
  2 \pi i^{-r-s} \Gamma_{\text{tot}}^{-3 - r - s} 
  \frac{(r + s + 2)!}{(r + 1)! (s + 1)!}.
\end{eqnarray}

\newpage
\subsection{The expression for the Fig.~\ref{fig:twotwotwo last diag}}

\begin{figure*}[hbt]
    \centering
    \begin{tikzpicture}
      % Horizontal lines
      \draw[thick] (-2, 1) -- (2, 1) node[right]{$p_1$};
      \draw[thick] (-2, 0) -- (2, 0) node[right]{$p_2$};
      \draw[thick] (-2, -1) -- (2, -1) node[right]{$p_3=-p_1-p_2$};
      
      % Wavy lines
      \draw[thick, decorate,
        decoration={snake, amplitude=1mm, segment length=4mm}]
        (-1, 1) -- (-1, 0);
      \draw[thick, decorate,
        decoration={snake, amplitude=1mm, segment length=4mm}]
        (0,1) -- (0,0);
      \draw[thick, decorate,
        decoration={snake, amplitude=1mm, segment length=4mm}]
        (1, 0) -- (1, -1);
      
      % Dots
      \filldraw[black] (0, 1) circle (2pt);
      \filldraw[black] (0, 0) circle (2pt);
      \filldraw[black] (-1, 0) circle (2pt); % Middle dot
      \filldraw[black] (1, 0) circle (2pt);
      \filldraw[black] (-1, 1) circle (2pt); % Top dot
      \filldraw[black] (1, -1) circle (2pt); % Bottom dot
      \node at (-0.5,1.3) {$l$};
      \node at (-0.5,-0.3) {$-l$};
      \node at (0.4,-0.3) {$-p_1$};
    \end{tikzpicture}
\end{figure*}

The contribution to the outgoing wavefunction of this diagram is,

\begin{eqnarray}
  && \sum_{j=0}^{M-3}\sum_{m=0}^{M-j-3}\sum_{q=0}^{M-j-m-3} \int_{-\infty}^{\infty} dl\, S_{l(-l),00}^C S_{p_1(-p_1),l(-l)}^C S_{p_2 (-p_1-p_2),(-p_1)0}^C   \nonumber\\
  &&\times (t_l t_{-l})^m t_0^{3j+2+m+q} t_{-p_1}^q t_{p_1}^{M-j-m-2} (t_{p_2}t_{-p_1-p_2})^{M-j-m-q-3}  \nonumber\\
  &&= \frac{2 \beta^6 (p_1 - i \Gamma_{\text{tot}}) \Gamma_{\text{tot}}^2}
  {\pi^2  (p_1 + i \Gamma_{\text{tot}}/2) 
  (-p_1 - p_2 + i \Gamma_{\text{tot}}/2) (p_2 + i \Gamma_{\text{tot}}/2)} \times \frac{1}{(-p_1 + i \Gamma_{\text{tot}}/2)^2} \nonumber\\
  &&  \times \sum_{j=0}^{M-3} \sum_{m=0}^{M-j-3} \sum_{q=0}^{M-j-m-3} t_0^{3j + m + q + 2} t_{-p_1}^q t_{p_1}^{M - j - m - 2} 
  (t_{p_2} t_{-p_1 - p_2})^{M - j - m - q - 3} \nonumber\\
  && \times \sum_{r=0}^{m} \sum_{s=0}^{m} 
  \binom{m}{r} \binom{m}{s} (-\beta)^{r+s} 
  \frac{(r + s + 2)!}{(r + 1)! (s + 1)!}.
\end{eqnarray}

\subsection{The expression for the Fig.~\ref{fig:twotwotwo non pla diag}}

\begin{figure*}[hbt]
    \centering
    \begin{tikzpicture}
      % Horizontal lines
      \draw[thick] (-2, 1) -- (2, 1) node[right]{$p_1$};
      \draw[thick] (-2, 0) -- (2, 0) node[right]{$p_2$};
      \draw[thick] (-2, -1) -- (2, -1) node[right]{$p_3=-p_1-p_2$};
      
      % Wavy lines
      \draw[thick, decorate,
        decoration={snake, amplitude=1mm, segment length=4mm}]
        (-1, 1) -- (-1, 0);
      \draw[thick, decorate,
        decoration={snake, amplitude=1mm, segment length=4mm}]
        (0,0) -- (0,-1);
      \draw[thick, decorate,
        decoration={snake, amplitude=1mm, segment length=4mm}]
        (1,1) .. controls (2,1) and (2,-1) .. (1,-1);
      
      % Dots
      \filldraw[black] (-1, 0) circle (2pt); % Middle dot
      \filldraw[black] (-1, 1) circle (2pt); % Top dot
      \filldraw[black] (0, -1) circle (2pt);
      \filldraw[black] (0, 0) circle (2pt);
      \filldraw[black] (1, -1) circle (2pt);
      \filldraw[black] (1, 1) circle (2pt); % (Label comment unchanged)
      \node at (-0,1.3) {$-l$};
      \node at (-0.5,-0.3) {$l$};
      \node at (0.5,-1.3) {$l-p_2$};
    \end{tikzpicture}
  \end{figure*}

\begin{eqnarray} \label{eq:C6}
  && \sum_{j=0}^{M-3} \sum_{m=0}^{M-j-3} \sum_{q=0}^{M-j-m-3} \int_{-\infty}^{\infty}dl\,S_{(-l)l,00}^C S_{p_2(l-p_2),l0}^C S_{p_1(-p_1-p_2),(-l)(l-p_2)}^C \nonumber\\
  &&\times t_{-l}^{m+q+1} t_{l}^m t_{l-p_2}^q t_0^{3j+1+m} t_{p_2}^{M-j-m-2} (t_{p_1}t_{-p_1-p_2})^{M-j-m-q-3} \nonumber\\
  &&= - \frac{2 i\beta^6 (-p_2 + i \Gamma_{\text{tot}}) \Gamma_{\text{tot}}^4}
  {\pi^2 (p_1 + i \Gamma_{\text{tot}}/2) (p_2 + i \Gamma_{\text{tot}}/2) (-p_1 - p_2 + i \Gamma_{\text{tot}}/2)} \nonumber\\
  && \times \sum_{j=0}^{M-3} \sum_{m=0}^{M-j-3} \sum_{q=0}^{M-j-m-3} 
  t_0^{3j + m + 1} t_{p_2}^{M - j - m - 2} (t_{p_1} t_{-p_1 - p_2})^{M - j - m - q - 3} \nonumber\\
  && \times \sum_{r=0}^{m+q+1} \sum_{s=0}^{m} \sum_{t=0}^{q} 
  \binom{m+q+1}{r} \binom{m}{s} \binom{q}{t} 
  (-i \beta \Gamma_{\text{tot}})^{r+s+t} i^{-s}  \times (-p_2 + i \Gamma_{\text{tot}})^{-r-t-3} 
  \Gamma_{\text{tot}}^{-s-1} \binom{r+t+2}{r+1} \nonumber\\
  && \times \left( 
  {}_2F_1\left(-r-1, s+1, -r-t-2, \frac{-i (-p_2 + i \Gamma_{\text{tot}})}{\Gamma_{\text{tot}}} \right) \right.  + \left. \frac{1}{2} \, {}_2F_1\left(-r-1, s+2, -r-t-2, \frac{-i (-p_2 + i \Gamma_{\text{tot}})}{\Gamma_{\text{tot}}} \right) 
  \right). \nonumber\\
\end{eqnarray}

\section{\label{apx:Coherent state convertion}Scattering with a Coherent State}

\begin{figure}[h]
  \centering
    \begin{tabular}{@{\hspace{2em}}c@{\hspace{2em}}c@{\hspace{2em}}c@{\hspace{2em}}c@{\hspace{2em}}c@{\hspace{2em}}}
      \begin{tikzpicture}[scale=0.25, baseline=(current bounding box.center)]
        \draw (-2,-1.5) rectangle (2,1.5) node[midway] {A};
      \end{tikzpicture}
      &
      \begin{tikzpicture}[scale=0.25, baseline=(current bounding box.center)]
        \draw (-2,-1.5) rectangle (2,1.5) node[midway] {A};
        \draw[thick] (-2, 3) -- (2, 3);
      \end{tikzpicture}
      &
      \begin{tikzpicture}[scale=0.25, baseline=(current bounding box.center)]
        \draw (-2,-1.5) rectangle (2,1.5) node[midway] {A};
        \draw[thick] (-2, 3) -- (2, 3);
        
        \node[cross out, draw, line width=1.2pt, minimum size=1mm] at (-2, 3) {};
      \end{tikzpicture}
      &
      \begin{tikzpicture}[scale=0.25, baseline=(current bounding box.center)]
        \draw (-2,-1.5) rectangle (2,1.5) node[midway] {A};
        \draw[thick] (-2, 2) -- (2, 2);
         \draw[thick] (-2, 3) -- (2, 3);
      \end{tikzpicture}
      &
      \raisebox{-0.04em}{$\cdots$}
    \end{tabular}
    \caption{\label{fig:dressed diagram} (From left to right) Concatenated diagram $A$, and its dressed diagrams with one linearly transmitted photon, one linearly transmitted then lost photon, and two linearly transmitted photons. The dots represents all the diagrams with more non-interacting photons in coherent state scattering.}
\end{figure}
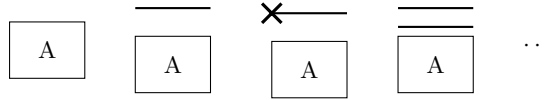

When we take the incident state to be a monochromatic, resonant coherent state, we can use the Fock-state expansion,
\begin{eqnarray*}
    \ket{\alpha}&=&e^{-|\alpha|^2/2}\left(\ket{0}+\alpha\ket{1}+\frac{\alpha^2}{\sqrt{2!}}\ket{2}+\frac{\alpha^3}{\sqrt{3!}}\ket{3}+\cdots\right)\\
    &=& e^{-|\alpha|^2/2}\left(1+\frac{\alpha}{L^{1/2}}\int dy \hat{a}^\dagger(y)+\frac{\alpha^2}{2!L}\int dy_1 dy_2 \hat{a}^\dagger(y_1)\hat{a}^\dagger(y_2)+\frac{\alpha^3}{3!L^{3/2}}\int dy_1 dy_2 dy_3 \hat{a}^\dagger(y_1)\hat{a}^\dagger(y_2)\hat{a}^\dagger (y_3)+\cdots \right)\ket{0}.
\end{eqnarray*}
Here, we recall that we are in a frame rotating at the resonance frequency of the atoms and resonant photons are thus constant-valued in space. The integral limit is from $-L/2$ to $L/2$. To calculate physical observables, we must sum over all sectors containing $0,1,2,\ldots,\infty$ photons. The purpose of this section is to show, under the assumption in Sec.~\ref{subsec:g3g3c}, the calculation of the $m$th-order correlator with $n$-photon incoming Fock state can be easily converted to the calculation with incoming coherent states.
First, we define \textit{dressed diagrams}. Take a concatenated diagram $A$ with $n$ incoming photons. Its \emph{dressed diagrams} are obtained by attaching any number of
non-interacting photons that either 1. Transmit via elastic individual scattering through the array or  
2. Are elastically scattered, but are scattered out of the array on one of the atoms. Fig~\ref{fig:dressed diagram} shows $A$ and its first few dressed diagrams. Our goal is to show that, any $L$-independent contribution $D(x_{1},\dots,x_{m})$, obtained by the matrix element with concatenated diagram $A$ and $B$ (Fig.~\ref{fig:fock input}), to the $m$-th-order correlator $G_m(x_1,x_2,\dots,x_m)$ acquires a simple overall factor
$\exp(\alpha^{2})$ when we sum over all Fock sectors.  This factor cancels the
normalization $\exp(-\alpha^{2})$ of the coherent state, so that the final answer
is  the result obtained directly with an $n$-photon input multiplied by a factor of $P_{\rm in}^n$.

  To guarantee the existence of a term $D(x_1,\dots,x_m)$, diagrams $A$ and $B$ have to satisfy a constraint: when $n>m$, (1) diagram $A$ and $B$ do not simultaneously contain a linearly transmitted photon. (2) diagram $A$ and $B$ do not simultaneously contain a linearly transmitted then lost photon.
\begin{figure}[h]
  \centering
    \begin{tabular}{@{}c@{}c@{\hspace{0.5em}}c@{\hspace{0.5em}}c@{}}
      \raisebox{-0.04cm}{\small $e^{-|\alpha|^2}\frac{|\alpha|^{2n}}{L^{n}}\frac{1}{n!^2}D(x_1,x_2,\cdots,x_m)=$}
      &
      \begin{tikzpicture}[scale=0.25, baseline=(current bounding box.center)]
        \draw (-2,-1.5) rectangle (2,1.5) node[midway] {A};
        
        \draw[thick] (-2.5, 1.5) -- (-3.5, 0);
        \draw[thick] (-3.5, 0) -- (-2.5, -1.5);
        \draw[thick] (3, 1.5) -- (3, -1.5);
      \end{tikzpicture}
      &
      \raisebox{-0.15cm}{\small $\hat{a}^\dagger(x_1)\cdots\hat{a}^\dagger(x_m)\hat{a}(x_m)\cdots \hat{a}(x_1)$}
      &
      \begin{tikzpicture}[scale=0.25, baseline=(current bounding box.center)]
        \draw (-2,-1.5) rectangle (2,1.5) node[midway] {B};
        
        \draw[thick] (2.5, 1.5) -- (3.5, 0);
        \draw[thick] (3.5, 0) -- (2.5, -1.5);
        \draw[thick] (-3, 1.5) -- (-3, -1.5);
      \end{tikzpicture}
    \end{tabular}
    \label{fig:fock input}
    \caption{The $m$th order correlator computed by diagram $A$ and $B$.}
\end{figure}

Then, let us compute a term in $n+1$-photon sector from the dressed diagrams of $A$ and $B$ with one more pair of individually transmitted photons. In the calculation below, we only keep track of the terms involving $D(x_1,\dots,x_m)$.

\begin{figure}[h]
  \centering
    \begin{tabular}{@{}c@{}c@{\hspace{0.5em}}c@{}}
      \begin{tikzpicture}[scale=0.25, baseline=(current bounding box.center)]
        \draw[thick] (-2, 2) -- (2, 2); % Top line
        \draw (-2,-2) rectangle (2,1) node[midway] {A};
        
        \draw[thick] (-2.5, 1.5) -- (-3.5, 0);
        \draw[thick] (-3.5, 0) -- (-2.5, -1.5);
        \draw[thick] (3, 1.5) -- (3, -1.5);
      \end{tikzpicture}
      &
      \raisebox{-0.15cm}{\small $\hat{a}^\dagger(x_1)\cdots\hat{a}^\dagger(x_m)\hat{a}(x_m)\cdots \hat{a}(x_1)$}
      &
      \begin{tikzpicture}[scale=0.25, baseline=(current bounding box.center)]
        \draw[thick] (-2, 1) -- (2, 1); % Top line
        \draw[thick] (-2, 2) -- (2, 2); % Top line
        \draw (-2,-2) rectangle (2,1) node[midway] {B};
        
        \draw[thick] (2.5, 1.5) -- (3.5, 0);
        \draw[thick] (3.5, 0) -- (2.5, -1.5);
        \draw[thick] (-3, 1.5) -- (-3, -1.5);
      \end{tikzpicture}
    \end{tabular}
\end{figure}
\begin{eqnarray*}
    &=& e^{-|\alpha|^2} \frac{|\alpha|^{2n+2}}{L^{n+1}} \frac{1}{(n+1)!^2} (n+1)^2 D(x_1,x_2,\cdots,x_m)\bra{0}\int dy_1 dy_2\, t_0^{M}\hat{a}(y_{n+1}')\hat{a}^\dagger(y_{n+1})t_0^{M}\ket{0}+\cdots\\
    &=& e^{-|\alpha|^2} \frac{|\alpha|^{2n+2}}{L^{n+1}} \frac{1}{n!^2} L  D(x_1,x_2,\cdots,x_m) t_0^{2M}+\cdots
\end{eqnarray*}
The term including $D(x_1,x_2,\cdots,x_m)$ came from contracting $\hat{a}(y_{n+1}')$ and $\hat{a}^\dagger(y_{n+1})$ of the additional linearly transmitted photon. The $(n+1)^2$ factor is  because, with the presence of an additional linearly transmitted photon, the number of same diagrams (but with the different permutations of photons) is multiplied by $n+1$. The leftover term represented by the dots at the end is from the contraction of $\hat{a}(y_{n+1}')$ and $\hat{a}^\dagger(y_{n+1})$ with other creation and annihilation operators. 

Similarly, for the $n+1$-photon diagrams with $A$ and $B$ added with a pair of linearly transmitted lost photon.

\begin{figure}[h!]
  \centering
    \begin{tabular}{@{}c@{}c@{\hspace{0.5em}}c@{}}
      \begin{tikzpicture}[scale=0.25, baseline=(current bounding box.center)]
        \draw[thick] (-2, 2) -- (1, 2); % Top line
        \draw (-2,-2) rectangle (2,1) node[midway] {A};
        
        \draw[thick] (-2.5, 1.5) -- (-3.5, 0);
        \draw[thick] (-3.5, 0) -- (-2.5, -1.5);
        \draw[thick] (3, 1.5) -- (3, -1.5);

        \node[cross out, draw, line width=1.2pt, minimum size=1mm] at (1, 2) {};
      \end{tikzpicture}
      &
      \raisebox{-0.15cm}{\small $\hat{a}^\dagger(x_1)\cdots\hat{a}^\dagger(x_m)\hat{a}(x_m)\cdots \hat{a}(x_1)$}
      &
      \begin{tikzpicture}[scale=0.25, baseline=(current bounding box.center)]
        \draw[thick] (-1, 2) -- (2, 2); % Top line
        \draw (-2,-2) rectangle (2,1) node[midway] {B};
        
        \draw[thick] (2.5, 1.5) -- (3.5, 0);
        \draw[thick] (3.5, 0) -- (2.5, -1.5);
        \draw[thick] (-3, 1.5) -- (-3, -1.5);
        \node[cross out, draw, line width=1.2pt, minimum size=1mm] at (-1, 2) {};
      \end{tikzpicture}
    \end{tabular}
\end{figure}

\begin{eqnarray*}
    &=& e^{-|\alpha|^2} \frac{|\alpha|^{2n+2}}{L^{n/2+1}} \frac{1}{(n+1)!^2} (n+1)^2 D(x_1,x_2,\cdots,x_m)\bra{0}\int dy_1 dy_2\, \sum_{l'=0}^{M-1} r_0 t_0^{l'}\hat{b}_{l'+1}(y_{n+1}')\hat{b}_{l+1}^\dagger(y_{n+1})\sum_{l=0}^{M-1}r_0 t_0^{l}\ket{0}+\cdots\\
    &=& e^{-|\alpha|^2} \frac{|\alpha|^{2n+2}}{L^{n+1}} \frac{1}{n!^2} L  D(x_1,x_2,\cdots,x_m) \left(\sum_{l=0}^{M-1}r_0^2 t_0^{2l}\right)+\cdots
\end{eqnarray*}
For the $n+2$-photon diagrams with A and B added with two pairs of linearly transmitted photons.

\begin{figure}[h!]
  \centering
    \begin{tabular}{@{}c@{}c@{\hspace{0.5em}}c@{}}
      \begin{tikzpicture}[scale=0.25, baseline=(current bounding box.center)]
        \draw[thick] (-2, 3) -- (2, 3);
        \draw[thick] (-2, 2) -- (2, 2); % Top line
        \draw (-2,-2) rectangle (2,1) node[midway] {A};
        
        \draw[thick] (-2.5, 4) -- (-3.5, 1);
        \draw[thick] (-3.5, 1) -- (-2.5, -2);
        \draw[thick] (3, 4) -- (3, -2);
      \end{tikzpicture}
      &
      \raisebox{-0.15cm}{\small $\hat{a}^\dagger(x_1)\cdots\hat{a}^\dagger(x_m)\hat{a}(x_m)\cdots \hat{a}(x_1)$}
      &
      \begin{tikzpicture}[scale=0.25, baseline=(current bounding box.center)]
        \draw[thick] (-2, 3) -- (2, 3);
        \draw[thick] (-2, 2) -- (2, 2); % Top line
        \draw (-2,-2) rectangle (2,1) node[midway] {B};
        
        \draw[thick] (2.5, 4) -- (3.5, 1);
        \draw[thick] (3.5, 1) -- (2.5, -2);
        \draw[thick] (-3, 4) -- (-3, -2);
      \end{tikzpicture}
    \end{tabular}
\end{figure}

\begin{eqnarray*}
    &=& e^{-|\alpha|^2} \frac{|\alpha|^{2n+4}}{L^{n/2+2}} \frac{1}{(n+2)!^2} \left[\frac{(n+2)!}{2!n!}\right]^2 D(x_1,x_2,\cdots,x_m)\bra{0}\int dy_1'dy_2' dy_1 dy_2\, t_0^{2M}\hat{a}(y_1')\hat{a}(y_2')\hat{a}^\dagger(y_1)\hat{a}^\dagger(y_2) t_0^{2M}\ket{0}+\cdots\\
    &=& e^{-|\alpha|^2} \frac{|\alpha|^{2n+4}}{L^{n+2}}\frac{1}{n!^2}  \frac{L^2}{2!} D(x_1,x_2,\cdots,x_m)  t_0^{4M}+\cdots
\end{eqnarray*}

After doing more similar calculations with more additional photons, one will find the pattern of the contribution of the diagram A and B added with $p$ non interacting photons to the term including $D(x_1,x_2,\cdots,x_m)$ in the $m$th-order correlation function. If $q$ additional photons elastically go through the array while $p-q$ photons are lost, then this pattern reads
\begin{eqnarray*}
    e^{-|\alpha|^2} \frac{|\alpha|^{2n+2p}}{L^{n+p}} \frac{1}{n!^2} \frac{L^p}{p!}  D(x_1,x_2,\cdots,x_m)\sum_{q=0}^{p}\binom{p}{q} (t_0^{2M})^q \left(\sum_{l=0}^{M-1}r_0^2 t_0^{2l}\right)^{p-q} 
\end{eqnarray*}
Due to the easy-to-check identity $t_0^{2M}+\sum_{l=0}^{M-1}r_0^2 t_0^{2l}=1$, the last sum $\sum_{q=0}^{p}\binom{p}{q} (t_0^{2M})^q \left(\sum_{l=0}^{M-1}r_0^2 t_0^{2l}\right)^{p-q}=1$. In the scattering with a coherent state, we have to sum all the terms including $D(x_1,x_2,\cdots,x_m)$ in each $n+p$ Fock sector for $p=0,1,\cdots,\infty$. 
\begin{eqnarray*}
    &&\sum_{p=0}^{\infty}e^{-|\alpha|^2} \frac{|\alpha|^{2n+2p}}{L^{n+p}} \frac{1}{n!^2} \frac{L^p}{p!} D(x_1,x_2,\cdots,x_m)\\
    &=& \frac{P_{\rm in}^n}{n!^2} D(x_1,x_2,\cdots,x_m)
\end{eqnarray*}
Hence the coherent-state calculation reproduces exactly the
$L$-independent contribution obtained in the $n$-photon sector, as promised.

\section{Power Correction} \label{apx:power correc}
In this first half of the main text, we calculated the correlation functions $g_c^{(3)}$ and $g^{(2)}$ by neglecting contributions from diagrams of higher order in $P_{\rm in}/\Gamma_{\rm tot}$ (those involving additional photons) and higher order in $\beta$ (those containing additional interaction vertices). In the ultra-low driving limit where $\mathcal{O}(P_{\rm in}/\Gamma_{\rm tot})=\mathcal{O}(\beta^2)$, diagrams involving more than three photons provide negligible contributions to $g^{(3)}$, while diagrams with more than two photons provide negligible contributions to $g^{(2)}$. 

However, to achieve experimentally measurable signal strength for three-photon correlation effects, the driving strength must be increased to at least satisfy:
\begin{equation} \label{eq:power-coup rel}\mathcal{O}\left(\frac{P_{\rm in}}{\Gamma_{\rm tot}}\right) =\mathcal{O}(\beta).
\end{equation}
This increased driving power necessitates consideration of additional terms in our perturbative calculation. Given that the leading-order term in $g_c^{(3)}$ is of order $\mathcal{O}(\beta^2)$, consistency with relation (\ref{eq:power-coup rel}) requires us to examine contributions of orders $\mathcal{O}(P_{\rm in}/\Gamma_{\rm tot})$, $\mathcal{O}(\beta P_{\rm in}/\Gamma_{\rm tot})$, and $\mathcal{O}(P_{\rm in}^2/\Gamma_{\rm tot}^2)$. Since $\mathcal{O}(P_{\rm in}/\Gamma_{\rm tot})$ and $\mathcal{O}(P_{\rm in}^2/\Gamma_{\rm tot}^2)$ correspond to the diagrams without interaction, they are part of the series sum which cancels with the normalization $\exp(-\alpha^2)$ of coherent state. So we only need to examine the $\mathcal{O}(\beta P_{\rm in}/\Gamma_{\rm tot})$. 

First, we should examine the correction terms in the power $\expval{\hat{a}^\dagger(x)\hat{a}(x)}$, which consists of the denominator of $g^{(2)}$ and $g^{(3)}$. At the order of interest, the correction terms  consist of the following diagrams:
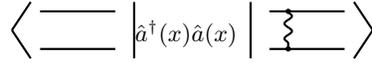
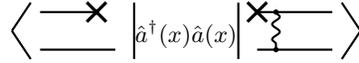
\begin{figure}[h]
  \centering
  % First row
  \subfloat[The $\mathcal{O}(\beta P_{\rm in}/\Gamma_{\rm tot})$ two photon-interacting diagram contributing to the output power.\label{fig:first_power_cor}]{%
    \begin{tabular}{@{}c@{}c@{\hspace{0.5em}}c@{}}
      \begin{tikzpicture}[scale=0.25, baseline=(current bounding box.center)]
        \draw[thick] (-2, 1) -- (2, 1); % Top line
        \draw[thick] (-2,-1) -- (2,-1); % Bottom line
        
        \draw[thick] (-2.5, 1.5) -- (-3.5, 0);
        \draw[thick] (-3.5, 0) -- (-2.5, -1.5);
        \draw[thick] (3, 1.5) -- (3, -1.5);
      \end{tikzpicture}
      &
      \raisebox{-0.15cm}{\small $\hat{a}^\dagger(x)\hat{a}(x)$}
      &
      \begin{tikzpicture}[scale=0.25, baseline=(current bounding box.center)]
        \draw[thick] (-2, 1) -- (2, 1); % Top line
        \draw[thick] (-2, -1) -- (2, -1); % Bottom line
        \draw[thick] (2.5, 1.5) -- (3.5, 0);
        \draw[thick] (3.5, 0) -- (2.5, -1.5);
        \draw[thick] (-3, 1.5) -- (-3, -1.5);
        \draw[thick, decorate,
        decoration={
                    snake,
                    amplitude=0.5mm,
                    segment length=2mm
        }] (-1, 1) -- (-1, -1); % Wavy line
        \fill[black] (-1, 1) circle (4pt) (-1, -1) circle (4pt);
      \end{tikzpicture}
    \end{tabular}
  }
  
  \vspace{1em}
  
  % Third row
  \subfloat[The $\mathcal{O}(\beta P_{\rm in}/\Gamma_{\rm tot})$ two photon-interacting diagram with one loss contributing to the output power.\label{fig:second_power_cor}]{%
    \begin{tabular}{@{}c@{}c@{}c@{\hspace{0.5em}}c@{}}
      % First diagram
      \begin{tikzpicture}[scale=0.25, baseline=(current bounding box.center)]
        \draw[thick] (-2, 1) -- (1, 1); % Top line
        \draw[thick] (-2,-1) -- (2,-1); % Bottom line
        
        \draw[thick] (-2.5, 1.5) -- (-3.5, 0);
        \draw[thick] (-3.5, 0) -- (-2.5, -1.5);
        \draw[thick] (3, 1.5) -- (3, -1.5);
        \fill[black] (1, 1) circle (4pt);
        \node[cross out, draw, line width=1.2pt, minimum size=1mm] at (1, 1) {};
      \end{tikzpicture}
      &
      \raisebox{-0.15cm}{\small $\hat{a}^\dagger(x)\hat{a}(x)$}
      &
      % Second diagram
      \begin{tikzpicture}[scale=0.25, baseline=(current bounding box.center)]
        \draw[thick] (-2, 1) -- (2, 1); % Top line
        \draw[thick] (-2, -1) -- (2, -1); % Bottom line
        \draw[thick] (2.5, 1.5) -- (3.5, 0);
        \draw[thick] (3.5, 0) -- (2.5, -1.5);
        \draw[thick] (-3, 1.5) -- (-3, -1.5);
  
        \draw[thick, decorate,
        decoration={
                    snake,
                    amplitude=0.5mm,
                    segment length=2mm
        }] (-1, 1) -- (-1, -1); % Wavy line
  
        \fill[black] (-1, 1) circle (4pt) (-2, 1) circle (4pt) (-1, -1) circle (4pt) ;
        \node[cross out, draw, line width=1.2pt, minimum size=1mm] at (-2, 1) {};
      \end{tikzpicture}
    \end{tabular}
  }
  
  \caption{Two-photon diagrams contributing to output power at order $\mathcal{O}(\beta P_{\rm in}/\Gamma_{\rm tot})$}
  \label{fig:power cor}
\end{figure}

The expression for Fig.~\ref{fig:first_power_cor} is,
\begin{eqnarray}
    &\left(\frac{1}{2!}\right)^2& e^{-|\alpha|^2} \frac{|\alpha|^4}{L^2} \bra{0} \int dy_1'dy_2' t_0^{2M} \hat{a}(y_1')\hat{a}(y_2')\hat{a}^\dagger(x) \hat{a}(x) \int dy_1 dy_2 \varphi(y_1,y_2) \hat{a}^\dagger(y_1)\hat{a}^\dagger(y_2)\ket{0}  \nonumber\\
    &=& -P_{\rm in}^2 t_0^{2M} \int dy \varphi(y,x) \nonumber\\
    &=& \frac{16M\beta^2 P_{\rm in}^2}{\Gamma_{\rm tot}}t_0^{4M-2}.
\end{eqnarray}

The expression for Fig.~\ref{fig:second_power_cor} is,
\begin{eqnarray}
    2\times 2\times &\left(\frac{1}{2!}\right)^2& e^{-|\alpha|^2} \frac{|\alpha|^4}{L^2} \int dy_1'dy_2'dy_1 dy_2 t_0^{M}\sum_{m'=0}^{M-1} r_0 t_0^{M+m'} \hat{b}_{m'+1}(y_1')\hat{a}(y_2')\hat{a}^\dagger(x) \hat{a}(x) \hat{b}_{m+1}^\dagger(y_1)\hat{a}^\dagger(y_2) \varphi_{m+1}(\slashed{y_1},y_2) \nonumber\\
    &=& -P_{\rm in}^2 \sum_{m=0}^{M-1}r_0 t_0^{M+m} \int dy \varphi_{m+1}(\slashed{y},x) \nonumber\\
    &=& \frac{16\beta P_{\rm in}^2}{\Gamma_{\rm tot}} \sum_{m=0}^{M-1} t_0^{2M+2m-1} (m\beta r_0+\sqrt{\beta(1-\beta)}),
\end{eqnarray}
where $\varphi_m(\slashed{y},x_i)$ is two-photon wavefunction with the photon-photon interaction happened only once, and one photon is lost on the  $m$th atom after interaction. 
The integrals over the real space wavefunction can be evaluated in momentum space,
\begin{eqnarray} \label{eq:2phint}
    && \int dy\, \varphi(y,x_i)\\
    &=& \int dy \int dp_1 dp_2 \, e^{-ip_1 y-i p_2 x_i} \sum_{j=0}^{M-1} t_0^{2j} \frac{i\Gamma_{\rm tot}^2}{2\pi} \frac{(p_1+p_2+i\Gamma_{\rm tot})\delta(p_1+p_2)}{(p_1+i\frac{\Gamma_{\rm tot}}{2})(p_2+i\frac{\Gamma_{\rm tot}}{2})(k_1+i\frac{\Gamma_{\rm tot}}{2})(k_2+i\frac{\Gamma_{\rm tot}}{2})}(t_{p_1}t_{p_2})^{M-j-1}\nonumber\\
    &=& -\frac{16M\beta^2}{\Gamma_{\rm tot}} t_0^{2M-2}, \nonumber
\end{eqnarray}
and
\begin{eqnarray} \label{eq:2phintwithoneloss}
    && \int dy\, \varphi_m(\slashed{y},x_i)  \\
    &=& \int dy \int dp_1 dp_2 \, e^{-i p_1 y-ip_2 x_i} \biggl(\sum_{j=0}^{m-2} t_0^{2j}\beta^2\frac{i\Gamma_{\rm tot}^2}{2\pi}\frac{(p_1+p_2+i\Gamma_{\rm tot})\delta(p_1+p_2)}{(p_1+i\frac{\Gamma_{\rm tot}}{2})(p_2+i\frac{\Gamma_{\rm tot}}{2})(k_1+i\frac{\Gamma_{\rm tot}}{2})(k_2+i\frac{\Gamma_{\rm tot}}{2})} t_{p_1}^{m-j-1} r_{p_1} t_{p_2}^{M-j-1}  \nonumber\\
    && +t_0^{2(m-1)}\beta\sqrt{\beta(1-\beta)}\frac{i\Gamma_{\rm tot}^2}{2\pi}\frac{(p_1+p_2+i\Gamma_{\rm tot})\delta(p_1+p_2)}{(p_1+i\frac{\Gamma_{\rm tot}}{2})(p_2+i\frac{\Gamma_{\rm tot}}{2})(k_1+i\frac{\Gamma_{\rm tot}}{2})(k_2+i\frac{\Gamma_{\rm tot}}{2})}t_{p_2}^{M-m}\biggr)   \nonumber\\
    &=&-(m-1)\frac{16\beta^2}{ \Gamma_{\rm tot}} r_0 t_0^{M+m-2}-\frac{16\beta}{ \Gamma_{\rm tot}}\sqrt{\beta(1-\beta)}t_0^{M+m-2}. \nonumber
\end{eqnarray}
In the first equality, we see that $\varphi_m(\slashed{y},x)$ includes two parts: The first term describes two photons individually scattering with $j$ atoms before two-photon interaction on the $j+1$th atom, after which the photon with $p_1$ scatter individually until getting lost on the $m$th atom, while the photon with $p_2$ transport until the end of the array. The second term describes two photons indivdually scatter with $m-1$ atoms before the photon with $p_1$ immediately lost after interacting with another photon on the $m$th atom.
Both integrals (\ref{eq:2phint}) and (\ref{eq:2phintwithoneloss}) are constants in position.

\subsection{Correction for the Correlators at the Order \texorpdfstring{$\mathcal{O}(\beta P_{\rm in}/\Gamma_{\rm tot})$}{O(betaPin/Gammatot)}}
Next, we consider the correction term at the order of $\mathcal{O}(\beta P_{\rm in}/\Gamma_{\rm tot})$ for $G^{(3)}(x_1,x_2,x_3)$. The presence of $\beta$ means that the diagrams sandwiching $\hat{a}^\dagger(x_1)\hat{a}^\dagger(x_2)\hat{a}^\dagger(x_3)\hat{a}(x_3)\hat{a}(x_2)\hat{a}(x_1)$ should have only one two-photon interaction in total. There are only three possible combinations shown in~Fig.\ref{fig:four_ph_interacting_diag_for_g3}.
\begin{figure}[h]
  \centering
  % First row
  \subfloat[The normalised third-order expectation value of four-photon diagram with two-photon interaction.\label{fig:four_ph_row1}]{%
    \begin{tabular}{@{}c@{}c@{}c@{\hspace{0.5em}}c@{}}
      \begin{tikzpicture}[scale=0.25, baseline=(current bounding box.center)]
        \draw[thick] (-2, 1.5) -- (2, 1.5); % Top line
        \draw[thick] (-2, 0.5) -- (2, 0.5); % Middle line
        \draw[thick] (-2,-0.5) -- (2,-0.5); % Bottom line
        \draw[thick] (-2, -1.5) -- (2, -1.5);
        
        \draw[thick] (-2.5, 1.5) -- (-3.5, 0);
        \draw[thick] (-3.5, 0) -- (-2.5, -1.5);
        \draw[thick] (3, 1.5) -- (3, -1.5);
      \end{tikzpicture}
      &
      \raisebox{-0.15cm}{\small $\hat{a}^\dagger(x_1)\hat{a}^\dagger(x_2)\hat{a}^\dagger(x_3)\hat{a}(x_3)\hat{a}(x_2)\hat{a}(x_1)$}
      &
      \begin{tikzpicture}[scale=0.25, baseline=(current bounding box.center)]
        \draw[thick] (-2, 1.5) -- (2, 1.5); % Top line
        \draw[thick] (-2, 0.5) -- (2, 0.5); % Middle line
        \draw[thick] (-2, -0.5) -- (2, -0.5); % Bottom line
        \draw[thick] (-2, -1.5) -- (2, -1.5); % Bottom line
        \draw[thick] (2.5, 1.5) -- (3.5, 0);
        \draw[thick] (3.5, 0) -- (2.5, -1.5);
        \draw[thick] (-3, 1.5) -- (-3, -1.5);
        \draw[thick, decorate,
        decoration={
                    snake,
                    amplitude=0.5mm,
                    segment length=1mm
        }] (-1, 1.5) -- (-1, 0.5); % Wavy line
        \fill[black] (-1, 1.5) circle (4pt) (-1, 0.5) circle (4pt);
      \end{tikzpicture}
      &
      \raisebox{-0.15cm}{\large / $\left(\expval{\hat{a}^\dagger (x_1)\hat{a}(x_1)}\expval{\hat{a}^\dagger (x_2)\hat{a}(x_2)}\expval{\hat{a}^\dagger (x_3)\hat{a}(x_3)}\right)$}
    \end{tabular}
  }
  
  \vspace{1em}
  
  % Second row
  \subfloat[The normalised third-order expectation value of four-photon diagram with two-photon interaction and one lost individual-scattering photon.\label{fig:four_ph_row2}]{%
    \begin{tabular}{@{}c@{}c@{}c@{\hspace{0.5em}}c@{}}
      \begin{tikzpicture}[scale=0.25, baseline=(current bounding box.center)]
        \draw[thick] (-2, 1.5) -- (1, 1.5); % Top line
        \draw[thick] (-2, 0.5) -- (2, 0.5); % Middle line
        \draw[thick] (-2,-0.5) -- (2,-0.5); % Bottom line
        \draw[thick] (-2, -1.5) -- (2, -1.5);
        
        \draw[thick] (-2.5, 1.5) -- (-3.5, 0);
        \draw[thick] (-3.5, 0) -- (-2.5, -1.5);
        \draw[thick] (3, 1.5) -- (3, -1.5);
        
        \fill[black] (1, 1.5) circle (4pt);
        \node[cross out, draw, line width=1.2pt, minimum size=1mm] at (1, 1.5) {};
      \end{tikzpicture}
      &
      \raisebox{-0.15cm}{\small $\hat{a}^\dagger(x_1)\hat{a}^\dagger(x_2)\hat{a}^\dagger(x_3)\hat{a}(x_3)\hat{a}(x_2)\hat{a}(x_1)$}
      &
      \begin{tikzpicture}[scale=0.25, baseline=(current bounding box.center)]
        \draw[thick] (-1, 1.5) -- (2, 1.5); % Top line
        \draw[thick] (-2, 0.5) -- (2, 0.5); % Middle line
        \draw[thick] (-2, -0.5) -- (2, -0.5); % Bottom line
        \draw[thick] (-2, -1.5) -- (2, -1.5); % Bottom line
        \draw[thick] (2.5, 1.5) -- (3.5, 0);
        \draw[thick] (3.5, 0) -- (2.5, -1.5);
        \draw[thick] (-3, 1.5) -- (-3, -1.5);
        \draw[thick, decorate,
        decoration={
                    snake,
                    amplitude=0.5mm,
                    segment length=1mm
        }] (-1, 0.5) -- (-1, -0.5); % Wavy line
        
        \fill[black] (-1, 1.5) circle (4pt) (-1, 0.5) circle (4pt) (-1, -0.5) circle (4pt);
        \node[cross out, draw, line width=1.2pt, minimum size=1mm] at (-1, 1.5) {};
      \end{tikzpicture}
      &
      \raisebox{-0.15cm}{\large / $\left(\expval{\hat{a}^\dagger (x_1)\hat{a}(x_1)}\expval{\hat{a}^\dagger (x_2)\hat{a}(x_2)}\expval{\hat{a}^\dagger (x_3)\hat{a}(x_3)}\right)$}
    \end{tabular}
  }
  
  \vspace{1em}
  
  % Third row
  \subfloat[The normalised third-order expectation value of four-photon diagram with two-photon interaction and one lost interacting photon.\label{fig:four_ph_row3}]{%
    \begin{tabular}{@{}c@{}c@{}c@{\hspace{0.5em}}c@{}}
      % First diagram
      \begin{tikzpicture}[scale=0.25, baseline=(current bounding box.center)]
        \draw[thick] (-2, 1.5) -- (1, 1.5); % Top line
        \draw[thick] (-2, 0.5) -- (2, 0.5); % Middle line
        \draw[thick] (-2,-0.5) -- (2,-0.5); % Bottom line
        \draw[thick] (-2, -1.5) -- (2, -1.5);
        
        \draw[thick] (-2.5, 1.5) -- (-3.5, 0);
        \draw[thick] (-3.5, 0) -- (-2.5, -1.5);
        \draw[thick] (3, 1.5) -- (3, -1.5);
        \fill[black] (1, 1.5) circle (4pt);
        \node[cross out, draw, line width=1.2pt, minimum size=1mm] at (1, 1.5) {};
      \end{tikzpicture}
      &
      \raisebox{-0.15cm}{\small $\hat{a}^\dagger(x_1)\hat{a}^\dagger(x_2)\hat{a}^\dagger (x_3)\hat{a}(x_3)\hat{a}(x_2)\hat{a}(x_1)$}
      &
      % Second diagram
      \begin{tikzpicture}[scale=0.25, baseline=(current bounding box.center)]
        \draw[thick] (-2, 1.5) -- (2, 1.5); % Top line
        \draw[thick] (-2, 0.5) -- (2, 0.5); % Middle line
        \draw[thick] (-2, -0.5) -- (2, -0.5); % Bottom line
        \draw[thick] (-2, -1.5) -- (2, -1.5); % Bottom line
        \draw[thick] (2.5, 1.5) -- (3.5, 0);
        \draw[thick] (3.5, 0) -- (2.5, -1.5);
        \draw[thick] (-3, 1.5) -- (-3, -1.5);
  
        \draw[thick, decorate,
        decoration={
                    snake,
                    amplitude=0.5mm,
                    segment length=1mm
        }] (-1, 1.5) -- (-1, 0.5); % Wavy line
  
        \fill[black] (-1, 1.5) circle (4pt) (-1, 0.5) circle (4pt) (-2, 1.5) circle (4pt);
        \node[cross out, draw, line width=1.2pt, minimum size=1mm] at (-2, 1.5) {};
      \end{tikzpicture}
      &
      \raisebox{-0.15cm}{\large / $\left(\expval{\hat{a}^\dagger (x_1)\hat{a}(x_1)}\expval{\hat{a}^\dagger (x_2)\hat{a}(x_2)}\expval{\hat{a}^\dagger (x_3)\hat{a}(x_3)}\right)$}
    \end{tabular}
  }
  
  \caption{Four-photon diagrams contributing to $G^{(3)}(x_1,x_2,x_3)$ at order $\mathcal{O}(\beta P_{\rm in}/\Gamma_{\rm tot})$}
  \label{fig:four_ph_interacting_diag_for_g3}
\end{figure}

The expression for Fig.~\ref{fig:four_ph_row1} is
\begin{eqnarray} \label{eq:four_ph_row1}
    6\times &\left(\frac{1}{4!}\right)^2&  e^{-|\alpha|^2} \frac{|\alpha|^8}{L^4} \bra{0}\int dy_1' dy_2' dy_3' dy_4' t_0^{4M} \hat{a}(y_1')\hat{a}(y_2')\hat{a}(y_3')\hat{a}(y_4') \hat{a}^\dagger(x_1)\hat{a}^\dagger(x_2)\hat{a}^\dagger(x_3)\hat{a}(x_3)\hat{a}(x_2)\hat{a}(x_1) \nonumber\\
    &\times& \int dy_1 dy_2 dy_3 dy_4 t_0^{2M} \varphi(y_3,y_4) \hat{a}^\dagger (y_1)\hat{a}^\dagger(y_2)\hat{a}^\dagger(y_3)\hat{a}^\dagger(y_4)\ket{0}/\left(\expval{\hat{a}^\dagger (x_1)\hat{a}(x_1)}\expval{\hat{a}^\dagger (x_2)\hat{a}(x_2)}\expval{\hat{a}^\dagger (x_3)\hat{a}(x_3)}\right) \nonumber\\
    &=& \frac{e^{-|\alpha|^2} P_{\rm in}^4 t_0^{6M} \Bigl[L\bigl(\varphi (x_1,x_2)+\varphi (x_2,x_3)+\varphi (x_1,x_3)\bigr)+\int dy\, \bigl(\varphi(y,x_1)+\varphi(y,x_2)+\varphi(y,x_3)\bigr)  \Bigr]}{\expval{\hat{a}^\dagger (x_1)\hat{a}(x_1)}\expval{\hat{a}^\dagger (x_2)\hat{a}(x_2)}\expval{\hat{a}^\dagger (x_3)\hat{a}(x_3)}} \\
    &=& \frac{e^{-|\alpha|^2} P_{\rm in}^4 t_0^{6M} \Bigl[L\bigl(\varphi (x_1,x_2)+\varphi (x_2,x_3)+\varphi (x_1,x_3)\bigr)-\frac{48M\beta^2}{\Gamma_{\rm tot}} t_0^{2M-2}  \Bigr]}{\expval{\hat{a}^\dagger (x_1)\hat{a}(x_1)}\expval{\hat{a}^\dagger (x_2)\hat{a}(x_2)}\expval{\hat{a}^\dagger (x_3)\hat{a}(x_3)}},
    \end{eqnarray}
$\varphi$ here denotes the two-photon entangled wavefunction with the entanglement induced by one two-photon interaction. It is symmetric under the permutations of its two variables. The first term in (\ref{eq:four_ph_row1}) scales with $L$. It is part of the series of the dressed diagrams, which eventually give $\frac{P_{\rm in}^3 t_0^{4M}\left(\varphi_2(x_1,x_2)+\varphi (x_2,x_3)+\varphi (x_1,x_3)\right)}{\expval{\hat{a}^\dagger (x_1)\hat{a}(x_1)}\expval{\hat{a}^\dagger (x_2)\hat{a}(x_2)}\expval{\hat{a}^\dagger (x_3)\hat{a}(x_3)}}$. 

The expression for Fig.~\ref{fig:four_ph_row2} is
\begin{eqnarray} \label{eq:four_ph_row2}
    4\times 12 \times &\left(\frac{1}{4!}\right)^2&  e^{-|\alpha|^2} \frac{|\alpha|^8}{L^4} \bra{0} \int dy_1' dy_2' dy_3' dy_4'\, t_0^{3M} \sum_{m'=0}^{M-1} r_0 t_0^{m'} \hat{b}_{m'}(y_1')\hat{a}(y_2')\hat{a}(y_3')\hat{a}(y_4') \hat{a}^\dagger(x_1)\hat{a}^\dagger(x_2)\hat{a}^\dagger(x_3)\hat{a}(x_3)\hat{a}(x_2)\hat{a}(x_1) \nonumber\\
    &\times& \int dy_1 dy_2 dy_3 dy_4\, t_0^{M} \sum_{m=0}^{M-1}\varphi(y_3,y_4) \hat{b}_m^\dagger (y_1)\hat{a}^\dagger(y_2)\hat{a}^\dagger(y_3)\hat{a}^\dagger(y_4)\ket{0}/\expval{\hat{a}^\dagger (x_1)\hat{a}(x_1)}\expval{\hat{a}^\dagger (x_2)\hat{a}(x_2)}\expval{\hat{a}^\dagger (x_3)\hat{a}(x_3)} \nonumber\\
    &=& \frac{e^{-|\alpha|^2} P_{\rm in}^4 L t_0^{4M} \sum_{m=0}^{M-1} r_0^2 t_0^{2m} \bigl(\varphi(x_1,x_2)+\varphi(x_2,x_3)+\varphi(x_1,x_3) \bigr)}{\expval{\hat{a}^\dagger (x_1)\hat{a}(x_1)}\expval{\hat{a}^\dagger (x_2)\hat{a}(x_2)}\expval{\hat{a}^\dagger (x_3)\hat{a}(x_3)}}.
\end{eqnarray}
It scales with $L$ because it is also a part of the series of the dressed diagrams, which eventually cancels out the normalization factor of coherent state and gives  $\frac{P_{\rm in}^3 t_0^{4M}\bigl(\varphi(x_1,x_2)+\varphi(x_2,x_3)+\varphi(x_1,x_3) \bigr)}{\expval{\hat{a}^\dagger (x_1)\hat{a}(x_1)}\expval{\hat{a}^\dagger (x_2)\hat{a}(x_2)}\expval{\hat{a}^\dagger (x_3)\hat{a}(x_3)}}$ after summing over all the photon number sectors in the coherent state.

The expression of Fig.~\ref{fig:four_ph_row3} reads
\begin{eqnarray} \label{eq:four_ph_row3}
    4\times 12\times &\left(\frac{1}{4!}\right)^2&  e^{-|\alpha|^2} \frac{|\alpha|^8}{L^4} \bra{0} \int dy_1' dy_2' dy_3' dy_4'\, t_0^{3M} \sum_{m'=0}^{M-1} r_0 t_0^{m'} \hat{b}_{m'+1}(y_1')\hat{a}(y_2')\hat{a}(y_3')\hat{a}(y_4') \hat{a}^\dagger(x_1)\hat{a}^\dagger(x_2)\hat{a}^\dagger(x_3)\hat{a}(x_3)\hat{a}(x_2)\hat{a}(x_1) \nonumber\\
    &\times& \int dy_1 dy_2 dy_3 dy_4\, t_0^{2M} \sum_{m=0}^{M-1}\varphi_{m+1}(\slashed{y}_1,y_2) \hat{b}_{m+1}^\dagger (y_1)\hat{a}^\dagger(y_2)\hat{a}^\dagger(y_3)\hat{a}^\dagger(y_4)\ket{0}/\left(\expval{\hat{a}^\dagger (x_1)\hat{a}(x_1)}\expval{\hat{a}^\dagger (x_2)\hat{a}(x_2)}\expval{\hat{a}^\dagger (x_3)\hat{a}(x_3)}\right) \nonumber\\
    &=& \frac{e^{-|\alpha|^2}P_{\rm in}^4 t_0^{5M}\sum_{m=0}^{M-1} r_0 t_0^{m}\int dy\,\bigl[\varphi_{m+1}(\slashed{y},x_1)+\varphi_{m+1}(\slashed{y},x_2)+\varphi_{m+1}(\slashed{y},x_3)\bigr]}{\expval{\hat{a}^\dagger (x_1)\hat{a}(x_1)}\expval{\hat{a}^\dagger (x_2)\hat{a}(x_2)}\expval{\hat{a}^\dagger (x_3)\hat{a}(x_3)}} \\
    &=& \frac{e^{-|\alpha|^2}P_{\rm in}^4 t_0^{5M}\sum_{m=0}^{M-1} r_0 t_0^{m}\bigl[-m\frac{48\beta^2}{ \Gamma_{\rm tot}} r_0 t_0^{M+m-1}-\frac{48\beta}{ \Gamma_{\rm tot}}\sqrt{\beta(1-\beta)}t_0^{M+m-1}\bigr]}{\expval{\hat{a}^\dagger (x_1)\hat{a}(x_1)}\expval{\hat{a}^\dagger (x_2)\hat{a}(x_2)}\expval{\hat{a}^\dagger (x_3)\hat{a}(x_3)}}.
\end{eqnarray}
After summing over all photon number sectors in the coherent state, the corrected $g^{(3)}$ reads
\begin{eqnarray} \label{eq:corrected g3}
    && g^{(3)}(x_1,x_2,x_3)= \\
    && \frac{P_{\rm in}^3 |t_0^{3M} + t_0^{M}\bigl[\phi_2(x_1,x_2) + \phi_2(x_2,x_3)+ \phi_2(x_1,x_3)\bigr] + \phi_3(x_1,x_2,x_3)|^2}{\{P_{\rm in}t_0^{2M}+2 \operatorname{Re}[P_{\rm in}^2\frac{16\beta}{\Gamma_{\rm tot}}t_0^{2M}(M\beta t_0^{2M-2}+\sum_{m=0}^{M-1} m\beta r_0^2 t_0^{2m-1}+r_0 \sqrt{\beta(1-\beta)}t_0^{2m-1})]\}^3} \nonumber\\
    && +\frac{2\operatorname{Re}[P_{\rm in}^4\frac{48\beta}{\Gamma_{\rm tot}}t_0^{6M}(M\beta t_0^{2M-2}+\sum_{m=0}^{M-1} m\beta r_0t_0^{2m-1}+\sqrt{\beta(1-\beta)}t_0^{2m-1})]}{\{P_{\rm in}t_0^{2M}+2\operatorname{Re}[P_{\rm in}^2\frac{16\beta}{\Gamma_{\rm tot}}t_0^{2M}(M\beta t_0^{2M-2}+\sum_{m=0}^{M-1} m\beta r_0^2 t_0^{2m-1}+r_0 \sqrt{\beta(1-\beta)}t_0^{2m-1})]\}^3} \nonumber\\
    && \approx \frac{|t_0^{3M} + t_0^{M}\bigl[\phi_2(x_1,x_2) + \phi_2(x_2,x_3)+ \phi_2(x_1,x_3)\bigr] + \phi_3(x_1,x_2,x_3)|^2}{t_0^{6M}} \nonumber\\
     &&- 3 (P_{\rm in} t_0^{2M})^2P_{\rm in}^2 t_0^{2M}\frac{32\beta}{\Gamma_{\rm tot}}(M\beta t_0^{2M-2}+\sum_{m=0}^{M-1} m\beta r_0t_0^{2m-1}+\sqrt{\beta(1-\beta)}t_0^{2m-1})/(P_{\rm in}t_0^{2M})^3 \nonumber\\
    && +P_{\rm in}^4\frac{96\beta}{\Gamma_{\rm tot}}t_0^{6M}(M\beta t_0^{2M-2}+\sum_{m=0}^{M-1} m\beta r_0t_0^{2m-1}+\sqrt{\beta(1-\beta)}t_0^{2m-1}) /(P_{\rm in}t_0^{2M})^3 \nonumber\\
    && =\frac{|t_0^{3M} + t_0^{M}\bigl[\phi_2(x_1,x_2) + \phi_2(x_2,x_3)+ \phi_2(x_1,x_3)\bigr] + \phi_3(x_1,x_2,x_3)|^2}{t_0^{6M}}. \nonumber
    \end{eqnarray}
In the second step, we expand the denominator of the first equality using a geometric series and retain only terms up to order $\mathcal{O}(\beta P_{\rm in}/\Gamma_{\rm tot}) = \mathcal{O}(\beta^2)$. We find two additional terms added to the original $g^{(3)}$ but they cancel each other. Therefore, the increased driving power with $\mathcal{O}(P_{\rm in}/\Gamma_{\rm tot})= \mathcal{O}(\beta)$ makes no additional correction at $\mathcal{O}(\beta P_{\rm in}/\Gamma_{\rm tot})$. 

For $g^{(2)}(x_1,x_2)$, the $\mathcal{O}(\beta P_{\rm in}/\Gamma_{\rm tot})$ correction is given by the following diagrams

\begin{figure}[h]
  \centering
  % First row
  \subfloat[The normalised second-order expectation value of three-photon diagram with two-photon interaction.\label{fig:g2_thr_ph_row1}]{%
    \begin{tabular}{@{}c@{}c@{}c@{\hspace{0.5em}}c@{}}
      \begin{tikzpicture}[scale=0.25, baseline=(current bounding box.center)]
        \draw[thick] (-2, 1) -- (2, 1); % Top line
        \draw[thick] (-2, 0) -- (2, 0); % Middle line
        \draw[thick] (-2,-1) -- (2,-1); % Bottom line
        
        \draw[thick] (-2.5, 1.5) -- (-3.5, 0);
        \draw[thick] (-3.5, 0) -- (-2.5, -1.5);
        \draw[thick] (3, 1.5) -- (3, -1.5);
      \end{tikzpicture}
      &
      \raisebox{-0.15cm}{\small $\hat{a}^\dagger(x_1)\hat{a}^\dagger(x_2)\hat{a}(x_2)\hat{a}(x_1)$}
      &
      \begin{tikzpicture}[scale=0.25, baseline=(current bounding box.center)]
        \draw[thick] (-2, 1) -- (2, 1); % Top line
        \draw[thick] (-2, 0) -- (2, 0); % Middle line
        \draw[thick] (-2, -1) -- (2, -1); % Bottom line 
        
        \draw[thick] (2.5, 1.5) -- (3.5, 0);
        \draw[thick] (3.5, 0) -- (2.5, -1.5);
        \draw[thick] (-3, 1.5) -- (-3, -1.5);
        \draw[thick, decorate,
        decoration={
                    snake,
                    amplitude=0.5mm,
                    segment length=1mm
        }] (-1, 1) -- (-1, 0); % Wavy line
        \fill[black] (-1, 1) circle (4pt) (-1, 0) circle (4pt);
      \end{tikzpicture}
      &
      \raisebox{-0.15cm}{\large / $\left(\expval{\hat{a}^\dagger (x_1)\hat{a}(x_1)}\expval{\hat{a}^\dagger (x_2)\hat{a}(x_2)}\right)$}
    \end{tabular}
  }
  
  \vspace{1em}
  
  % Second row
  \subfloat[The normalised second-order expectation value of four-photon diagram with two-photon interaction and one lost individual-scattering photon.\label{fig:g2_thr_ph_row2}]{%
    \begin{tabular}{@{}c@{}c@{}c@{\hspace{0.5em}}c@{}}
      \begin{tikzpicture}[scale=0.25, baseline=(current bounding box.center)]
        \draw[thick] (-2, 1) -- (1, 1); % Top line
        \draw[thick] (-2, 0) -- (2, 0); % Middle line
        \draw[thick] (-2,-1) -- (2,-1); % Bottom line
        
        \draw[thick] (-2.5, 1.5) -- (-3.5, 0);
        \draw[thick] (-3.5, 0) -- (-2.5, -1.5);
        \draw[thick] (3, 1.5) -- (3, -1.5);
        
        \fill[black] (1, 1) circle (4pt);
        \node[cross out, draw, line width=1.2pt, minimum size=1mm] at (1, 1) {};
      \end{tikzpicture}
      &
      \raisebox{-0.15cm}{\small $\hat{a}^\dagger(x_1)\hat{a}^\dagger(x_2)\hat{a}(x_2)\hat{a}(x_1)$}
      &
      \begin{tikzpicture}[scale=0.25, baseline=(current bounding box.center)]
        \draw[thick] (-1, 1) -- (2, 1); % Top line
        \draw[thick] (-2, 0) -- (2, 0); % Middle line
        \draw[thick] (-2, -1) -- (2, -1); % Bottom line
        
        \draw[thick] (2.5, 1.5) -- (3.5, 0);
        \draw[thick] (3.5, 0) -- (2.5, -1.5);
        \draw[thick] (-3, 1.5) -- (-3, -1.5);
        \draw[thick, decorate,
        decoration={
                    snake,
                    amplitude=0.5mm,
                    segment length=1mm
        }] (-1, 0) -- (-1, -1); % Wavy line
        
        \fill[black] (-1, 1) circle (4pt) (-1, 0) circle (4pt) (-1, 0) circle (4pt);
        \node[cross out, draw, line width=1.2pt, minimum size=1mm] at (-1, 1) {};
      \end{tikzpicture}
      &
      \raisebox{-0.15cm}{\large / $\left(\expval{\hat{a}^\dagger (x_1)\hat{a}(x_1)}\expval{\hat{a}^\dagger (x_2)\hat{a}(x_2)}\right)$}
    \end{tabular}
  }
  
  \vspace{1em}
  
  % Third row
  \subfloat[The normalised second-order expectation value of three-photon diagram with two-photon interaction and one lost interacting photon.\label{fig:g2_thr_ph_row3}]{%
    \begin{tabular}{@{}c@{}c@{}c@{\hspace{0.5em}}c@{}}
      % First diagram
      \begin{tikzpicture}[scale=0.25, baseline=(current bounding box.center)]
        \draw[thick] (-2, 1) -- (1, 1); % Top line
        \draw[thick] (-2, 0) -- (2, 0); % Middle line
        \draw[thick] (-2,-1) -- (2,-1); % Bottom line
        
        \draw[thick] (-2.5, 1.5) -- (-3.5, 0);
        \draw[thick] (-3.5, 0) -- (-2.5, -1.5);
        \draw[thick] (3, 1.5) -- (3, -1.5);
        \fill[black] (1, 1) circle (4pt);
        \node[cross out, draw, line width=1.2pt, minimum size=1mm] at (1, 1) {};
      \end{tikzpicture}
      &
      \raisebox{-0.15cm}{\small $\hat{a}^\dagger(x_1)\hat{a}^\dagger(x_2)\hat{a}(x_2)\hat{a}(x_1)$}
      &
      % Second diagram
      \begin{tikzpicture}[scale=0.25, baseline=(current bounding box.center)]
        \draw[thick] (-2, 1) -- (2, 1); % Top line
        \draw[thick] (-2, 0) -- (2, 0); % Middle line
        \draw[thick] (-2, -1) -- (2, -1); % Bottom line
        
        \draw[thick] (2.5, 1.5) -- (3.5, 0);
        \draw[thick] (3.5, 0) -- (2.5, -1.5);
        \draw[thick] (-3, 1.5) -- (-3, -1.5);
  
        \draw[thick, decorate,
        decoration={
                    snake,
                    amplitude=0.5mm,
                    segment length=1mm
        }] (-1, 1) -- (-1, 0); % Wavy line
  
        \fill[black] (-1, 1) circle (4pt) (-1, 0) circle (4pt) (-2, 1) circle (4pt);
        \node[cross out, draw, line width=1.2pt, minimum size=1mm] at (-2, 1) {};
      \end{tikzpicture}
      &
      \raisebox{-0.15cm}{\large /$\left(\expval{\hat{a}^\dagger (x_1)\hat{a}(x_1)}\expval{\hat{a}^\dagger (x_2)\hat{a}(x_2)}\right)$}
    \end{tabular}
  }
  
  \caption{Three-photon diagrams contributing to $g^{(2)}(x_1,x_2)$ at order $\mathcal{O}(\beta P_{\rm in}/\Gamma_{\rm tot})$}
  \label{fig:g2_thr_ph_interacting_diag_for_g2}
\end{figure}
The expression for Fig.~\ref{fig:g2_thr_ph_row1} is,
\begin{eqnarray} \label{eq:g2_thr_ph_row1}
    3\times &\left(\frac{1}{3!}\right)^2& e^{-|\alpha|^2} \frac{|\alpha|^6}{L^3} \bra{0} \int dy_1' dy_2' dy_3'\, t_0^{3M}\hat{a}(y_1') \hat{a}(y_2') \hat{a}(y_3') \hat{a}^\dagger (x_1) \hat{a}^\dagger (x_2) \hat{a}(x_2) \hat{a} (x_1) \nonumber \\
    &\times& \int dy_1 dy_2 dy_3\, \hat{a}^\dagger (y_1) \hat{a}^\dagger (y_2) \hat{a}^\dagger (y_3) t_0^{M} \varphi(y_2,y_3)\ket{0}/\left(\expval{\hat{a}^\dagger (x_1)\hat{a}(x_1)}\expval{\hat{a}^\dagger (x_2)\hat{a}(x_2)}\right) \nonumber\\
    &=& \frac{e^{-|\alpha|^2}P_{\rm in}^3 t_0^{4 M}\left[L \varphi(x_1,x_2)+\int dy\,\bigl( \varphi(y,x_1)+\varphi(y,x_2)\bigr)\right]}{\expval{\hat{a}^\dagger (x_1)\hat{a}(x_1)}\expval{\hat{a}^\dagger (x_2)\hat{a}(x_2)}}
\end{eqnarray}
The expression for Fig.~\ref{fig:g2_thr_ph_row2} is,
\begin{eqnarray} \label{eq:g2_thr_ph_row2}
    3^2\times &\left(\frac{1}{3!}\right)^2& \frac{|\alpha|^6}{L^3} \bra{0} t_0^{2M} \sum_{m'=0}^{M-1} \int dy_1' dy_2' dy_3'\,r_0 t_0^{m'} \hat{b}_{m'+1}(y_1')\hat{a}(y_2')\hat{a}(y_3') \hat{a}^\dagger (x_1) \hat{a}^\dagger (x_2) \hat{a}(x_2) \hat{a} (x_1) \nonumber\\
    &\times& \int dy_1 dy_2 dy_3 \sum_{m=0}^{M-1} \hat{b}_{m+1}^\dagger (y_1) \hat{a}^\dagger(y_2) \hat{a}^\dagger(y_3) r_0 t_0^{m}\varphi(y_2,y_3)\ket{0} /\left( \expval{\hat{a}^\dagger (x_1)\hat{a}(x_1)}\expval{\hat{a}^\dagger (x_2)\hat{a}(x_2)}\right)\nonumber\\
    &=& \frac{e^{-|\alpha|^2} P_{\rm in}^3 L t_0^{2M}\varphi(x_1,x_2) \sum_{m=0}^{M-1} r_0^2 t_0^{2m}}{\expval{\hat{a}^\dagger (x_1)\hat{a}(x_1)}\expval{\hat{a}^\dagger (x_2)\hat{a}(x_2)}}   
\end{eqnarray}

Again, the first term in the eq.(\ref{eq:g2_thr_ph_row1}) and the eq.(\ref{eq:g2_thr_ph_row2}) are part of the series of the dressed diagrams which will be canceled out. The surviving correction terms come from the second term in (\ref{eq:four_ph_row1}) and the second term in (\ref{eq:g2_thr_ph_row1}). They do not scale with $L$ and do not cancel out with the normalization factor.

The expression of Fig.~\ref{fig:g2_thr_ph_row3} reads
\begin{eqnarray} \label{eq:g2_thr_ph_row3}
    3\times 6\times &\left(\frac{1}{3!}\right)^2&  e^{-|\alpha|^2} \frac{|\alpha|^6}{L^3} \bra{0}\int dy_1' dy_2' dy_3'\, t_0^{2M} \sum_{m'=0}^{M-1} r_0 t_0^{m'} \hat{b}_{m'+1}(y_1')\hat{a}(y_2')\hat{a}(y_3')\hat{a}^\dagger(x_1)\hat{a}^\dagger(x_2)\hat{a}(x_2)\hat{a}(x_1) \nonumber\\
    &\times& \int dy_1 dy_2 dy_3 \, t_0^{M} \sum_{m=0}^{M-1}\varphi_m(\slashed{y}_1,y_2) \hat{b}_{m+1}^\dagger (y_1)\hat{a}^\dagger(y_2)\hat{a}^\dagger(y_3)\ket{0}/\left(\expval{\hat{a}^\dagger (x_1)\hat{a}(x_1)}\expval{\hat{a}^\dagger (x_2)\hat{a}(x_2)}\right)\nonumber\\
    &=& \frac{e^{-|\alpha|^2}P_{\rm in}^3 t_0^{3M}\sum_{m=0}^{M-1} r_0 t_0^{m}\int dy\,\bigl[\varphi_{m+1}(\slashed{y},x_1)+\varphi_{m+1}(\slashed{y},x_2)\bigr]}{\expval{\hat{a}^\dagger (x_1)\hat{a}(x_1)}\expval{\hat{a}^\dagger (x_2)\hat{a}(x_2)}}
\end{eqnarray}
By following the similar calculation as we did for $g^{(3)}$ in Eq.~(\ref{eq:corrected g3}), the correction term at $\mathcal{O}(\beta P_{\rm in}/\Gamma_{\rm tot})$ for $g^{(2)}$ also cancels. Therefore, the increased driving power at  $\mathcal{O}(P_{\rm in}/\Gamma_{\rm tot})=\mathcal{O}(\beta)$ makes no change to $g_c^{(3)}$ at $\mathcal{O}(\beta P_{\rm in}/\Gamma_{\rm tot})$.

\section{\label{apx:DXDXDX}Quadrature Cumulant Operator}
In this section, we derive the relation (\ref{eq:DXDXDX relation}) in the main text. We first expand the left-hand side of (\ref{eq:DXDXDX relation}) by definition of $\Delta \hat{X}_{\theta}(t)$ and $\hat{X}_{\theta}(t)$
\begin{eqnarray} \label{eq:DXDXDX first step}
  \expval{:\Delta \hat{X}_{\theta}(x_1) \Delta \hat{X}_{\theta}(x_2) \Delta \hat{X}_{\theta}(x_3):} =&& \expval{: \hat{X}_{\theta}(x_1)  \hat{X}_{\theta}(x_2)  \hat{X}_{\theta}(x_3):} -\expval{\hat{X}_{\theta}(x_1)}\expval{:\Delta \hat{X}_{\theta}(x_2) \Delta \hat{X}_{\theta}(x_3):} \nonumber\\ 
  -&&\expval{\hat{X}_{\theta}(x_2)}\expval{:\Delta \hat{X}_{\theta}(x_1) \Delta \hat{X}_{\theta}(x_3):}
  -\expval{\hat{X}_{\theta}(x_3)}\expval{:\Delta \hat{X}_{\theta}(x_1) \Delta \hat{X}_{\theta}(x_2):}
\end{eqnarray}
Recall that we are in the weak coherent driving. Previous work \cite{Mahmoodian2021} has shown that 
\begin{eqnarray} \label{eq:X and DXDX}
  \expval{\hat{X}_{\theta}(x)} \approx\,&& \sqrt{P_{\rm in}} \operatorname{Re}(e^{i\theta}t_{0}^M)\nonumber\\
  \expval{:\Delta \hat{X}_{\theta}(x_1) \Delta \hat{X}_{\theta}(x_2):} \approx&& \frac{P_{\rm in}}{2} \operatorname{Re}(e^{2i\theta}\phi(x_1,x_2))
\end{eqnarray}

We note that the symbol $\approx$, used here and throughout this section, indicates that higher-order correction terms of $\mathcal{O}(\beta\frac{P_{\rm in}}{\Gamma_{\rm tot}})$ beyond the leading order are omitted. We now wish to find out the three-photon leading-order term with the prefactor $P_{\rm in}^{3/2}$ in $\expval{:\Delta \hat{X}_{\theta}(x_1) \Delta \hat{X}_{\theta}(x_2) \Delta \hat{X}_{\theta}(x_3):}$. We find from the second to the fourth term in the expression (\ref{eq:DXDXDX first step}) already have the desired prefactor if we substitute the approximation (\ref{eq:X and DXDX}) into (\ref{eq:DXDXDX first step}). We here just need to work out the leading order term in $\expval{: \hat{X}_{\theta}(x_1)  \hat{X}_{\theta}(x_2)  \hat{X}_{\theta}(x_3):}$. By the definition of quadrature operator, we have:
\begin{eqnarray*}
  \expval{: \hat{X}_{\theta}(x_1)  \hat{X}_{\theta}(x_2)  \hat{X}_{\theta}(x_3):} =&& \frac{1}{8}\Bigl[e^{3i\theta}\expval{\hat{a}(x_1)\hat{a}(x_2)\hat{a}(x_3)}+e^{i\theta}\Bigl(\expval{\hat{a}^\dagger (x_1)\hat{a}(x_2)\hat{a}(x_3)}+\expval{\hat{a}^\dagger (x_2)\hat{a}(x_1)\hat{a}(x_3)}\\
  +&& \expval{\hat{a}^\dagger (x_3)\hat{a}(x_1)\hat{a}(x_2)}\Bigr)\Bigr]+c.c.
\end{eqnarray*}
where $\hat{a}(x)$ is the annihilation operator for the outgoing steady state field after scattering. In the following text, we temporarily use a simplified notation: $\ket{m_a;n_b}$ to label the outgoing state with $m$ photons in the propagating channel and $n$ photons in all $M$ loss channels. For the first term in the bracket, we see its leading order term is at $\mathcal{O}\left((P_{\rm in}/\Gamma_{\rm tot})^{3/2}\right)$
\begin{eqnarray} \label{eq:aaa}
  \expval{\hat{a}(x_1)\hat{a}(x_2)\hat{a}(x_3)}=&&\mel{0_a;0_b}{\hat{a}(x_1)\hat{a}(x_2)\hat{a}(x_3)}{3_a;0_b}+\mathcal{O}\left((P_{\rm in}/\Gamma_{\rm tot})^2\right) \\
  \approx&& P_{\rm in}^{3/2}\psi_3(x_1,x_2,x_3) \nonumber\\
  =&& P_{\rm in}^{3/2}\Bigl\{t_0^{3M} + t_0^{M} [\phi_2(x_1,x_2) + \phi_2(x_1,x_3) + \phi_2(x_2,x_3)] + \phi_3(x_1,x_2,x_3)\Bigr\} \nonumber
\end{eqnarray}
The leading order in the rest terms is also at $\mathcal{O}(P_{\rm in}^{3/2}/\Gamma_{\rm tot}^{3/2})$. 
\begin{eqnarray} \label{eq:adaa}
  \expval{\hat{a}^\dagger(x_1)\hat{a}(x_2)\hat{a}(x_3)}=&&\mel{1_a;0_b}{\hat{a}^\dagger(x_1)\hat{a}(x_2)\hat{a}(x_3)}{2_a;0_b}+\mathcal{O}\left((P_{\rm in}/\Gamma_{\rm tot})^2\right)) \nonumber\\
  \approx && P_{\rm in}^{3/2} t_0^{M}\psi_2(x_2,x_3) = P_{\rm in}^{3/2}\Bigl\{ t_0^{3M}+t_0^M \phi_2(x_2,x_3)\Bigr\}
\end{eqnarray}
Substituting the expression (\ref{eq:X and DXDX})-(\ref{eq:adaa}) into the expression (\ref{eq:DXDXDX first step}), we obtain the relation (\ref{eq:DXDXDX relation}) in the main text.
\begin{equation*} 
  \expval{:\Delta \hat{X}_\theta (x_1) \Delta \hat{X}_\theta (x_2)\Delta \hat{X}_\theta (x_3):}\approx\frac{P_{\rm in}^{3/2}}{4}\operatorname{Re}[e^{i3\theta} \phi_3(x_1,x_2,x_3)]
\end{equation*}

Apart from the analytical expression above, the expectation value $\expval{:\Delta \hat{X}_{\theta}(x_1) \Delta \hat{X}_{\theta}(x_2) \Delta \hat{X}_{\theta}(x_3):}$ also can be computed by numerical simulations. Since it is expanded in terms of the product of the product of creation and annihilation operators, such as  $\expval{\hat{a}(x_1)\hat{a}(x_2)\hat{a}(x_3)}$. Using the input-output relation $\hat{a}_{\rm out}(x)=\hat{a}_{\rm in}(x)-i\sqrt{\beta \Gamma_{\rm tot}}\sum_{m=1}^M \hat{\sigma}_m^-$, we can further expand $\expval{\hat{a}(x_1)\hat{a}(x_2)\hat{a}(x_3)}$ in terms of the product of $\hat{a}_{\rm in}(x)$ and $\hat{\sigma}_m^-(x)$. Since the input is a coherent field, the action of $\hat{a}_{\rm in}$ on the steady state gives $\hat{a}_{\rm in}(x)\ket{\text{out}}=\frac{\alpha}{L}\ket{\text{out}}$. Therefore, the product like $\expval{\hat{a}(x_1)\hat{a}(x_2)\hat{a}(x_3)}$ is just a linear combination of the expectation value of the product of $\hat{\sigma}_m^\pm (x)$'s, which can be computed by quantum regression theorem (QRT) with the master equation derived in~\cite{Mahmoodian2023},
\begin{eqnarray*}
  \frac{1}{\Gamma_{\rm tot}} \frac{d\rho_N}{dt} &= -i \sum_{j=1}^{N} \sqrt{\frac{P_{\text{in}}}{P_{\text{sat}}}} \left[ \hat{\sigma}_j^- + \hat{\sigma}_j^+, \rho_N \right] + (1 - \beta) \sum_{j=1}^{N} D \left[ \hat{\sigma}_j^- \right] \rho_N \\
  &\quad + \frac{\beta}{2} \sum_{\substack{j,l=1 \\ j>l}}^{N} \left[ \hat{\sigma}_l^+ \hat{\sigma}_j^- - \hat{\sigma}_j^+ \hat{\sigma}_l^-, \rho_N \right]  + \beta D \left[ \sum_{j=1}^{N} \hat{\sigma}_j^- \right] \rho_N.
\end{eqnarray*}
Here, the master equation is written in the rotating frame with respect to the input laser frequency $\omega_0$ and the decay operator is defined as $D[x] \rho = x \rho x^\dagger - \frac{1}{2} x^\dagger x \rho - \frac{1}{2} \rho x^\dagger x$.

\end{document}